\documentclass[useAMS,usenatbib,usegraphicx]{mn2e}
\usepackage{ txfonts, color, soul,ulem,rotating,subfig,fixltx2e,caption} 

    \def\independenT#1#2{\mathrel{\setbox0\hbox{$#1#2$}%
    \copy0\kern-\wd0\mkern4mu\box0}}

\title[000]{Coronal Structure of Low-Mass Stars}
\author[P. Lang et al]
{Pauline Lang$^{1}$\thanks{E-mail:
pl42@st-andrews.ac.uk},  Moira Jardine$^{1}$, Jean-Fran\c{c}ois Donati$^{2}$, Julien Morin$^{3,4}$, Aline Vidotto$^{1}$\\
1.SUPA, School of Physics \& Astronomy, University of St Andrews, North Haugh, St Andrews,  KY16 9SS, UK\\
2.IRAP-UMR 5277, CNR $\&$ Univ. de Toulouse, 14 Av. E. Belin, F-31400 Toulouse, France\\
3.Institut f$\ddot{u}$r Astrophysik, Friedrich-Hund-platz 1, 370777 G$\ddot{o}$ttingen\\
4.Dublin Institute for Advanced Studies, School of Cosmic Physics, 31 Fitzwilliam Place, Dublin 2, Ireland\\
}

\begin{document}

\date{Accepted 2012 May 9. Received 2012 May 9; in original form 2012 January 13}

\pagerange{\pageref{firstpage}--\pageref{lastpage}} \pubyear{2011}

\maketitle

\label{firstpage}

\begin{abstract}

We investigate the change in stellar magnetic topology across the fully-convective boundary and its effects on coronal properties.  We consider both the magnitude of the open flux that influences angular momentum loss in the stellar wind and X-ray emission measure.  We use reconstructed maps of the radial magnetic field at the stellar surface and the potential-field source surface method to extrapolate a 3D coronal magnetic field for a sample of early-to-mid M dwarfs.  During the magnetic reconstruction process it is possible to force a solution towards field geometries that are symmetric or antisymmetric about the equator but we demonstrate that this has only a modest impact on the coronal tracers mentioned above. We find that the dipole component of the field, which governs the large-scale structure, becomes increasingly strong as the stellar mass decreases, while the magnitude of the open (wind-bearing)  magnetic flux is proportional to the magnitude of the reconstructed magnetic flux.  By assuming a hydrostatic and isothermal corona we calculate X-ray emission measures (in magnitude and rotational modulation) for each star and, using observed stellar densities as a constraint, we reproduce the observed X-ray saturation at $Ro \le 0.1$.  We find that X-ray rotational modulation is not a good indicator of magnetic structure as it shows no trend with Rossby number but can be useful in discriminating between different assumptions on the field geometry.

\end{abstract}


\section{Introduction}\label{sec.Introduction}

Stellar coronal magnetic activity is predominantly investigated in the X-ray and radio bands.  A linear relationship between the X-ray luminosity, $L_X$, and the radio luminosity, $L_R$, established by \citet{Guedel_X-ray_Microwave_1993}, pointed to a close connection between the hot coronal plasma responsible for the thermal X-ray emission and the free electrons causing the non-thermal radio emission,

\begin{equation}
LogL_{X} \approx LogL_{R} + 15.5   .
\label{eq.Lx-Lr}
\end{equation}

While this linear relationship holds for several classes of active, main sequence star, between types F and early-M, and is also verified for T Tauri stars and solar flares, the assumption that the $L_X - L_R$ relationship would continue into the realm of the very low-mass stars proved to be short lived; the detection of radio emission from an M9 dwarf, LP944-20, defied this relation by almost four orders of magnitude \citep{Berger_Discovery_2001}.  Since the discovery of radio emission from LP944-20, many more observations have been carried out on low-mass stars to reveal that they also show the same radio phenomenon \citep{Berger_flaring_2002,Berger_Magnetic_2005,Burgasser_Quiescent_2005,Hallinan_RadioObservations_2006,McLean_RadioSurvey_2011}.

Radio observations are important diagnostic tools as they can help determine the magnetic field configuration \citep{Hallinan_ECMI_2008} and the nature of plasma emitting regions.  X-ray observations on the other hand help provide a deeper understanding of the generation of the magnetic field.  Originating principally in the magnetically confined plasma of the hot stellar corona - at temperatures greater than $10^{6}K$ \citep{Guedel_XrayReview_2004}- X-ray emissions prove useful for determining stellar parameters, such as the extent of the stellar corona, which also aids in establishing the structure and evolution of the magnetic field.  X-ray emission has shown to be prevalent throughout the main sequence, with the maximum value of X-ray luminosity observed for each spectral type decreasing with bolometric luminosity, (e.g. \citealt{Barbera_LxLbol_1993}) - for spectral type down to early-M.

K and M stars are the most numerous in the Galaxy, and coupled with their high levels of X-ray emission, they are prime candidates for investigating the dependence of X-ray emission on physical parameters such as mass, rotation rate, and more recently, magnetic topology. So far, the X-ray luminosity has been shown to correlate well with either rotational velocity or Rossby number (the ratio of the stellar rotation period, P, to the convective turnover time, $\tau_{c}$)  (e.g.,\citealt{Mangeney_XrayRotation_1984,Marilli_Chromospherid-coronal_1984,Schmidt_EinsteinXray_1985, Fleming_RelationXrayRotation_1989}); in general, $L_{X}/L_{bol}$ increases and then saturates ($L_{X}/L_{bol} \approx10^{-3}$; \citep{Delfosse_LxLbolM7_1998}) with increasing rotation rate.  This behaviour is attributed to coronal saturation \citep{Vilhu_Chromospheric_1987,Stauffer_radial_1994}; however, it is not clear whether saturation is a reflection of the dynamo itself or the total filling of the stellar surface with active regions \citep{Vilhu_Magnetic_1984}.\\
For a small sample of stars later than M5, the $L_{X}/L_{bol}$ ratio was also shown to decline once again at the lowest Rossby numbers within the sample stars (e.g. \citealt{Golub_Quiescent_1983,Bookbinder_PhDThesis_1985,Rosner_X-ray_1985,Jeffries_coronalSaturation_2011}) and  \citet{Berger_radio_2006} found that the value drops to $L_{X}/L_{bol} \approx10^{-3.5}$ at spectral type M7.  This indicates that outwith flaring, the emission from the thermal plasma cannot exceed a certain fraction of the total stellar flux.  This \textit{super-saturation} has been attributed to negative dynamo feedback \citep{Kitchatinov_NegativeFeedback_1994}, lack of coverage of active regions \citep{Stepien_Supersaturation_2001}, or centrifugal stripping of the corona \citep{Jardie_CoronalStripping_2004}.

It is important to note that there is a change in internal structure between the early-M dwarfs (dM) and mid-M dwarfs.  The former have a solar-like internal structure: a turbulent, electrically conducting convective envelope, surrounding a radiative core. The interface between these two zones i.e. \textit{the tachocline}, is the site where amplification of the magnetic field is believed to take place \citep{Spiegel_tachocline_1992}.  The relative size of the radiative core dramatically drops with decreasing temperature for early M dwarfs, and at spectral type M4 ($\approx 0.35M_\odot$), the stars are fully convective (e.g. \citealt{Chabrier_Structure_1997}).  This change in internal structure coincides with a change in magnetic topology (\citealt{Donati_EarlyM_2008,Morin_MidM_2008,Morin_LateM_2010} hereafter D08, M08 and M10, respectively).  In the higher mass stars, $M \ge 0.45M_\odot$, the field configuration is more complex, i.e. a non-axisymmetric poloidal field with a strong toroidal component, in comparison to the nature of the field structure at spectral types later than M4, which is mainly axisymmetric and poloidal.
  
As for magnetic activity, it is crucial to look at coronal properties such as emission measure, especially for fast rotators, to explore whether saturation and super-saturation occur at a fixed rotation period or Rossby number.  This could potentially help in determining the physical mechanism that is responsible for the saturation, i.e. an effect of the dynamo efficiency, or a consequence of the fast rotation rates exhibited by these stars.

Stellar models (e.g. \citealt{Gilliland_Chromospheric_1986,Kim_Rossby_1996}) suggest that the convective turnover time is longer in lower mass stars; if one considers the Sun and an M-dwarf with an equivalent rotation period, then the M dwarf has a smaller Rossby number and yet is more active in terms of $L_{X}/L_{bol}$.  Although coronal saturation occurs at $L_{X}/L_{bol} \approx10^{-3}$ for spectral type G,K and M, the rotation period at which it sets in is larger for the lower masses.  Therefore,  this results in a Rossby number of 0.1 being the point where coronal saturation sets in \citep{Patten_Evolution_1996,Pizzolato_ActivityRotation_2003,Jeffries_coronalSaturation_2011}.  However, \citet{Jeffries_coronalSaturation_2011} show that the super-saturation is more clear as a function of rotation period within each spectral type, rather than Rossby number, with the effect occurring at periods of $P \le 0.3$days for K-dwarfs and $P \le 0.2$days for M dwarfs.

Magnetic activity and rotation are well correlated \citep{Stauffer_Rotation-Activity_1997} and both are strongly dependent on stellar age for main-sequence solar-like stars \citep{Skumanich_AgeActivityRotation_1972}, 
\begin{equation}
\Omega_{0} \propto t^{-1/2}    ,
\label{eq.age-rotation}
\end{equation}
where $\Omega_{0}$ is stellar angular velocity.
During their lifetime, rotational evolution of stars can be governed by disc braking, pre-main-sequence contraction i.e. \textit{spin-up}, and/or magnetic winds.  From observations of young open clusters (e.g.\citealt{Irwin_rotation_2009}), it is clear that rapidly rotating dM stars are a common occurrence; however, in older open clusters, this number is reduced.  \citet{Scholz_RotationPeriods_2011} find a mass-dependent exponential rotational braking law 
\begin{equation}
P \propto exp[t/\tau]    ,
\label{eq.rotational-braking}
\end{equation}
where towards lower masses the spin-down timescale $\tau$ increases, with $\tau \approx 0.5$Gyr for $0.3M_{\odot}$ and $\tau > 1$Gyr for $0.1M_{\odot}$.  This indicates that at an age of 600Myr stars of $0.3M_{\odot}$ are almost exclusively fast rotators.

As well as X-ray emission, another coronal tracer of magnetic activity is radio emission.  The relation that correlates the X-ray luminosity with the radio luminosty (eqn. ~\ref{eq.Lx-Lr}), suggests that thermal X-ray emission, assumed to be due to hot coronal plasma, and non-thermal radio emission, possibly generated by the electron cyclotron maser instability \citep{Melrose_Dulk_EMC_1982} or gyrosynchrotron, are both consequences of the magnetic field.  Throughout the subtypes of dM stars, the radio luminosity remains approximately constant, while $L_{X}$ tracks the bolometric luminosity (i.e. $L_{X}/L_{bol} \approx10^{-3}$).
The sharp deviation from the $L_{X}-L_{R}$ relation, first noted by \citet{Berger_Discovery_2001}, does not present itself until beyond spectral subtype M7 \citep{Berger_Basri_2010}, where the relation evolves from $L_{R}/L_{X} \approx 10^{-15.5}$ to $\approx 10^{-11.5}$.  This would indicate that the deviation must not directly relate to the transition to full convection.  Along with this increase in radio luminosity, it is still unclear as to why the emission is present on these stars during one set of observations and then absent during the next (e.g. \citealt{Berger_Basri_2010}).

In order to investigate the role of field topology on the coronal structure and emission properties of M dwarfs, we use the observed surface magnetic field maps of their coronae.  From this we can predict the X-ray emission measure.

\section{Sample Selection}\label{sec.SampleSelection}

Spectropolarimetry along with multi-line extraction methods (e.g. least-squares deconvolution), has the power to detect Zeeman signatures in the polarisation spectra of low-mass stars.  With a significant improvement in instrumentation and imaging capabilities, we are now able to recover information on the distribution of large-scale magnetic fields on the surface of cool active stars \citep{Donati_DynamoProcesses_2003a}.  This has allowed for a much more in-depth investigation of dynamo processes in low-mass stars.  Using rotationally modulated circular polarisation signatures, the photospheric magnetic field can be reconstructed from a series of polarised spectra, through Zeeman-Doppler Imaging (ZDI).  The magnetic field is decomposed into its toroidal and poloidal components and described as a set of spherical harmonic coefficients \citep{Donati_Sph_harm_2006b}.

The first low-mass star on which this was attempted was the M4 dwarf V374 Peg, which exhibited a very strong, large-scale, mainly axisymmetric poloidal field \citep{Donati_LargeScale_2006a}, which is stable on timescales of $\approx 1yr$ \citep{Morin_V374Peg_2008a}.  This result lead D08, M08 and M10 to investigate dM stars further: a survey was conducted using data from ESPaDOnS and NARVAL on 23 active, main-sequence, dM stars spanning a broad range of masses and rotation rates.  The mass range embodied dM stars located both above and below the fully convective boundary and the rotation periods were derived using ZDI, which provides a constraint on the rotation period as well as the projected equatorial velocity $(v\sin i)$. Rossby number are from \citet{Donati_EarlyM_2008,Morin_MidM_2008} and were computed from empirical $\tau_{c}$ suited to the stellar mass from \citet{Kiraga_Stepien_MDwarfs_2007}.

In this paper, we concentrate on the M-dwarf sample with masses ranging from $0.25M_\odot$ to $0.75M_\odot$ (D08, M08), where $0.75M_\odot$ refers to a very young star GJ182, and compare results to stellar parameters such as mass and Rossby number as these can provide the best insight into dynamo action and efficiency.

\begin{figure*}
	\begin{center}
	\subfloat(a) GJ 182{\label{fig.edge-a}\includegraphics[width=44mm]{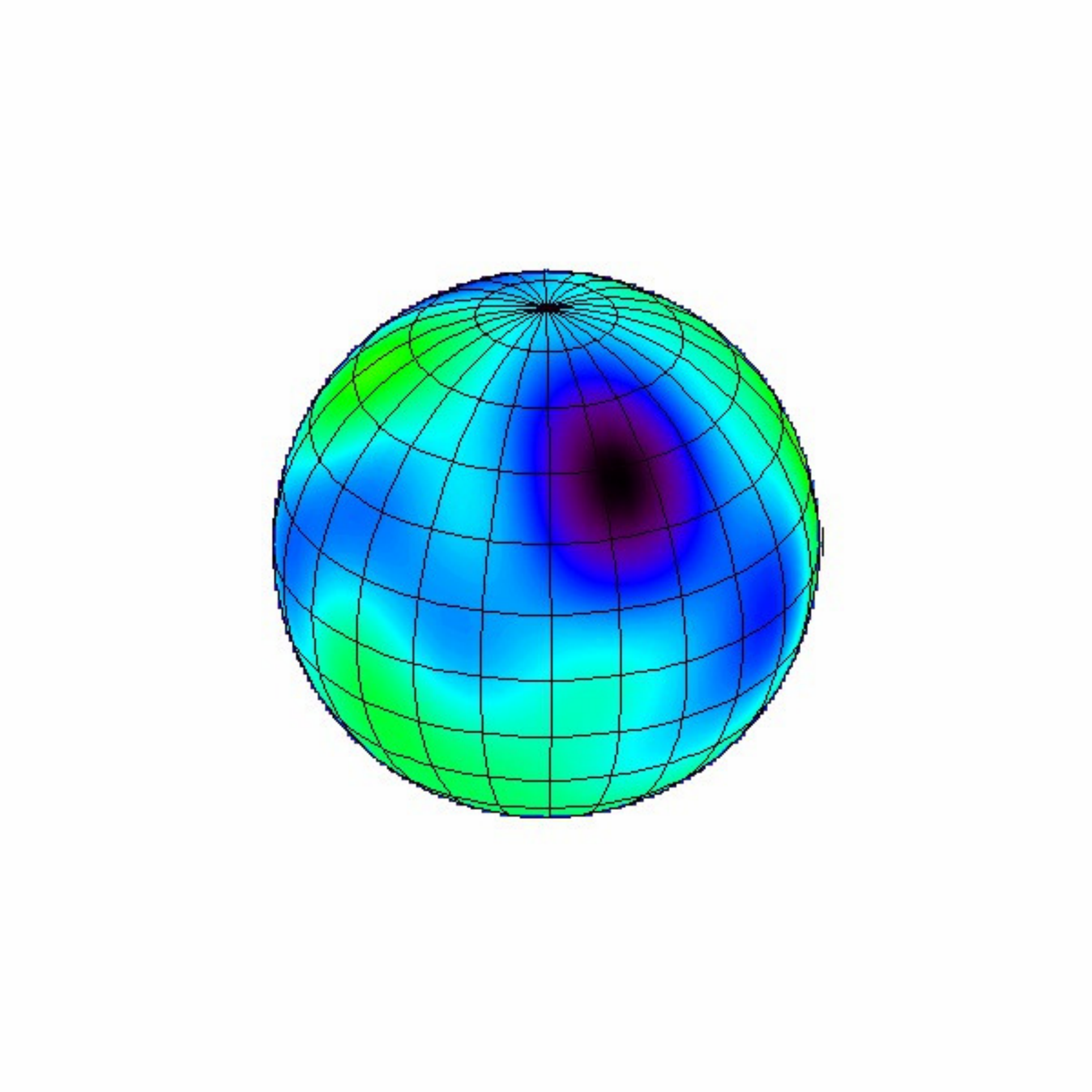}}
	\subfloat(b){\label{fig.edge-b}	\includegraphics[width=44mm]{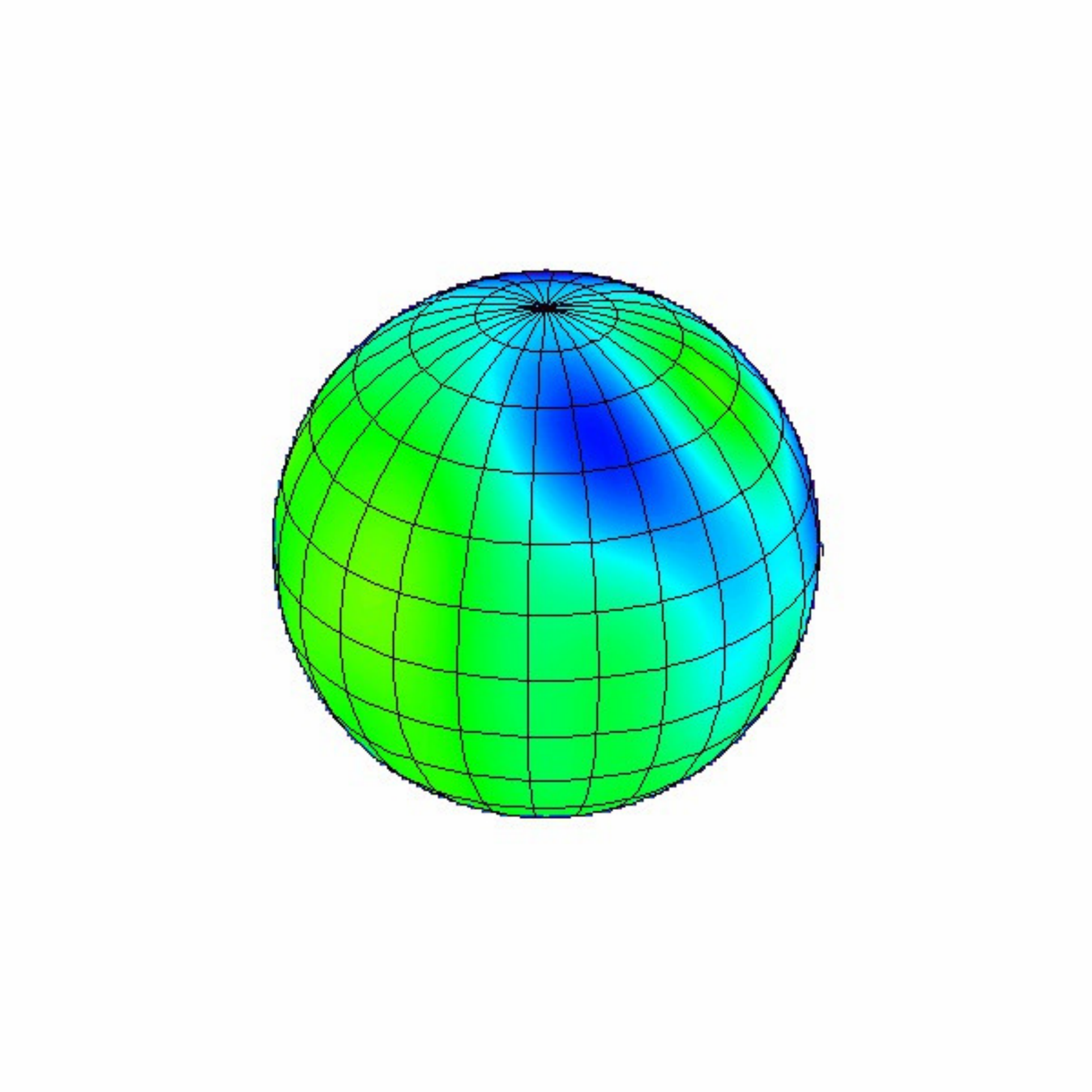}}
	\subfloat(c){\label{fig.edge-c}	\includegraphics[width=44mm]{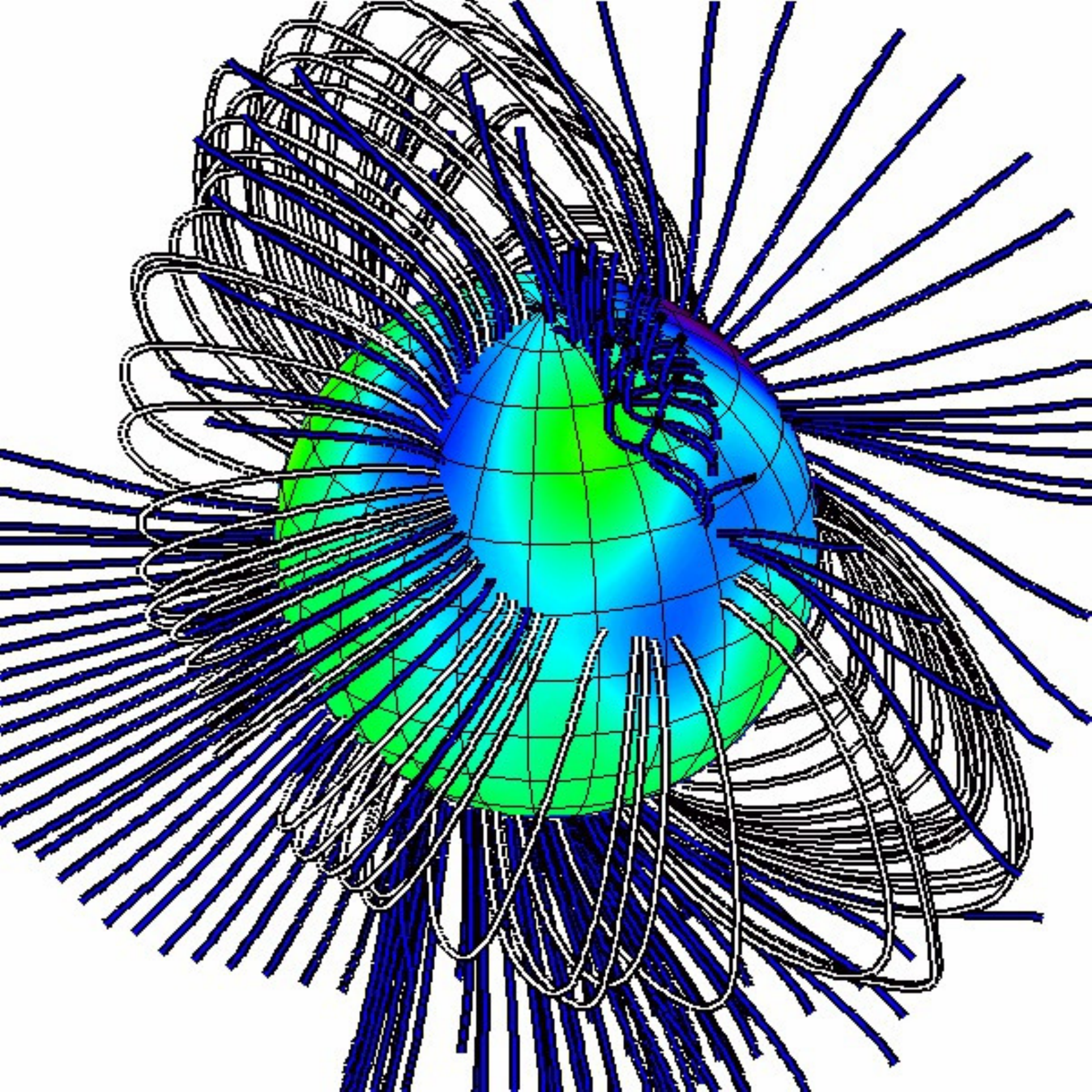}}\\
	\subfloat(a) DT Vir{\label{fig.edge-a}\includegraphics[width=44mm]{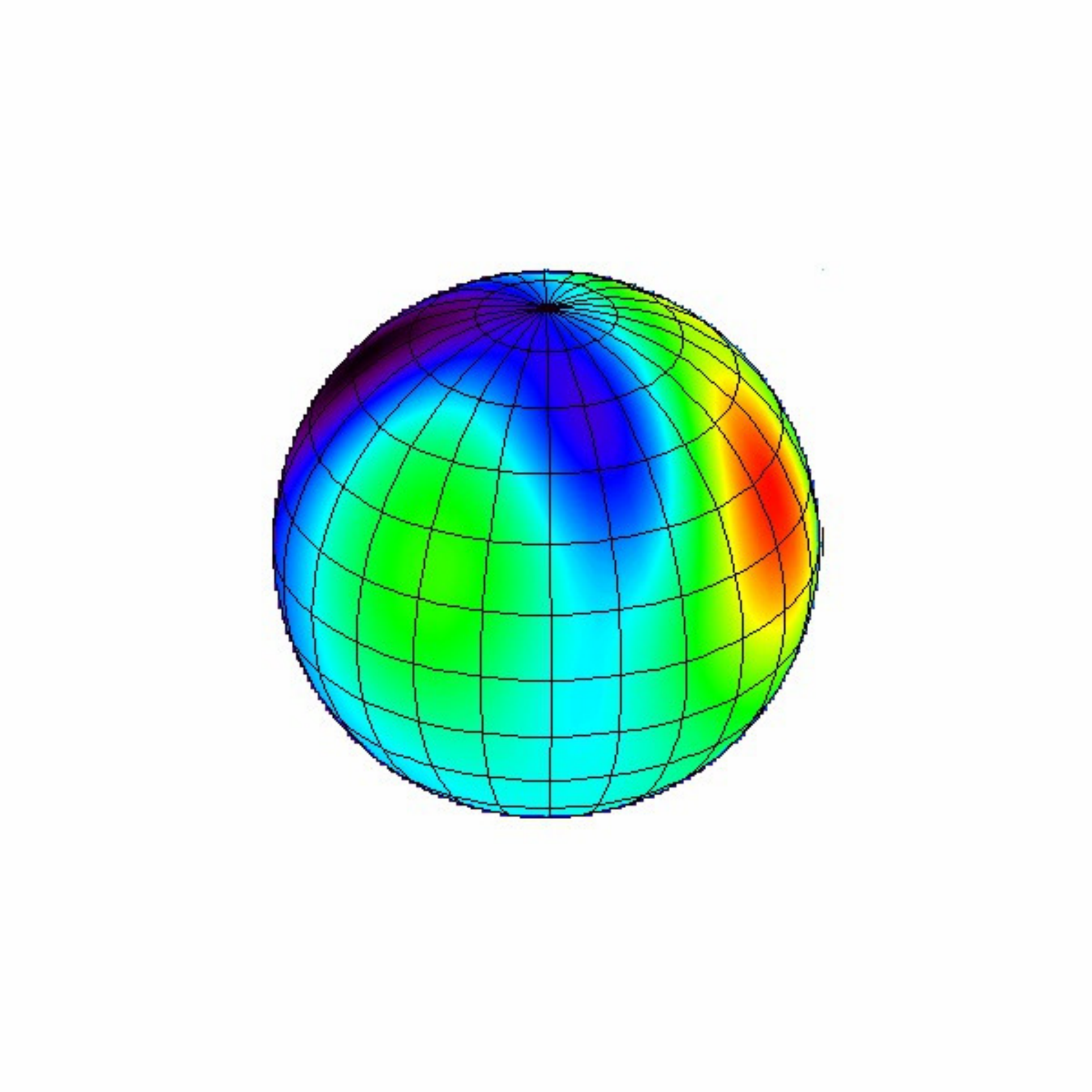}}
	\subfloat(b){\label{fig.edge-b}\includegraphics[width=44mm]{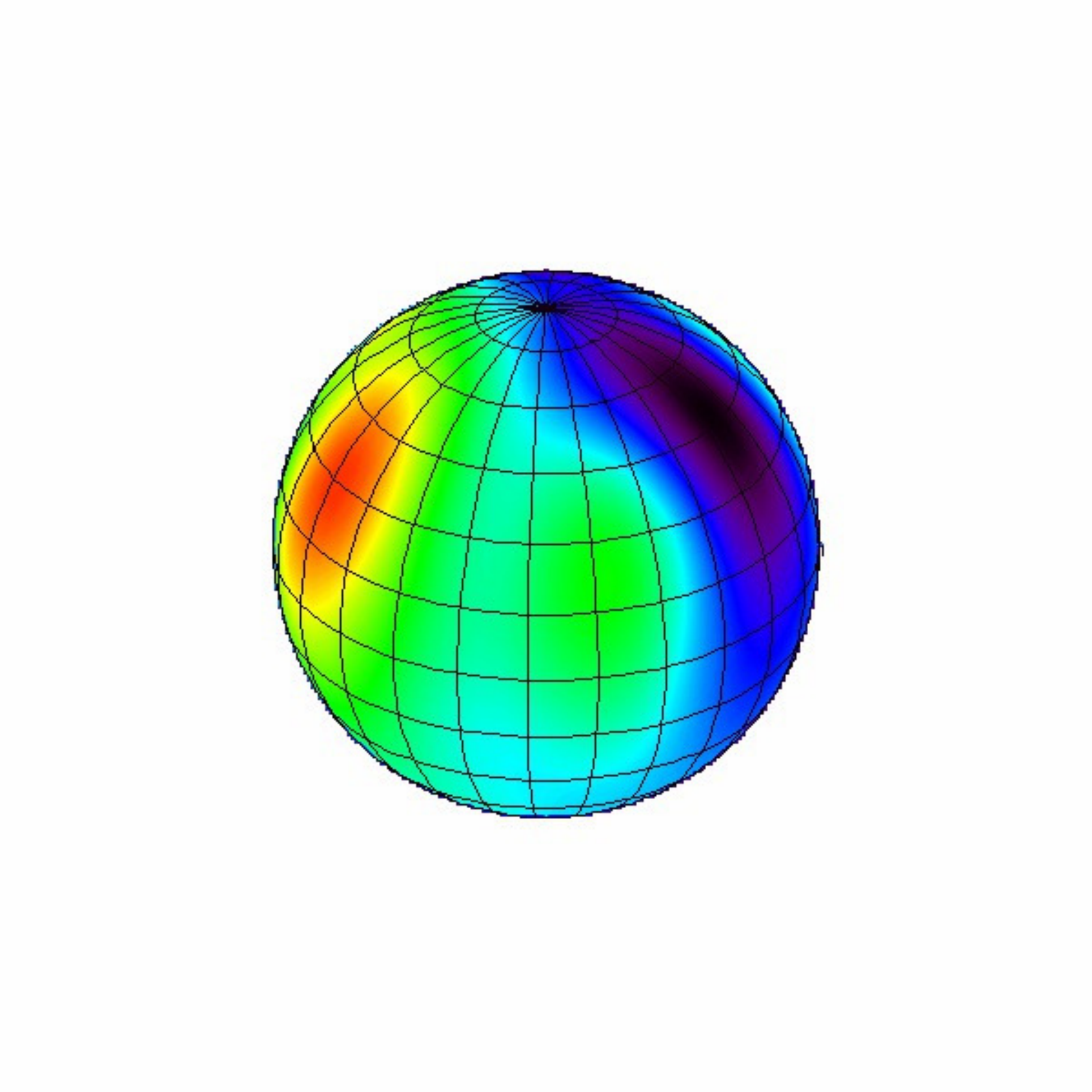}}
	\subfloat(c){\label{fig.edge-c}\includegraphics[width=44mm]{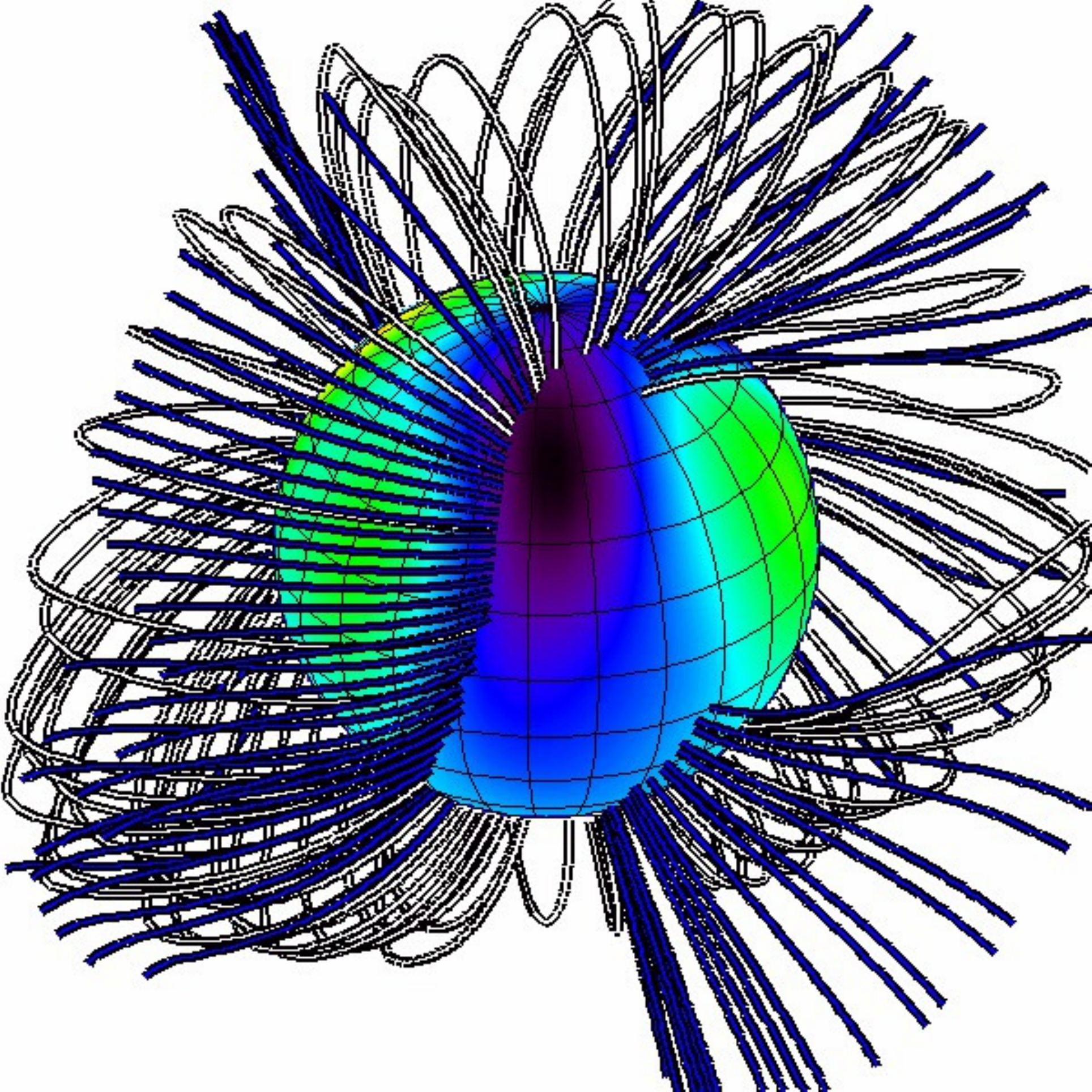}}\\
	\subfloat(a) DS Leo{\label{fig.edge-a}\includegraphics[width=44mm]{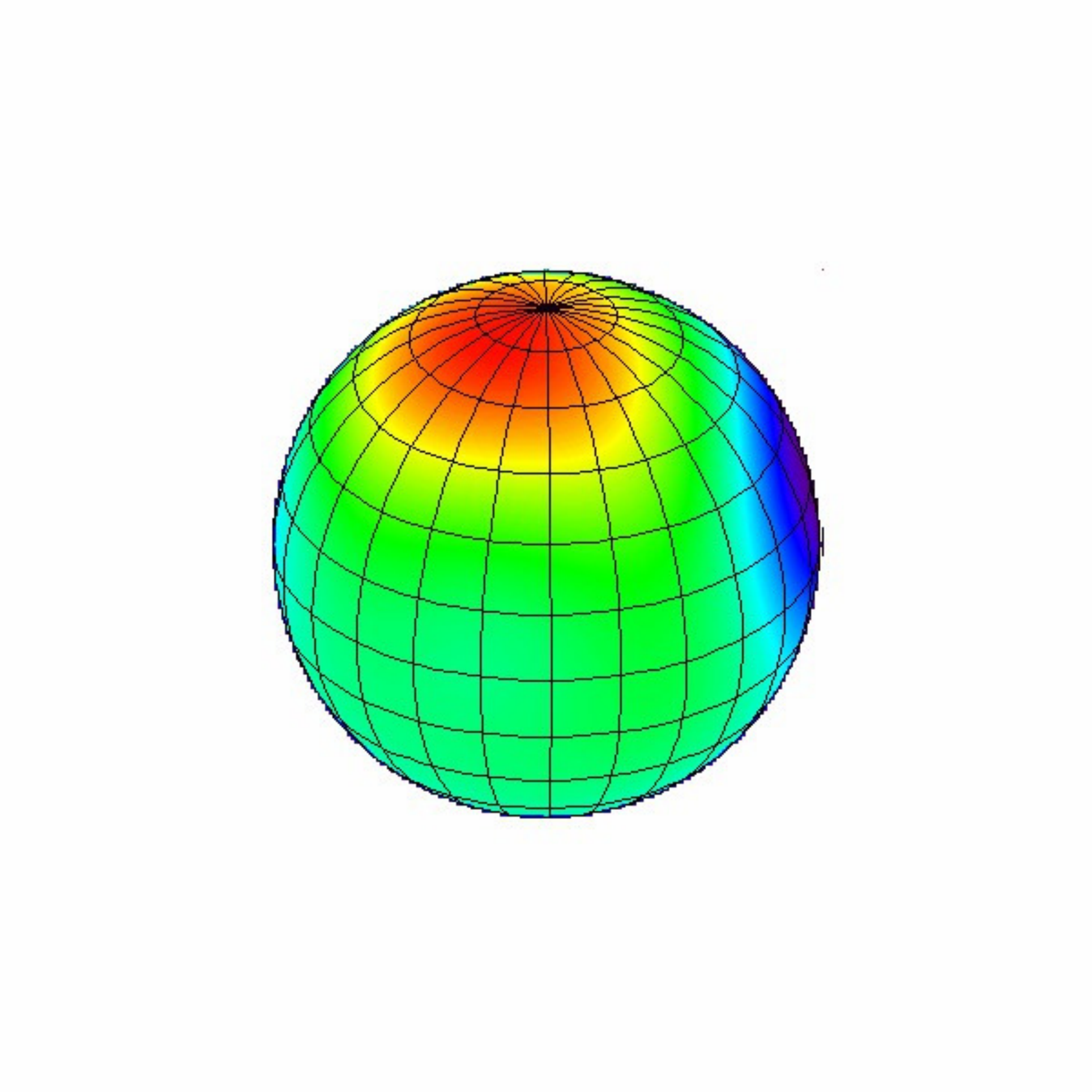}}
	\subfloat(b){\label{fig.edge-b}\includegraphics[width=44mm]{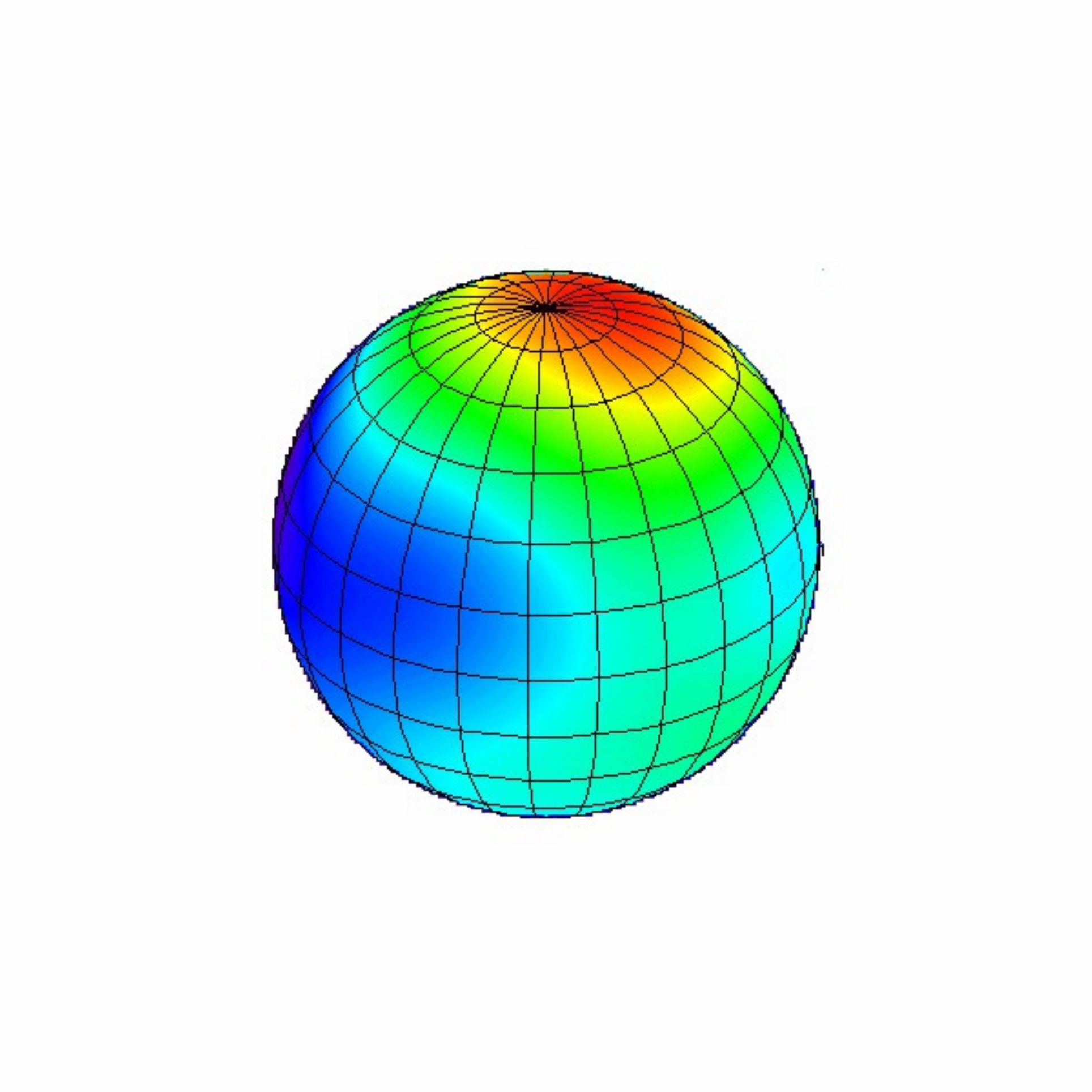}}
	\subfloat(c){\label{fig.edge-c}\includegraphics[width=44mm]{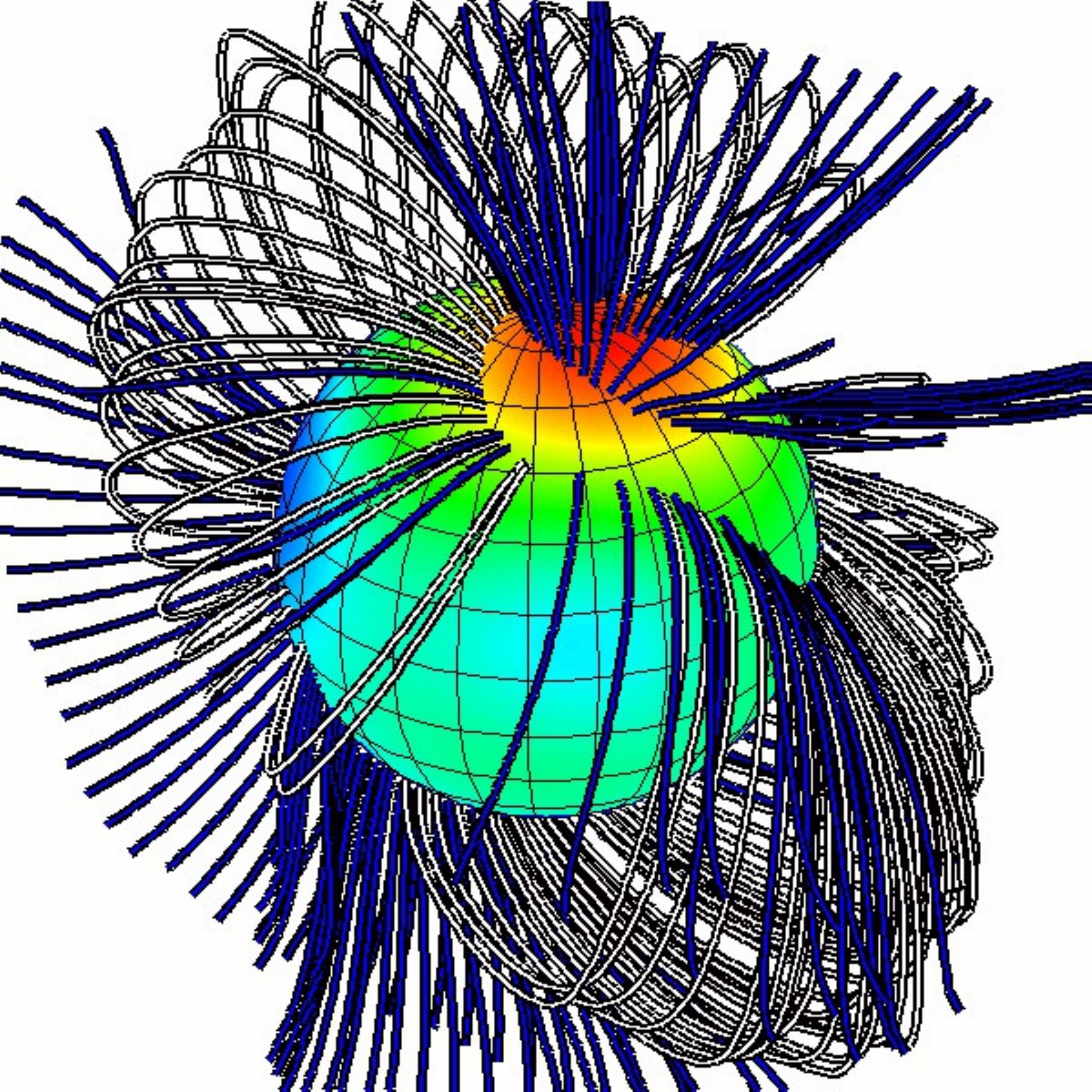}}\\
	\subfloat(a) GJ 49{\label{fig.edge-a}\includegraphics[width=44mm]{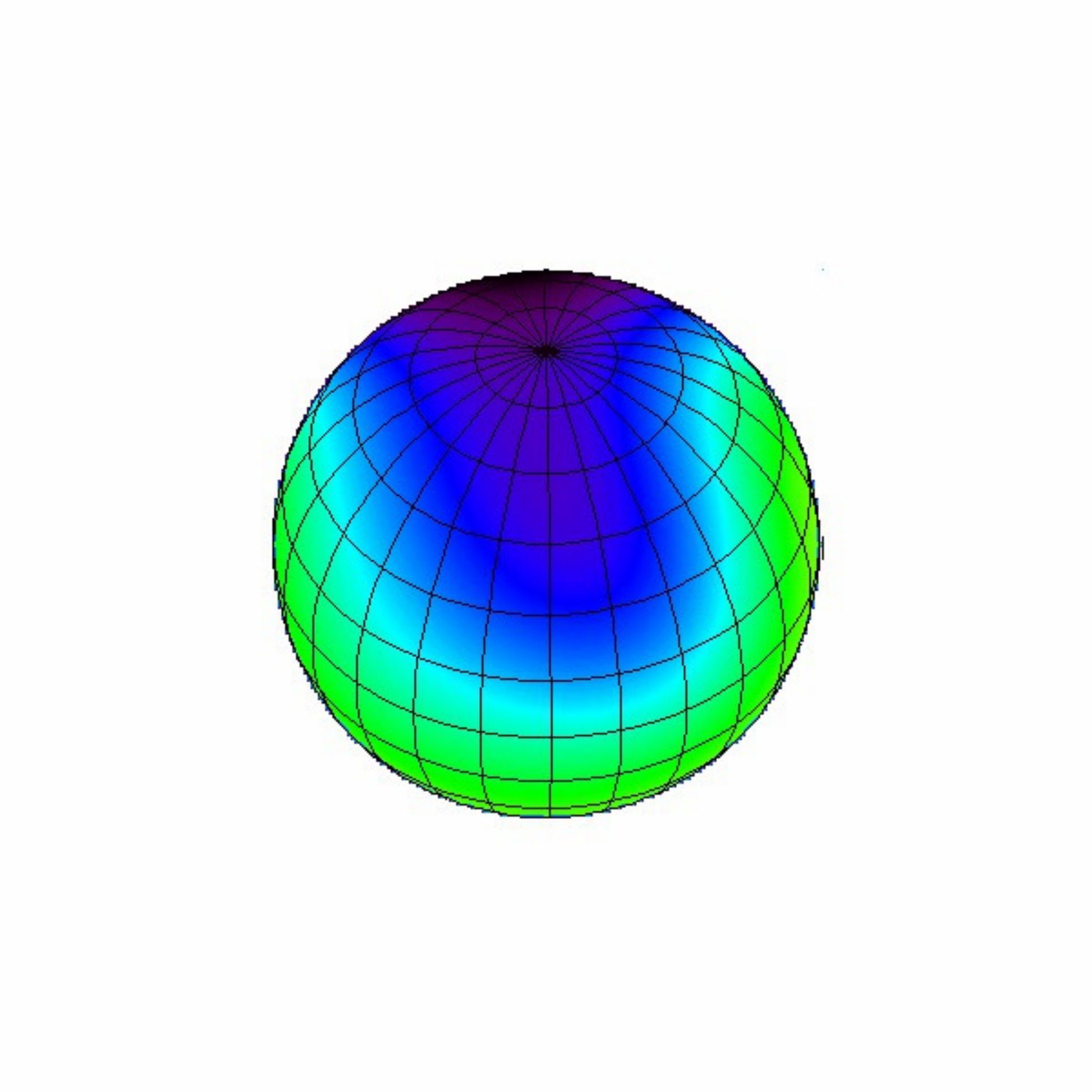}}
	\subfloat(b){\label{fig.edge-b}\includegraphics[width=44mm]{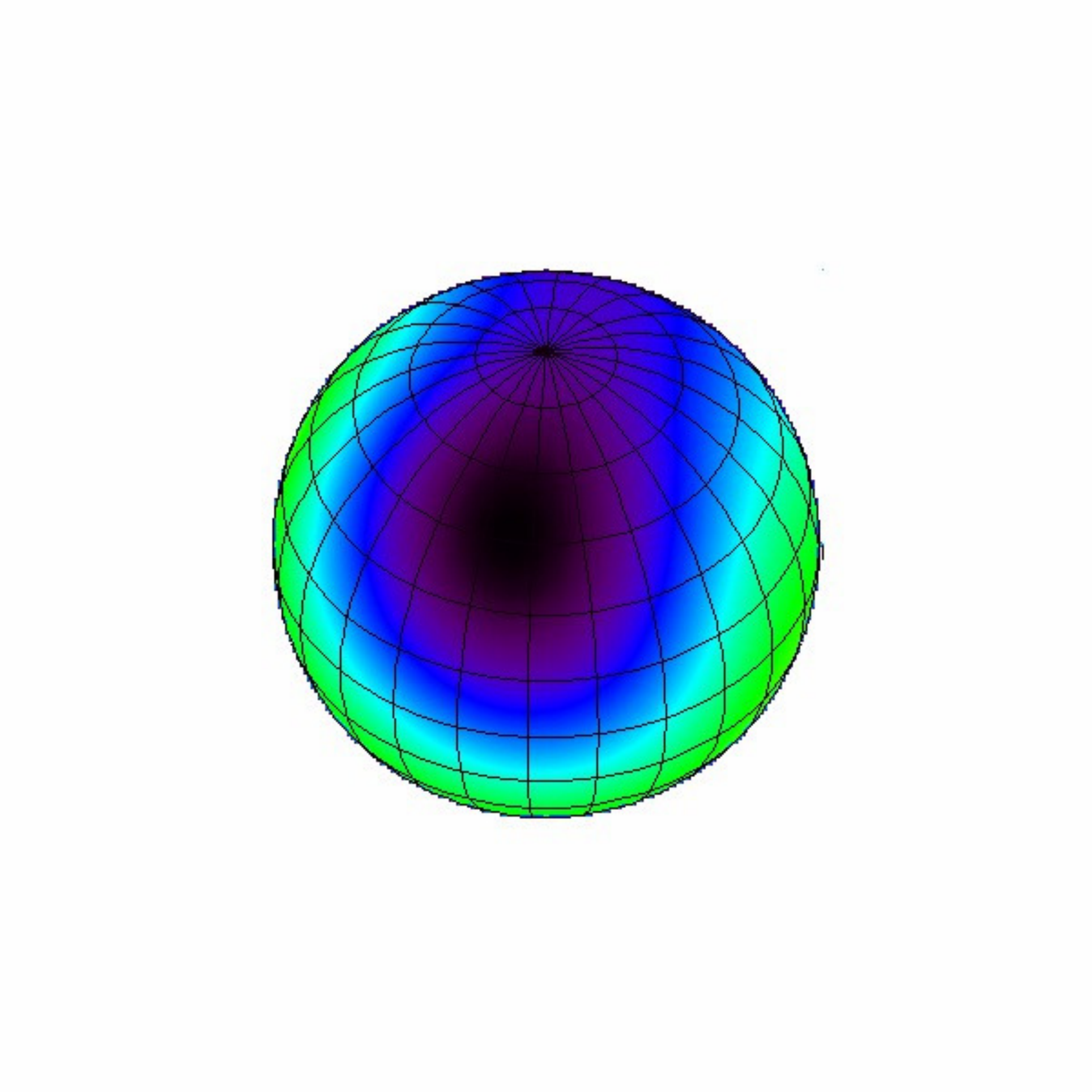}}
	\subfloat(c){\label{fig.edge-c}\includegraphics[width=44mm]{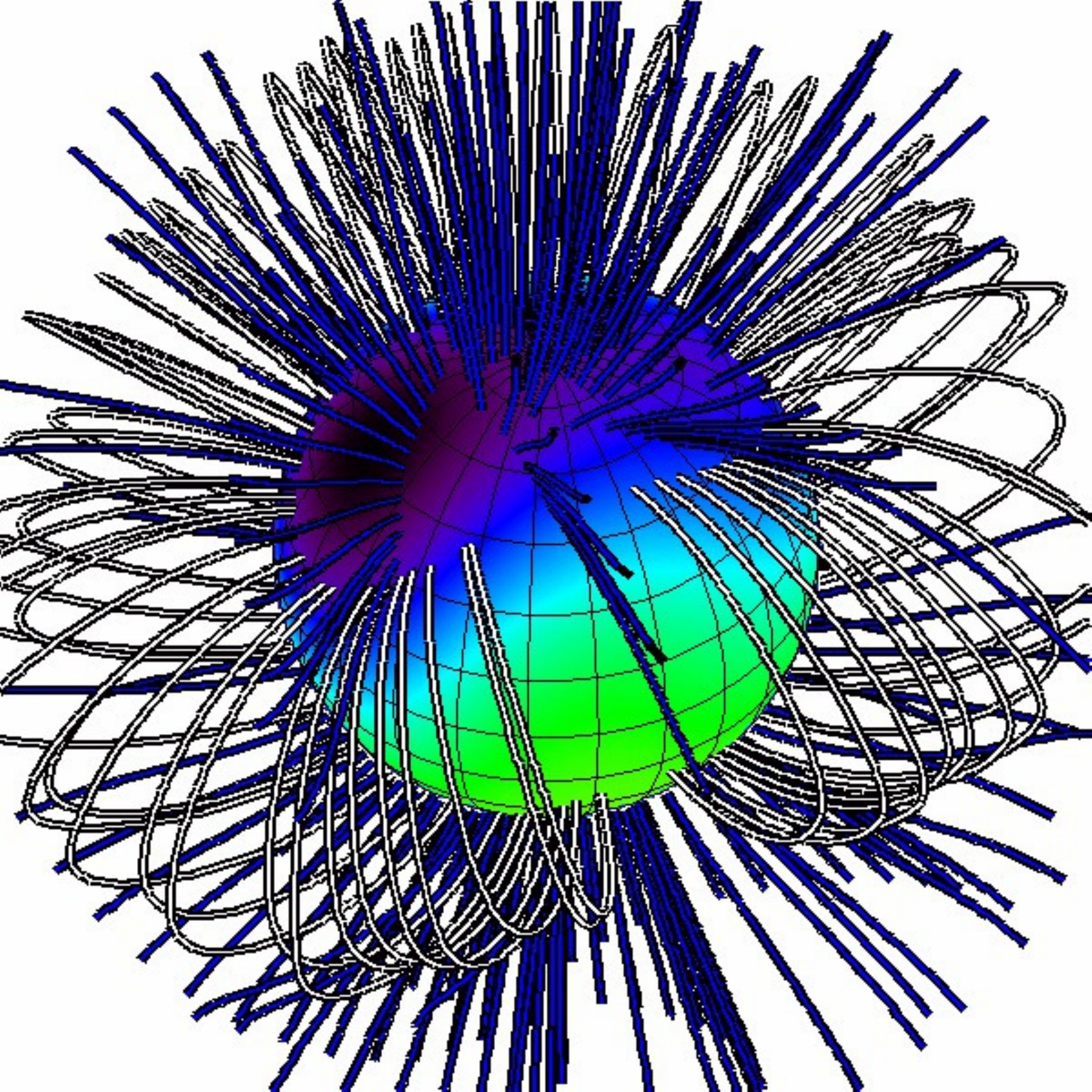}}\\
	\subfloat(a) OT Ser{\label{fig.edge-a}\includegraphics[width=44mm]{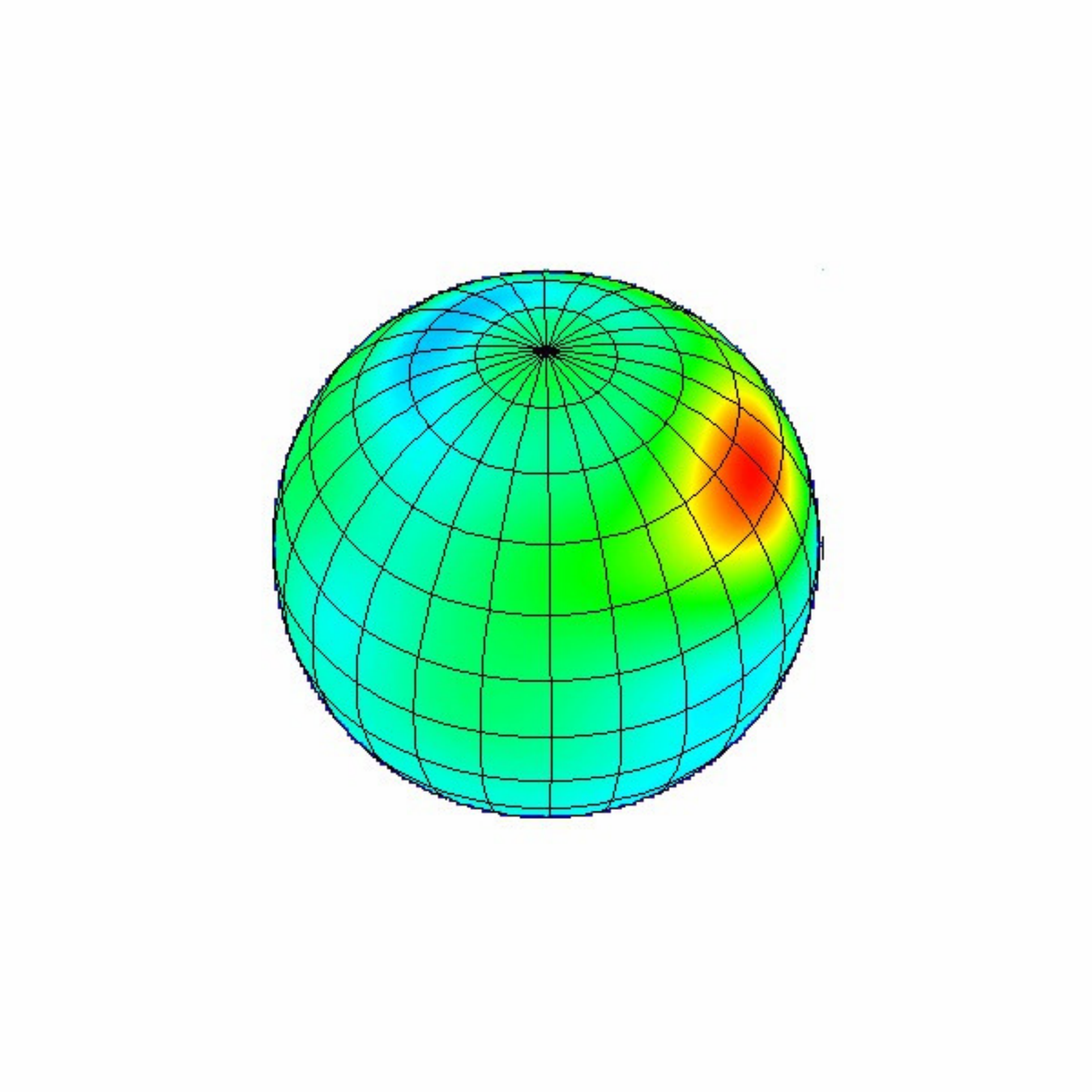}}
	\subfloat(b){\label{fig.edge-b}\includegraphics[width=44mm]{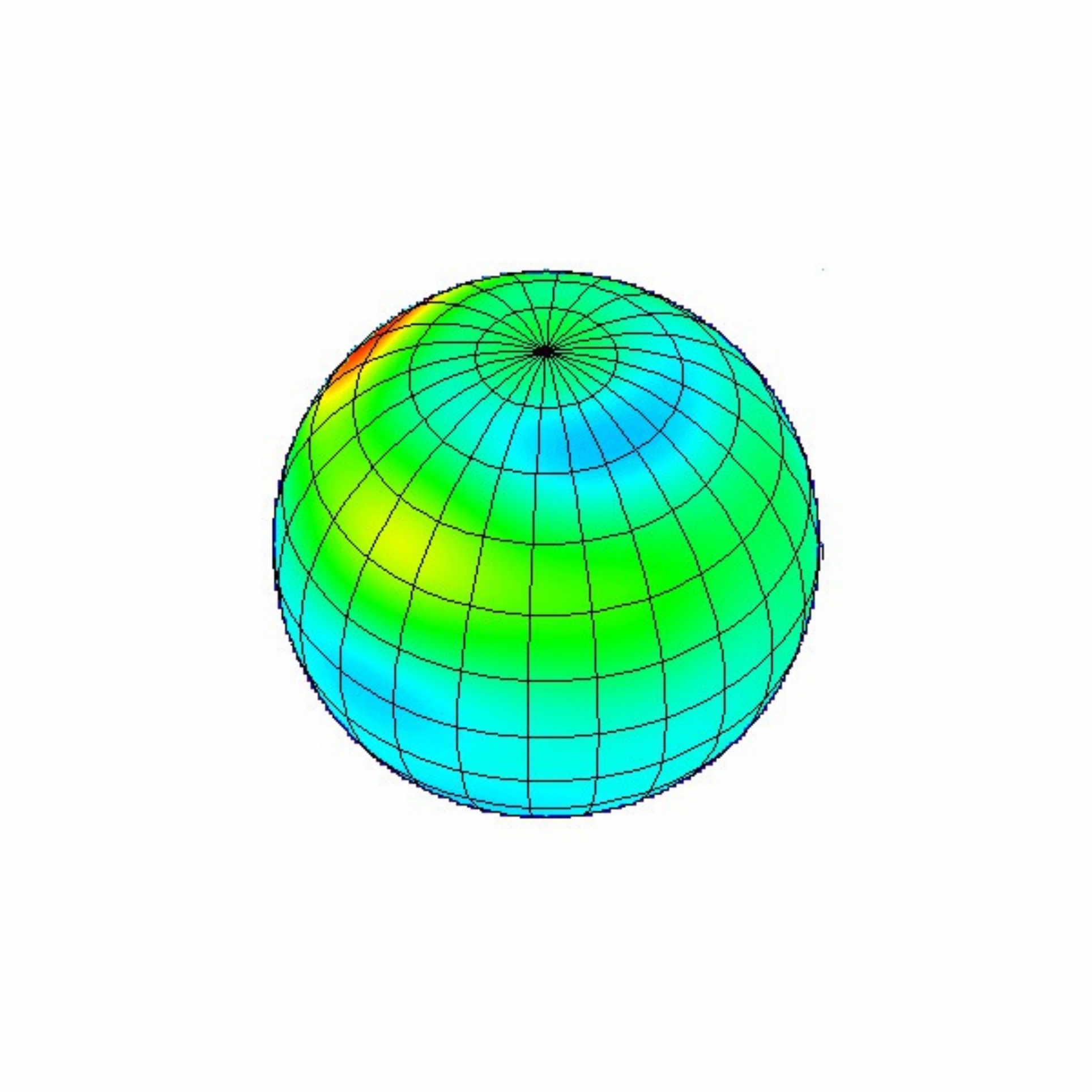}}
	\subfloat(c){\label{fig.edge-c}\includegraphics[width=44mm]{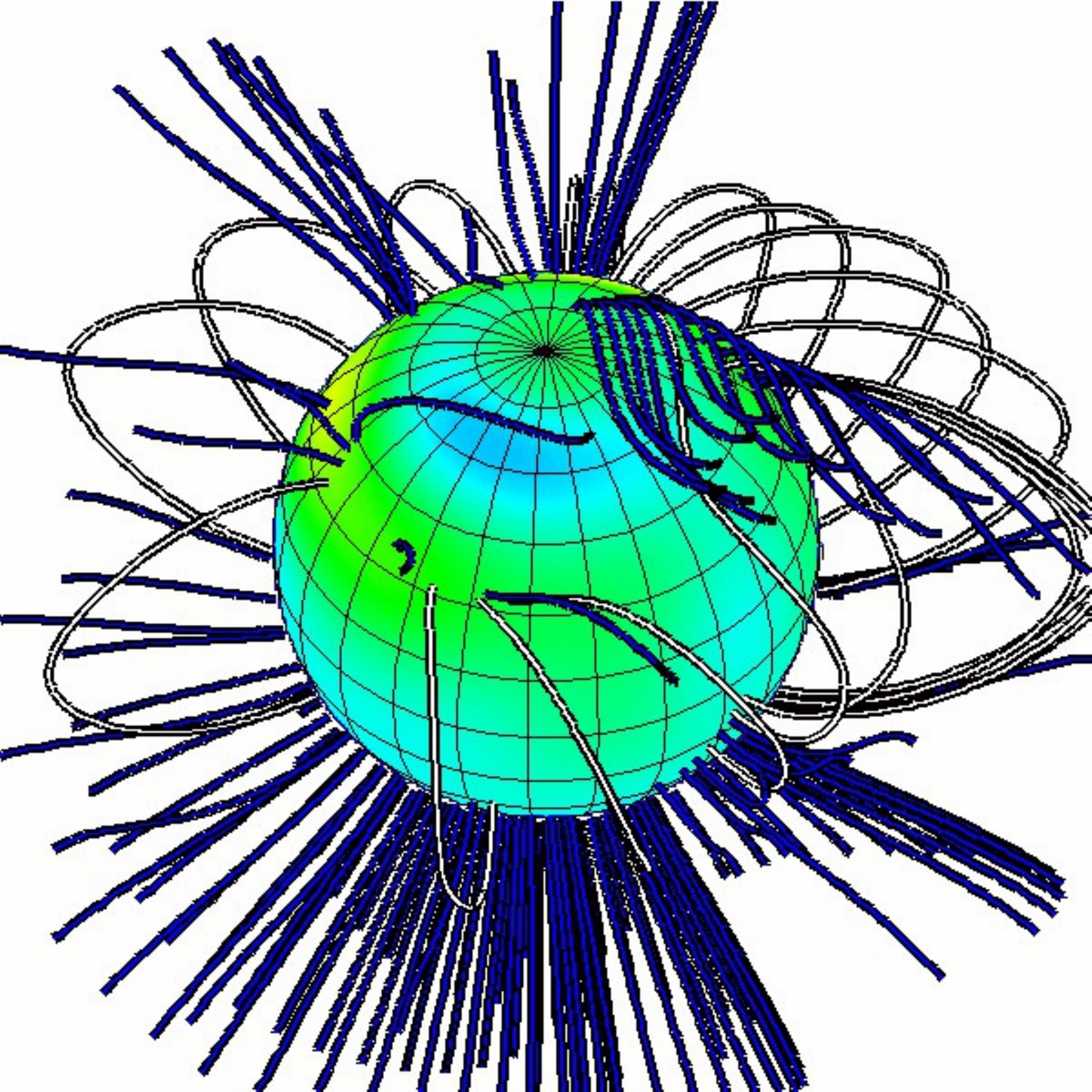}}\\
	\pagebreak
	\caption{(a) and (b): The reconstructed surface radial maps shown at longitudes $180^{\circ}$ apart to give the best viewing angle of both magnetic poles.  (c): The extrapolated field, using the PFSS method with $R_{ss}=2.5R_*$.  For each star, colours are scaled to the maximum and minimum values of the radial magnetic field component: blue represents negative flux and red positive flux.}
	\label{fig.surface_get_1}
	\end{center}
\end{figure*}

\begin{figure*}
	\begin{center}
	\subfloat(a) CE Boo{\label{fig.edge-a}\includegraphics[width=44mm]{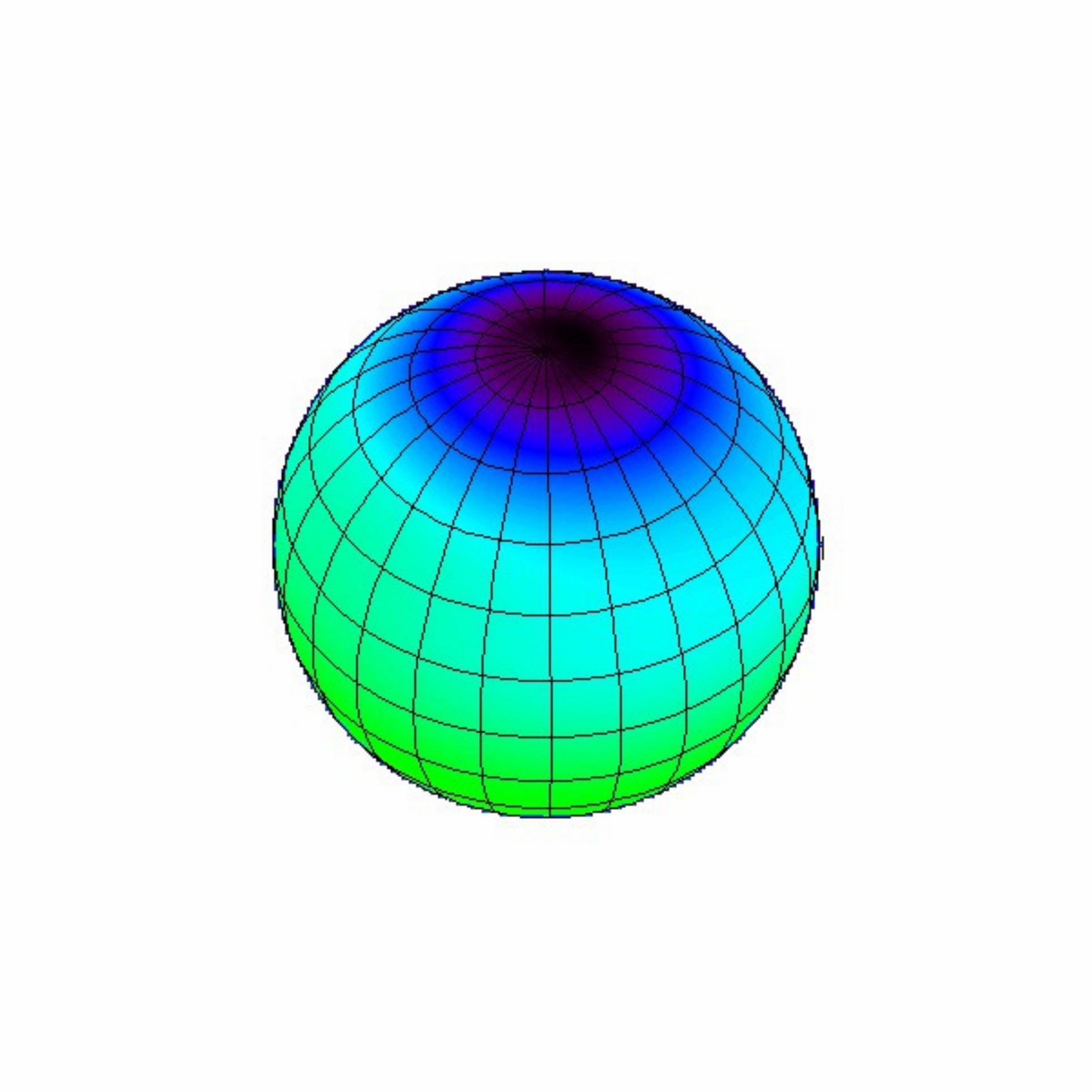}}
	\subfloat(b){\label{fig.edge-b}	\includegraphics[width=44mm]{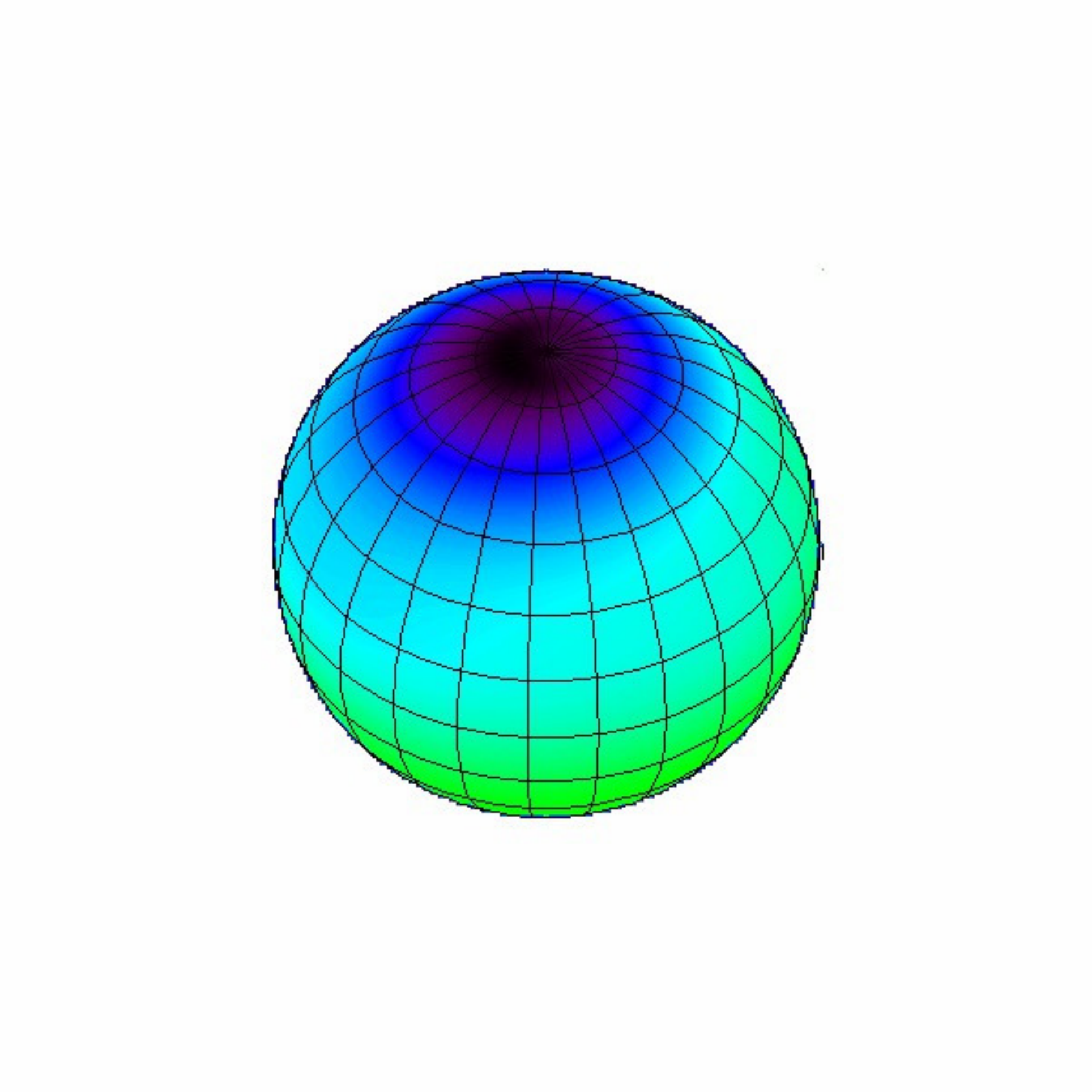}}
	\subfloat(c){\label{fig.edge-c}	\includegraphics[width=44mm]{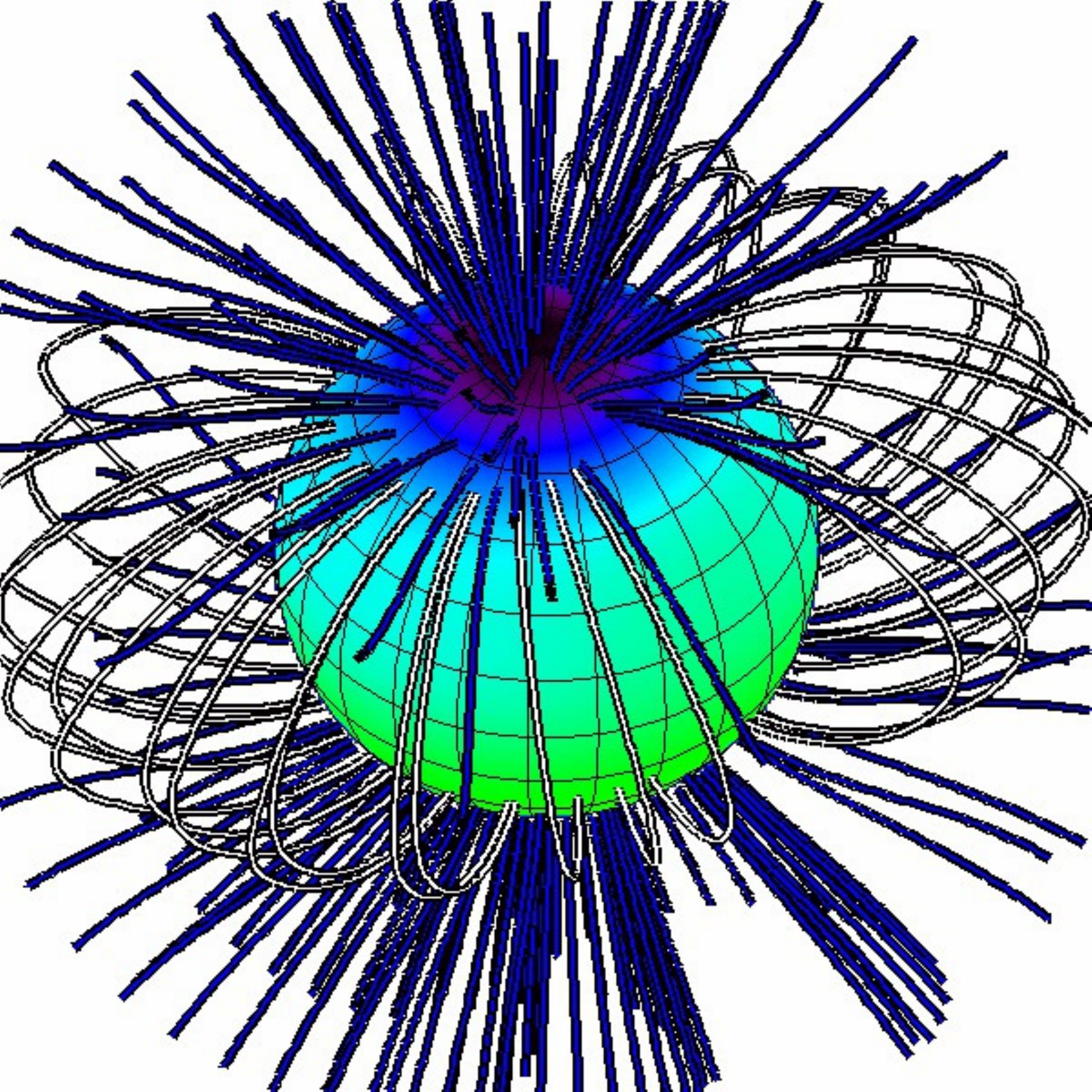}}\\
	\subfloat(a) AD Leo{\label{fig.edge-a}\includegraphics[width=44mm]{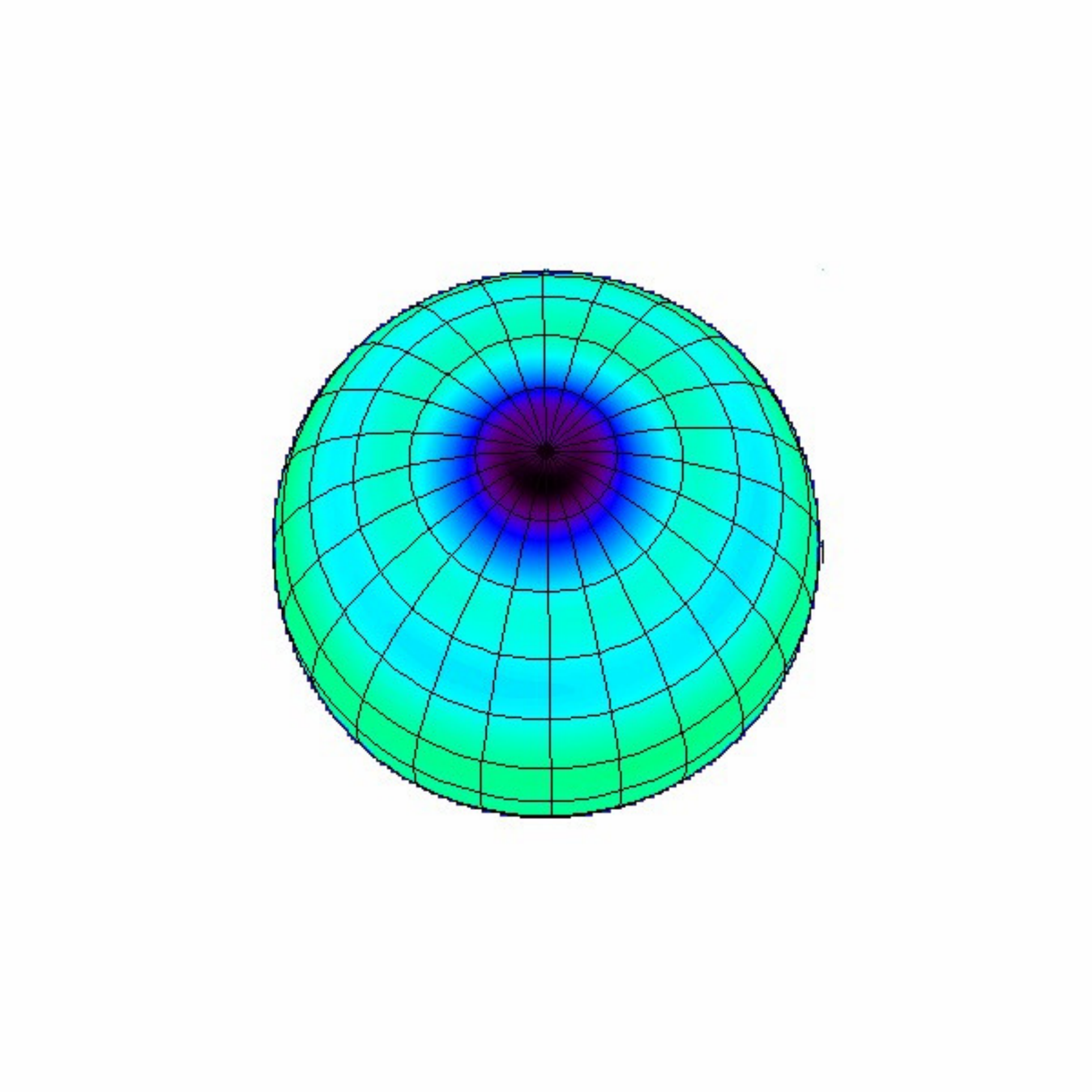}}
	\subfloat(b){\label{fig.edge-b}\includegraphics[width=44mm]{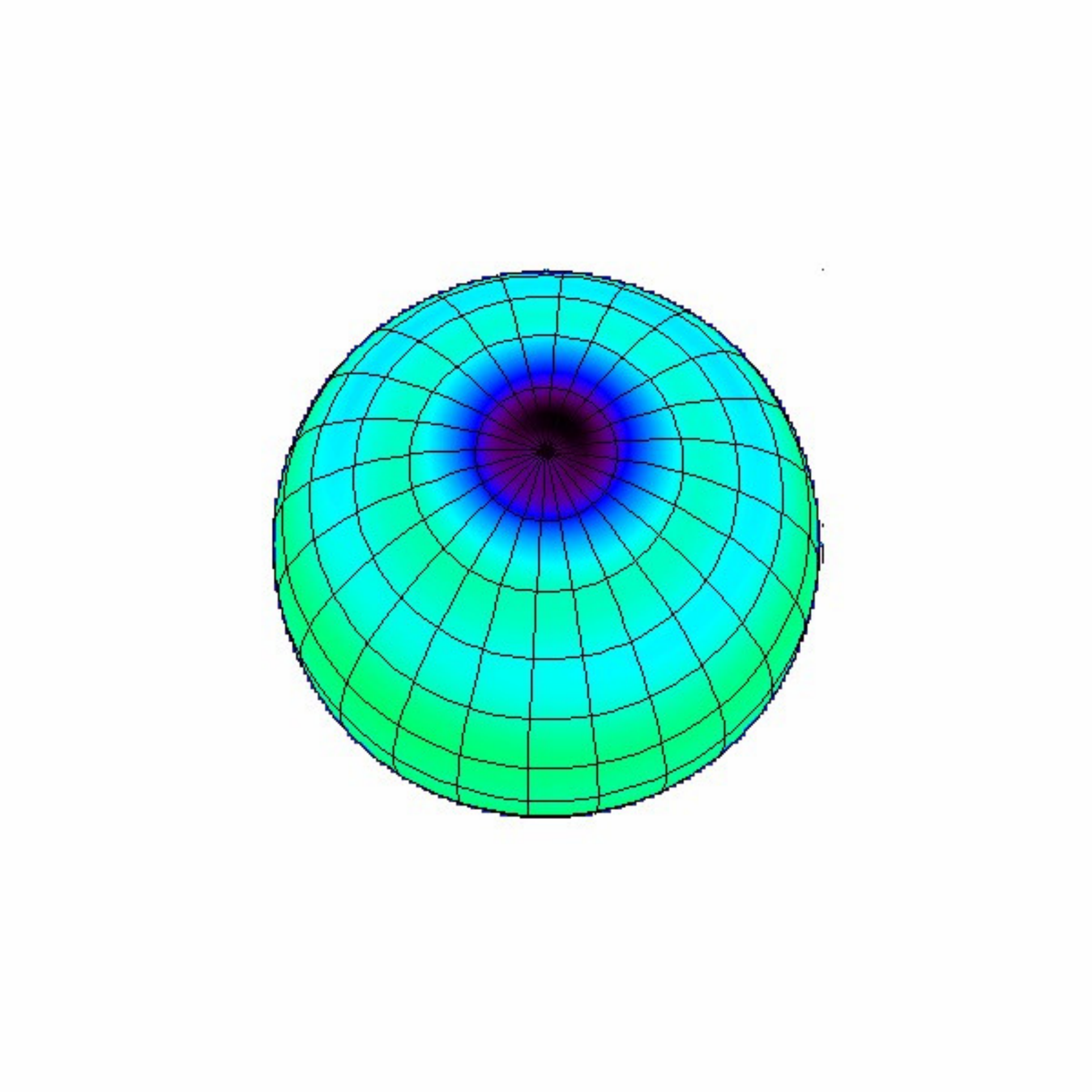}}
	\subfloat(c){\label{fig.edge-c}\includegraphics[width=44mm]{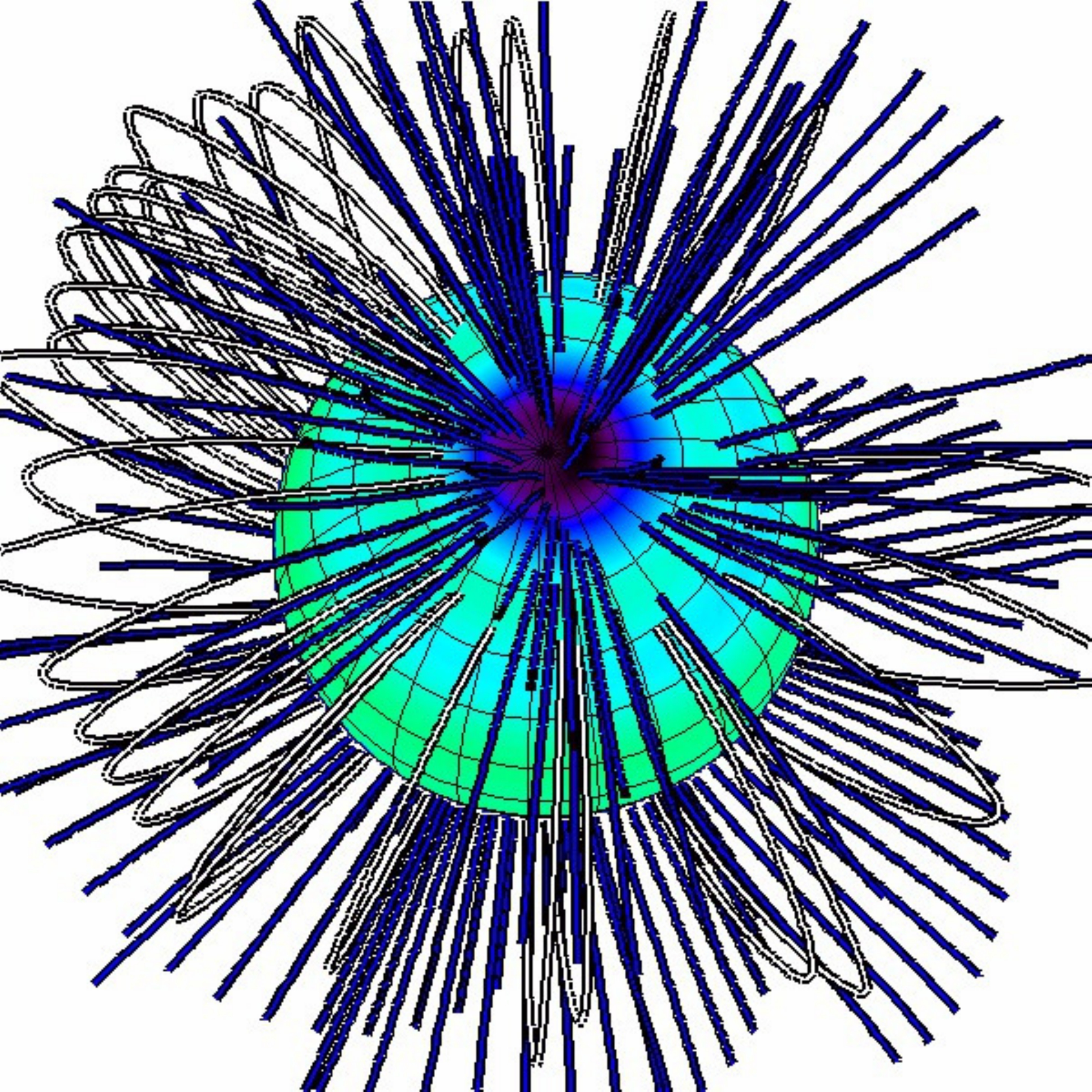}}\\
	\subfloat(a) EQ Peg A{\label{fig.edge-a}\includegraphics[width=44mm]{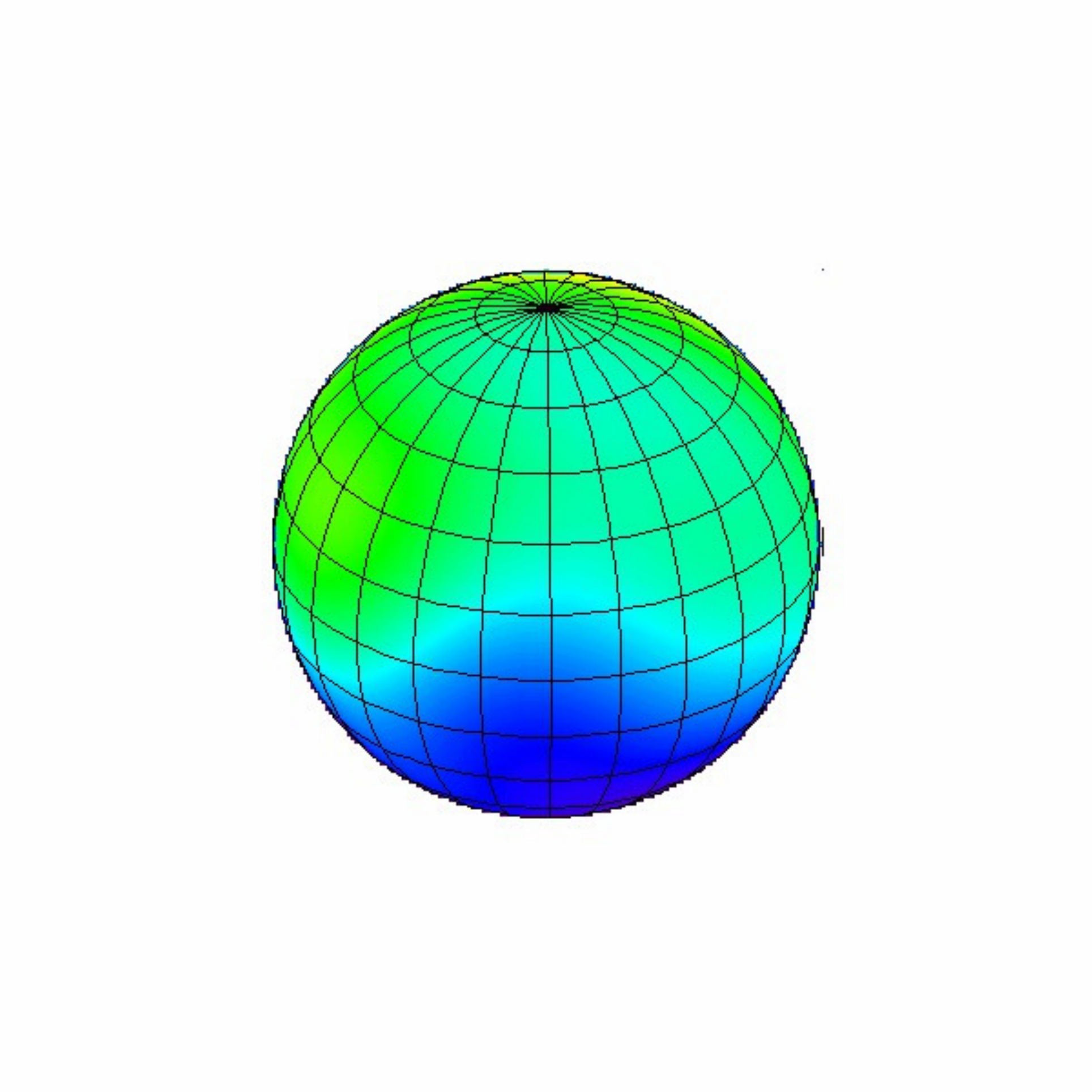}}
	\subfloat(b){\label{fig.edge-b}\includegraphics[width=44mm]{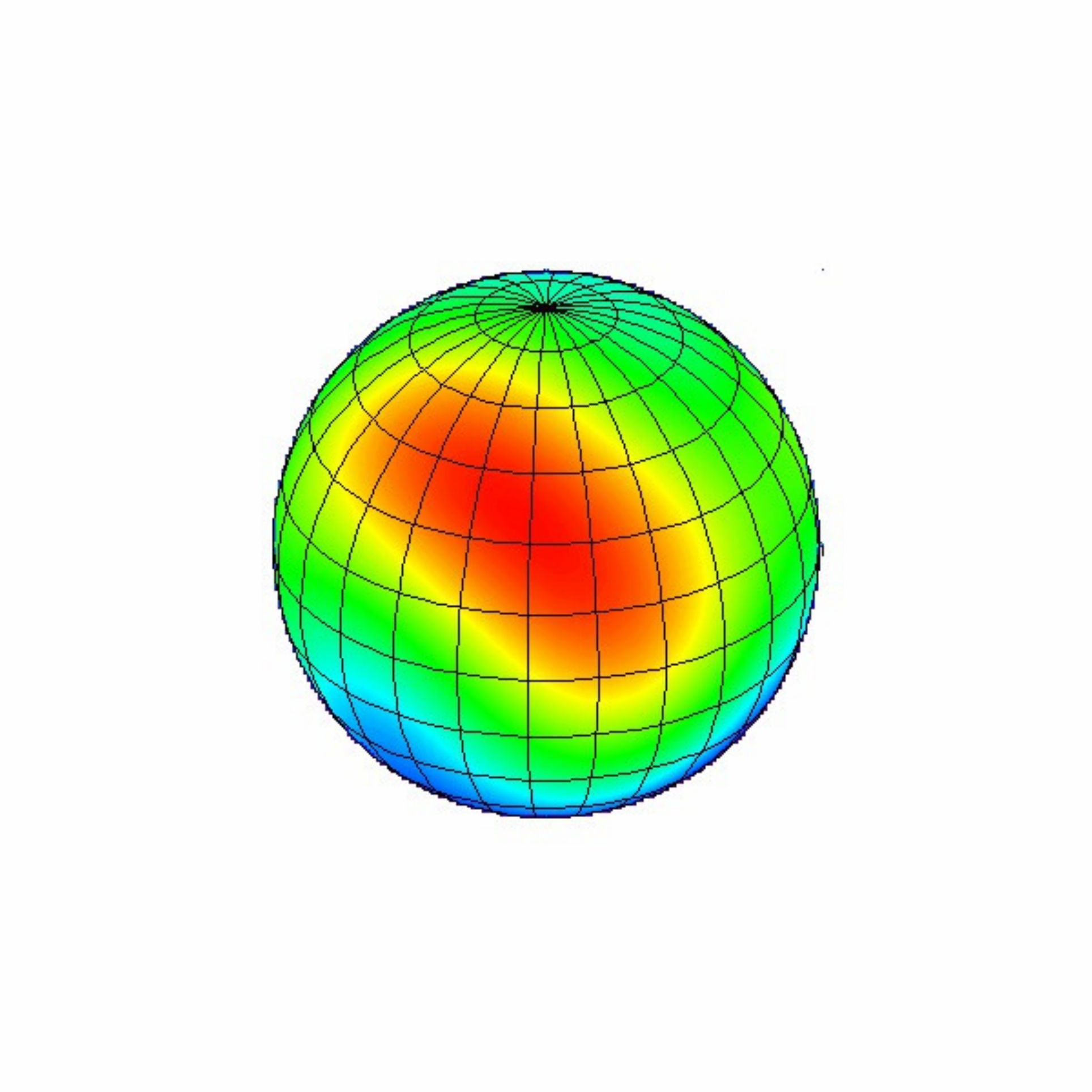}}
	\subfloat(c){\label{fig.edge-c}\includegraphics[width=44mm]{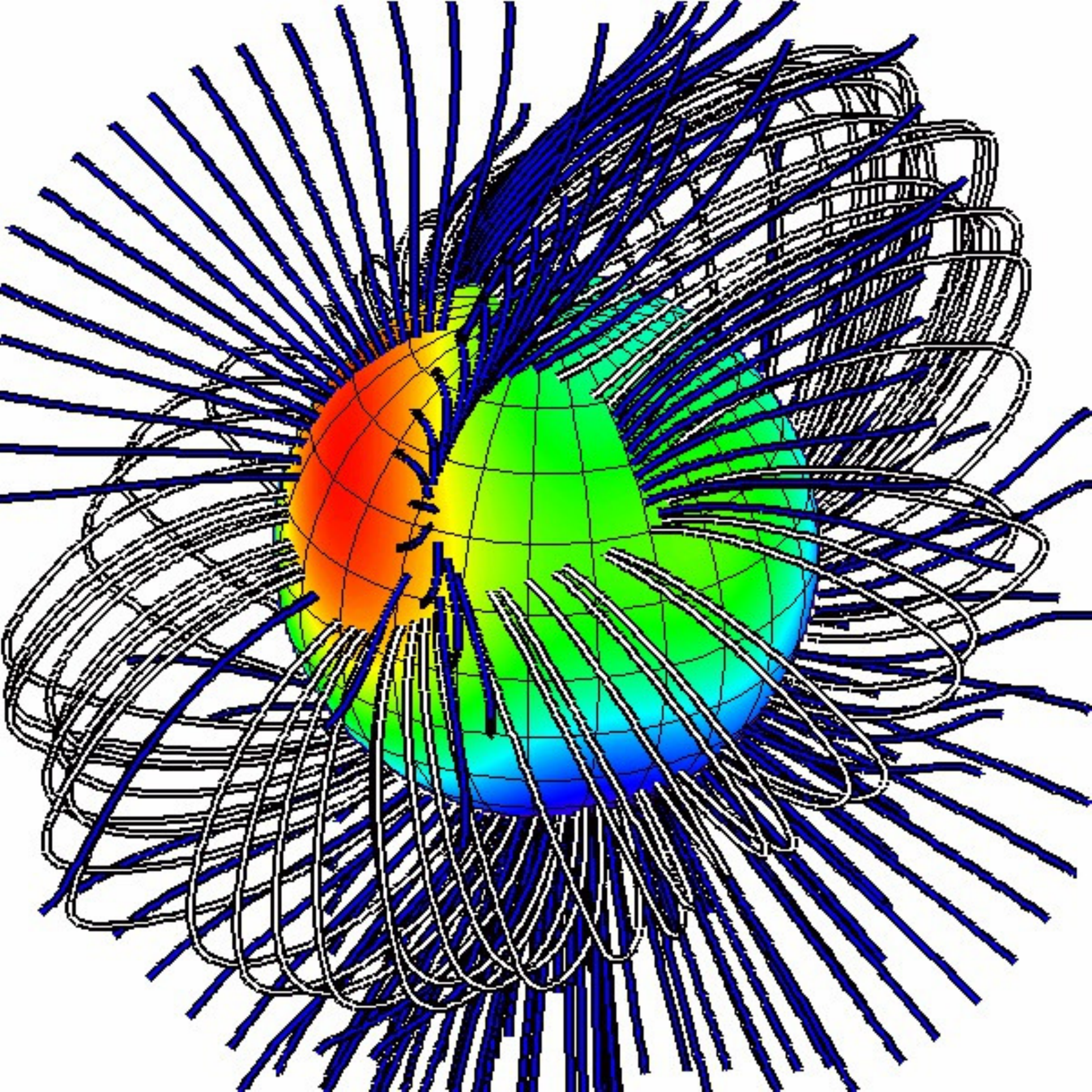}}\\
	\subfloat(a) EV Lac{\label{fig.edge-a}\includegraphics[width=44mm]{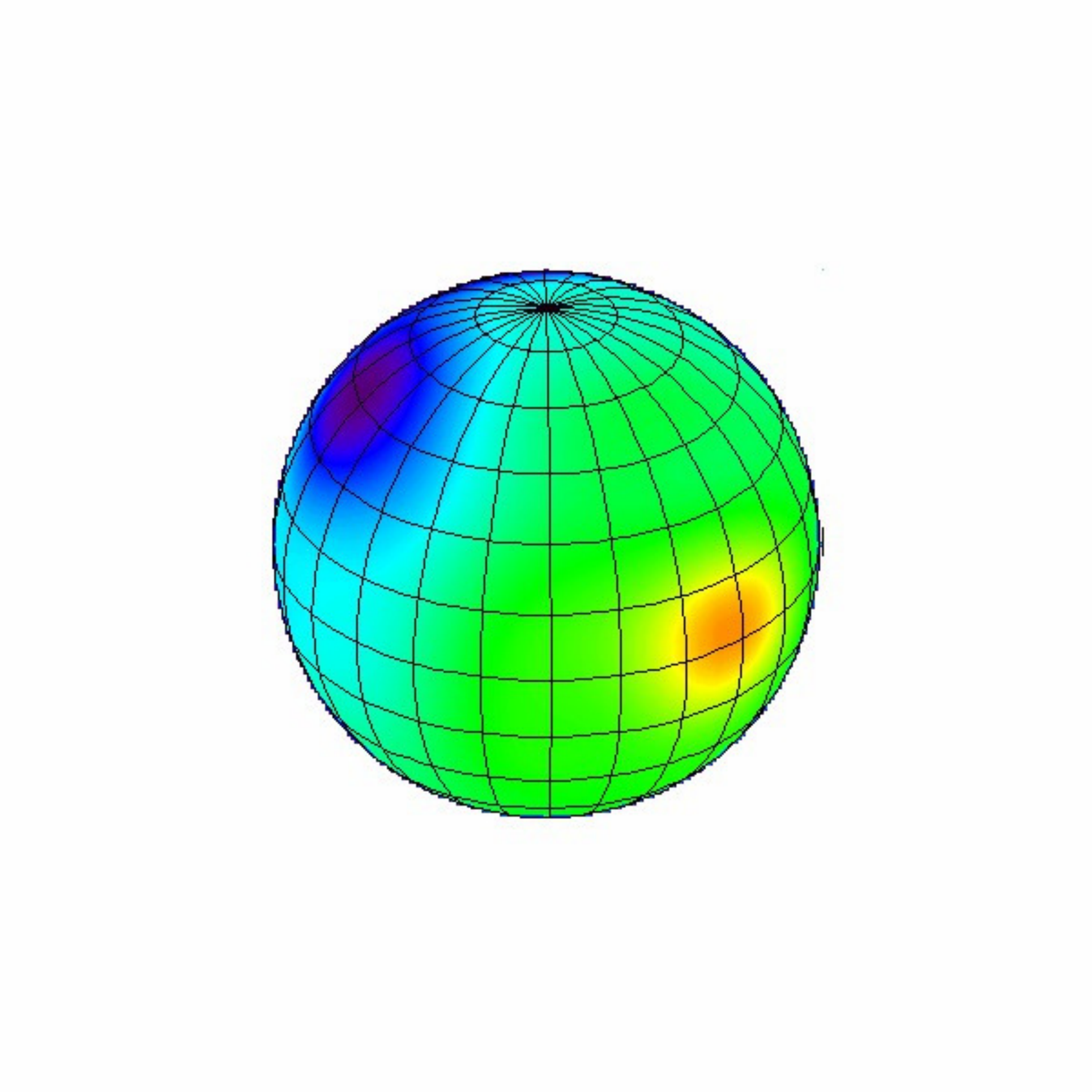}}
	\subfloat(b){\label{fig.edge-b}\includegraphics[width=44mm]{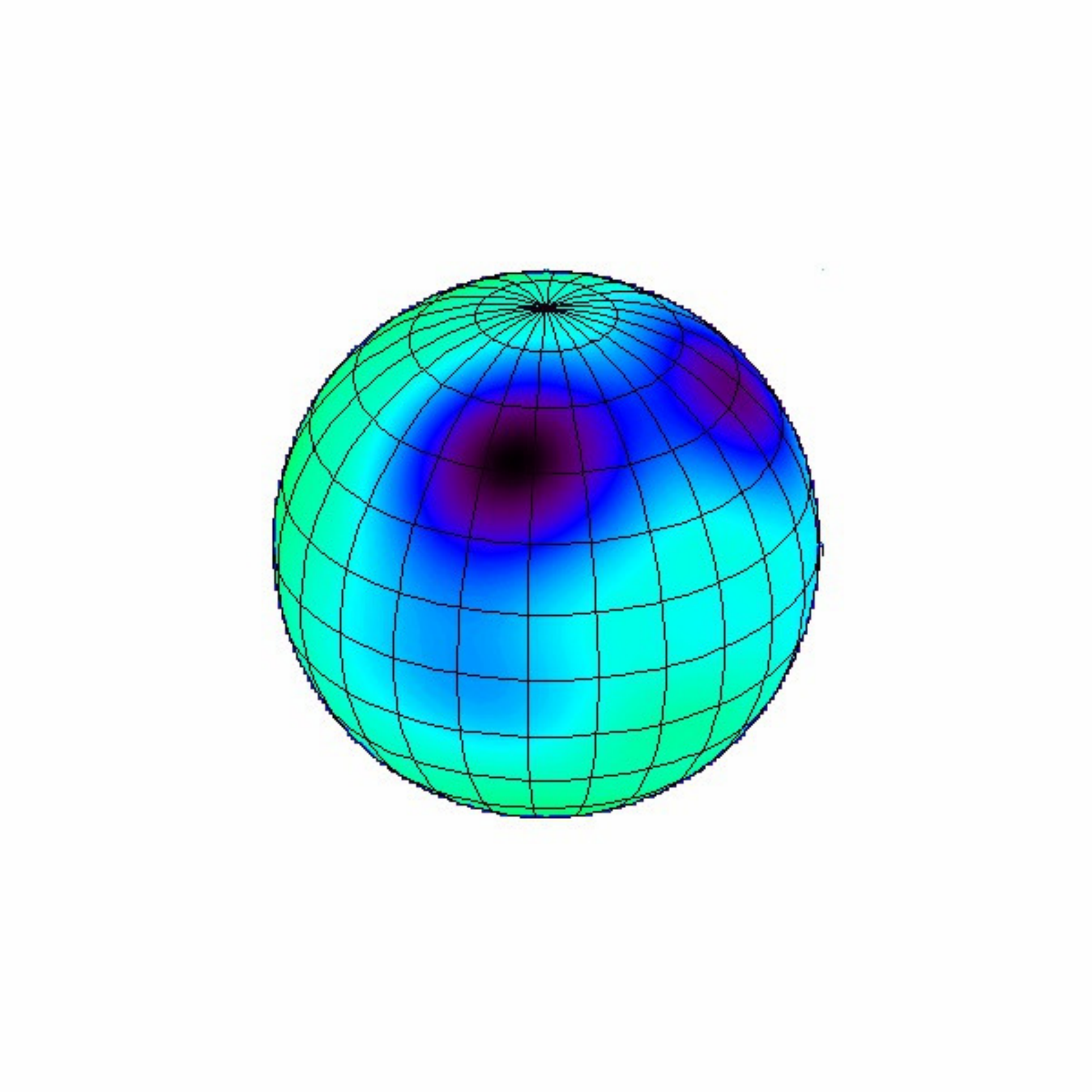}}
	\subfloat(c){\label{fig.edge-c}\includegraphics[width=44mm]{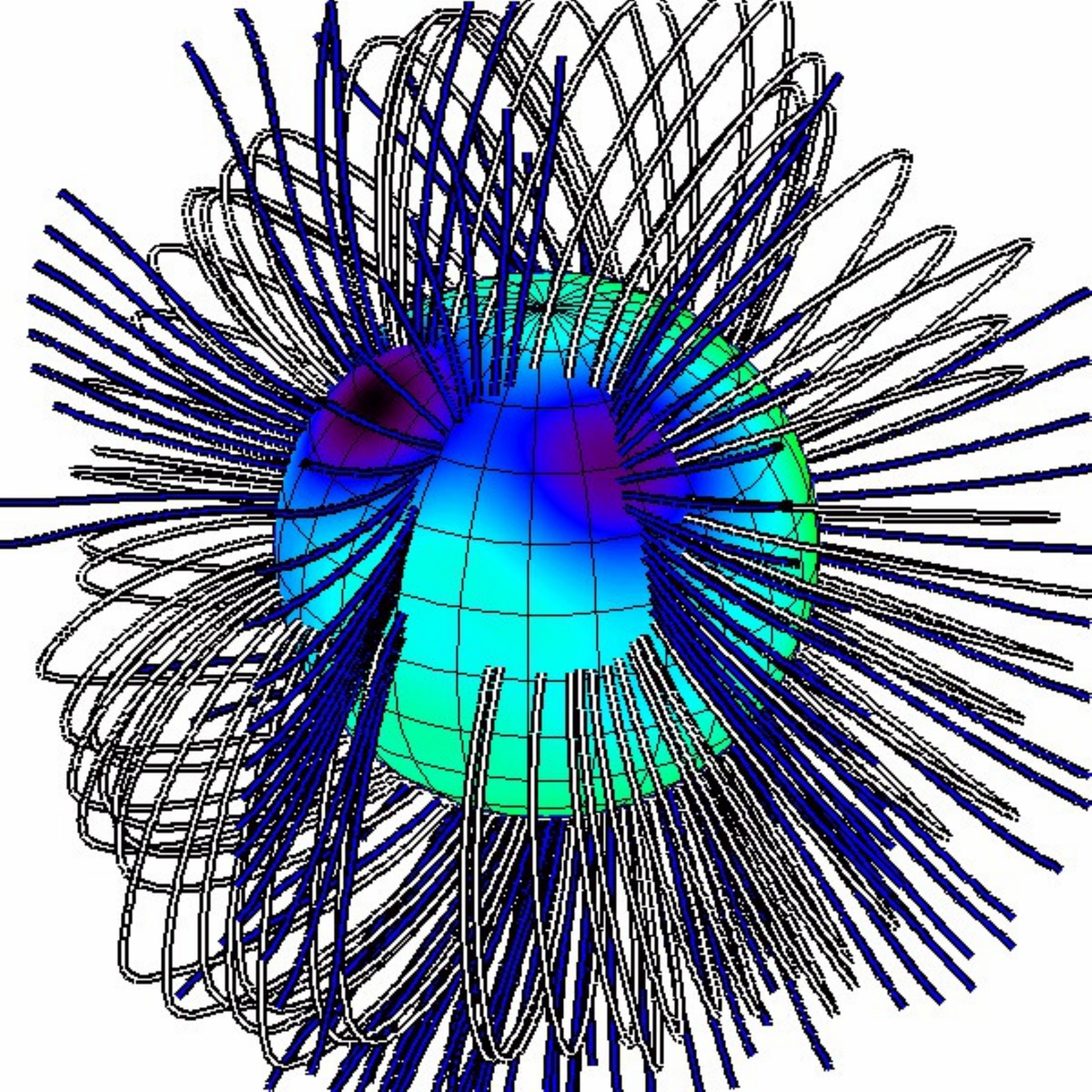}}\\
	\subfloat(a) YZ CMi{\label{fig.edge-a}\includegraphics[width=44mm]{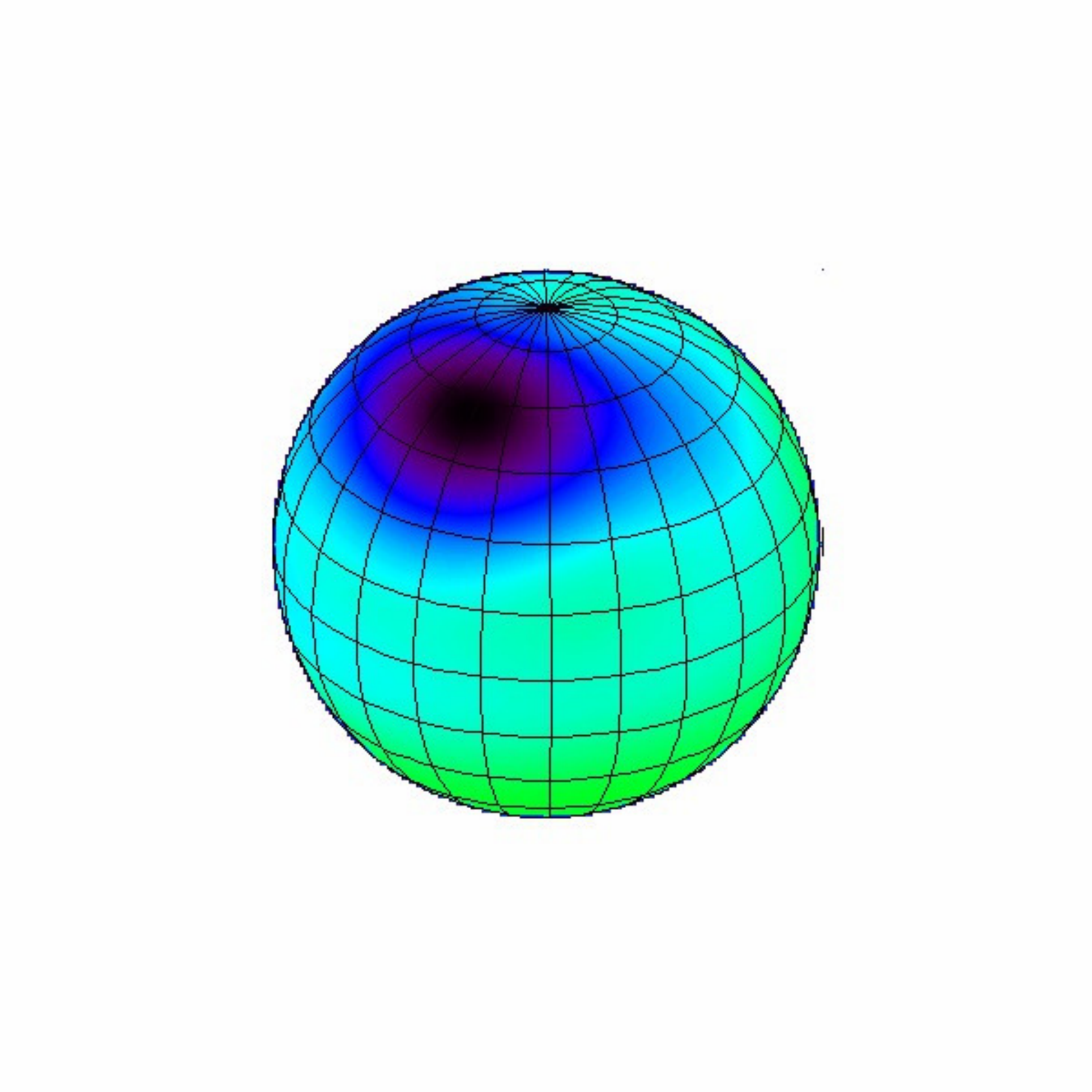}}
	\subfloat(b){\label{fig.edge-b}\includegraphics[width=44mm]{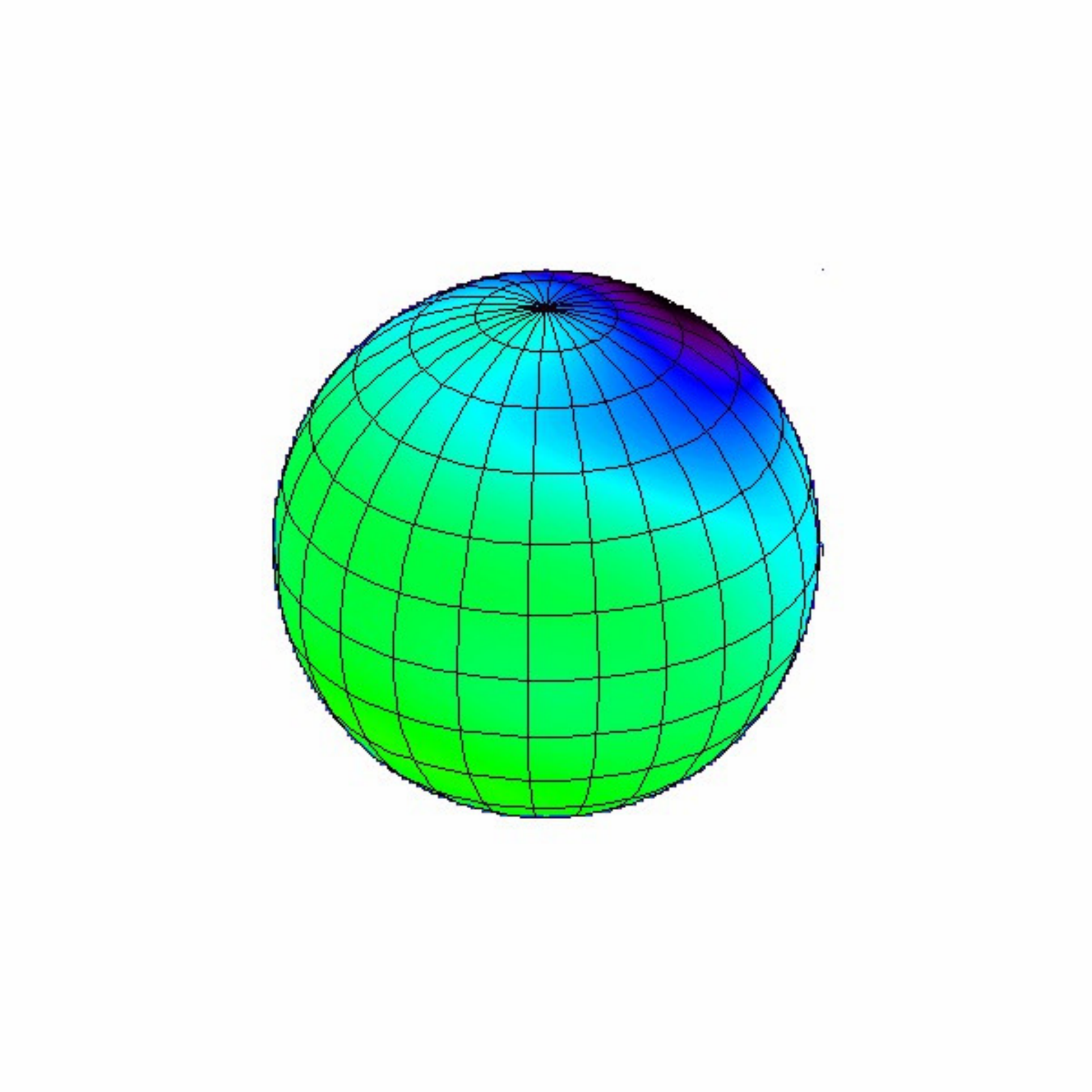}}
	\subfloat(c){\label{fig.edge-c}\includegraphics[width=44mm]{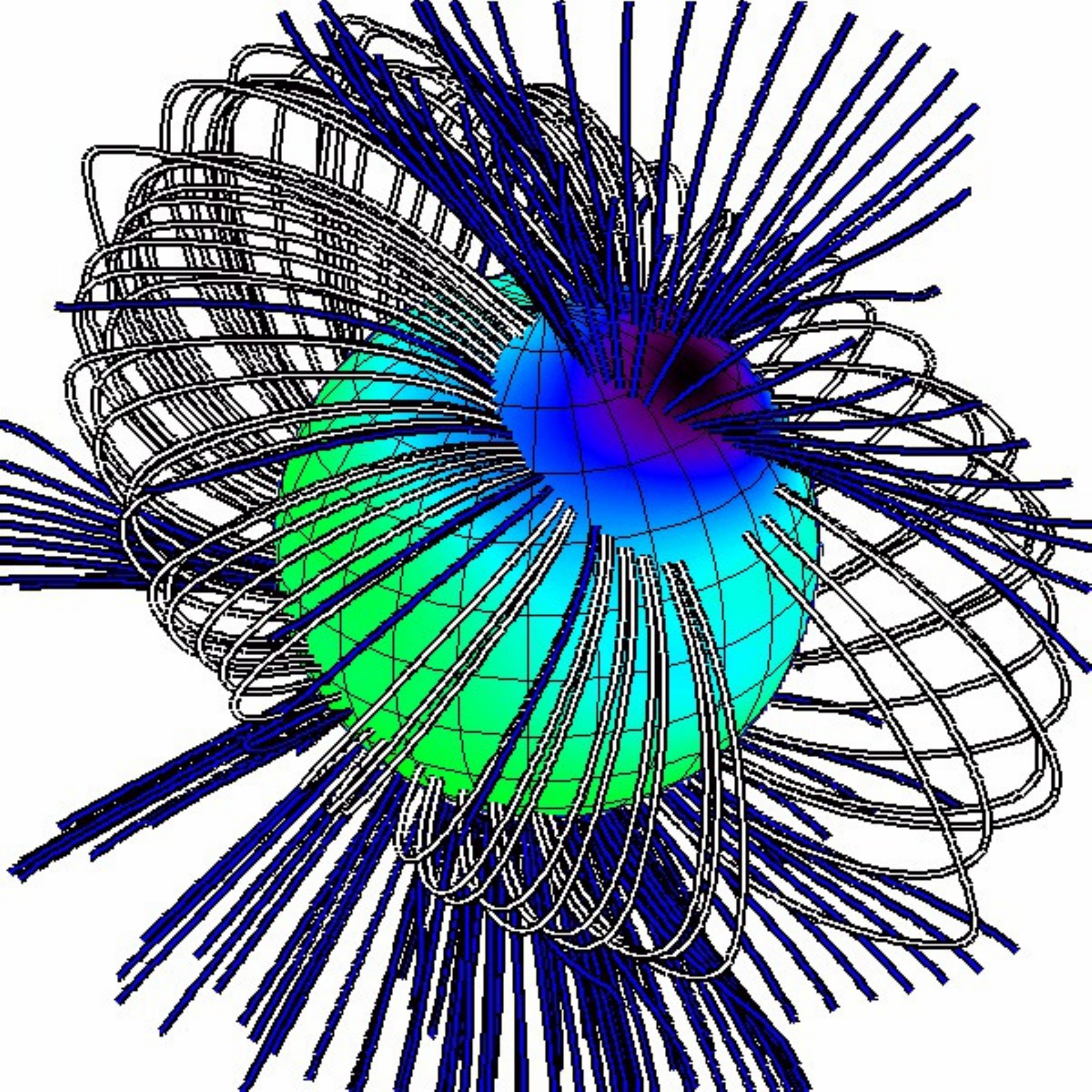}}\\
	\caption{(a) and (b): The reconstructed surface radial maps shown at longitudes $180^{\circ}$ apart to give the best viewing angle of both magnetic poles.  (c): The extrapolated field, using the PFSS method with $R_{ss}=2.5R_*$.  For each star, colours are scaled to the maximum and minimum values of the radial magnetic field component: blue represents negative flux and red positive flux.  (cont.)}
	\label{fig.surface_get_2}
	\end{center}
\end{figure*}

\begin{figure*}
	\begin{center}
	\subfloat(a) V374 Peg{\label{fig.edge-a}\includegraphics[width=44mm]{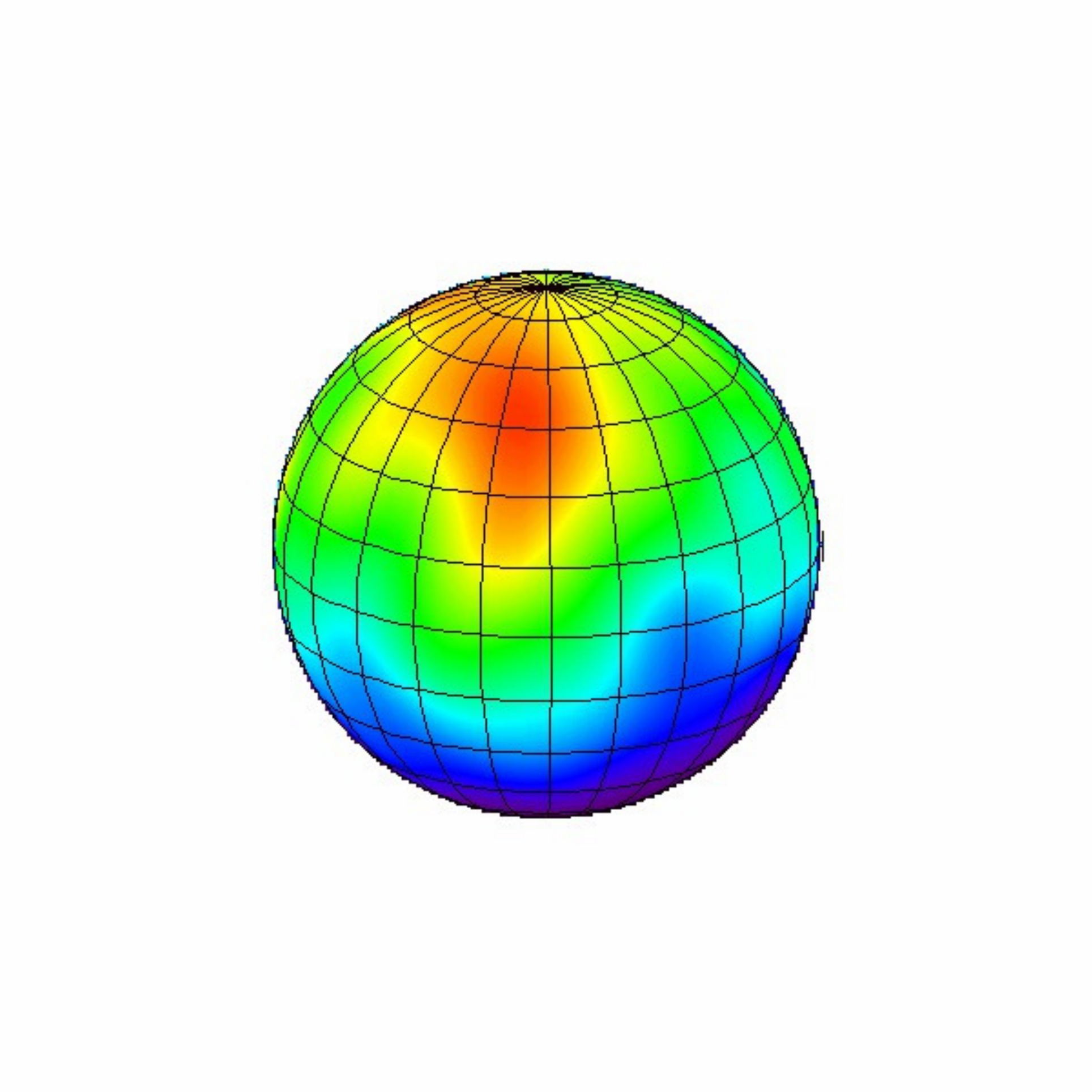}}
	\subfloat(b){\label{fig.edge-b}	\includegraphics[width=44mm]{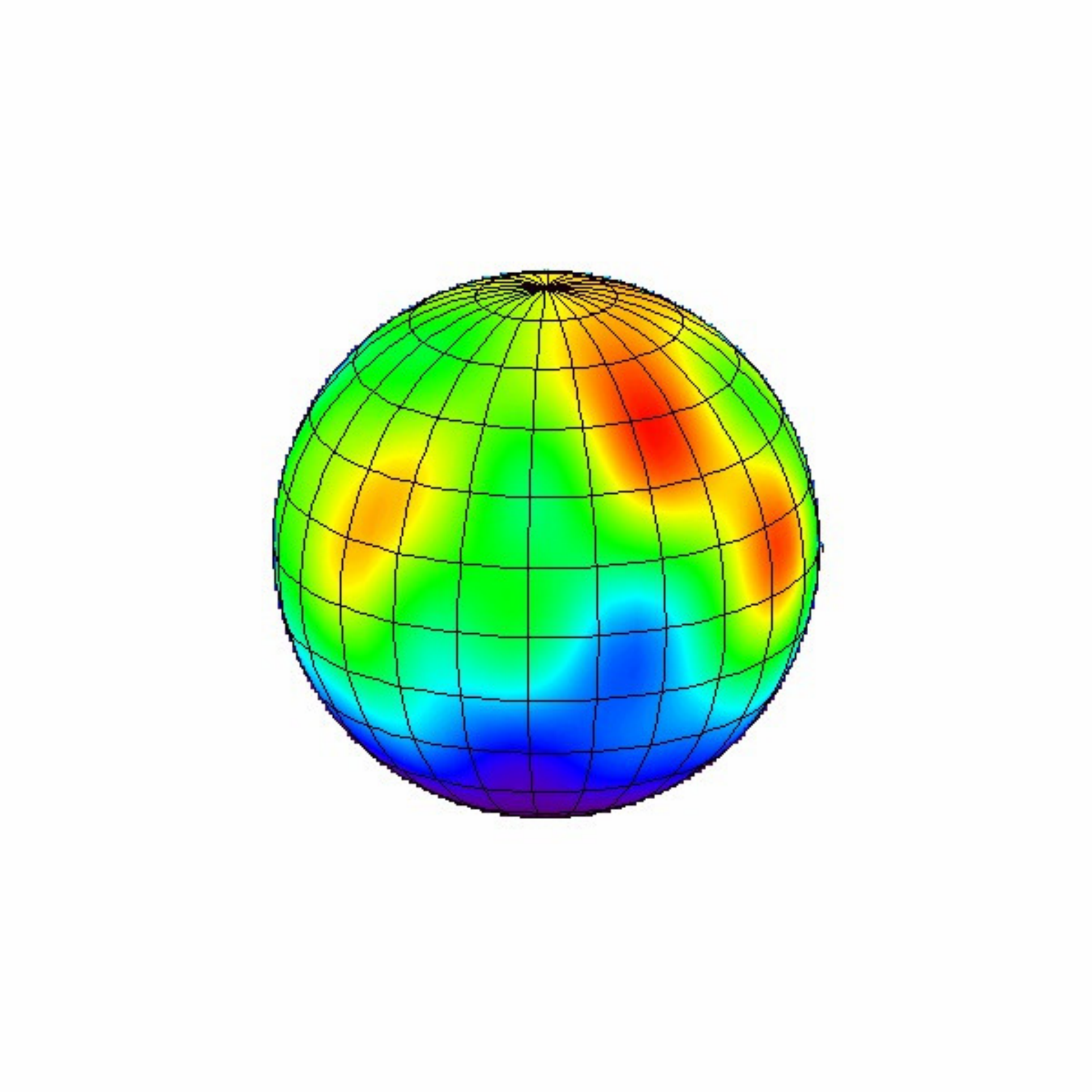}}
	\subfloat(c){\label{fig.edge-c}	\includegraphics[width=44mm]{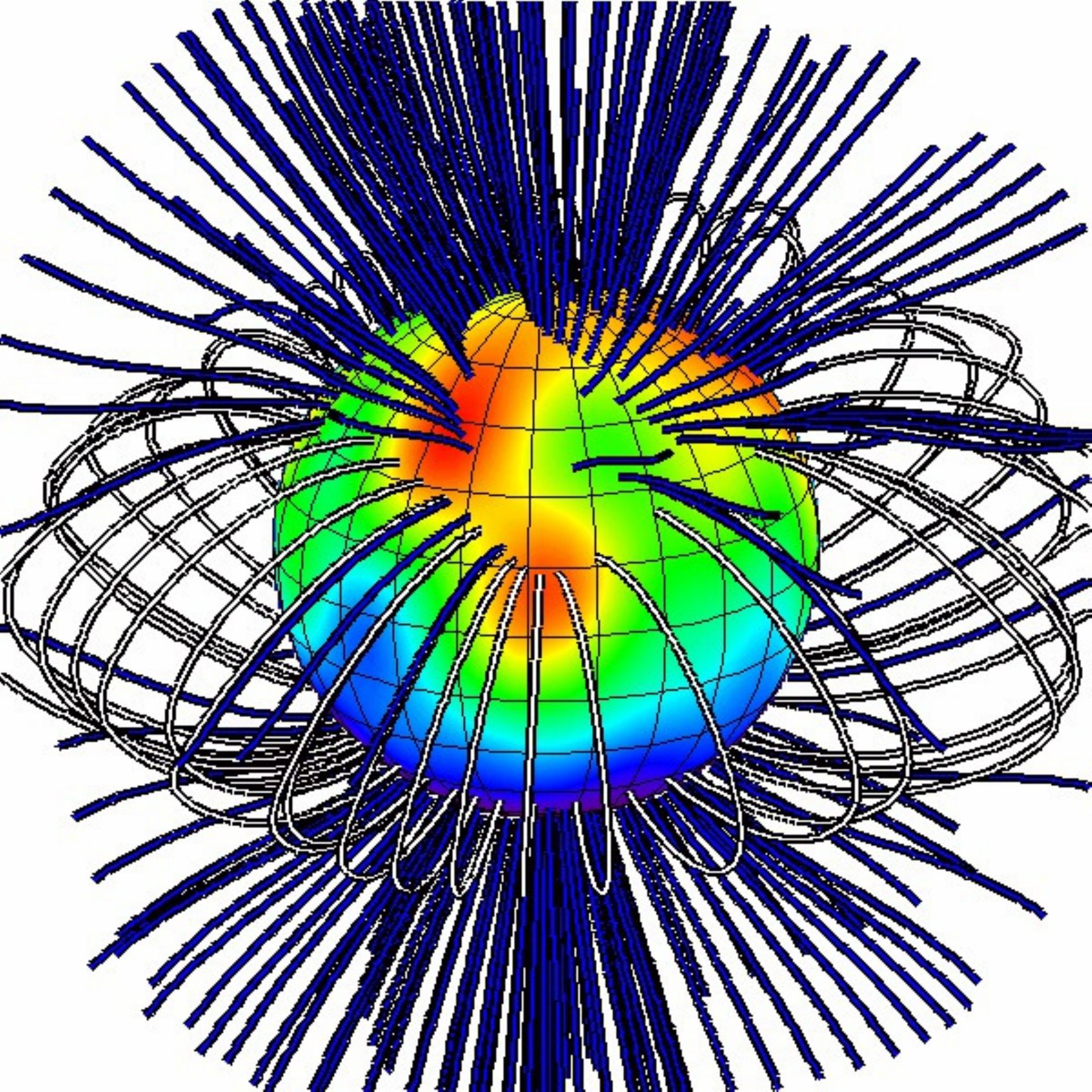}}\\
	\subfloat(a) EQ Peg B{\label{fig.edge-a}\includegraphics[width=44mm]{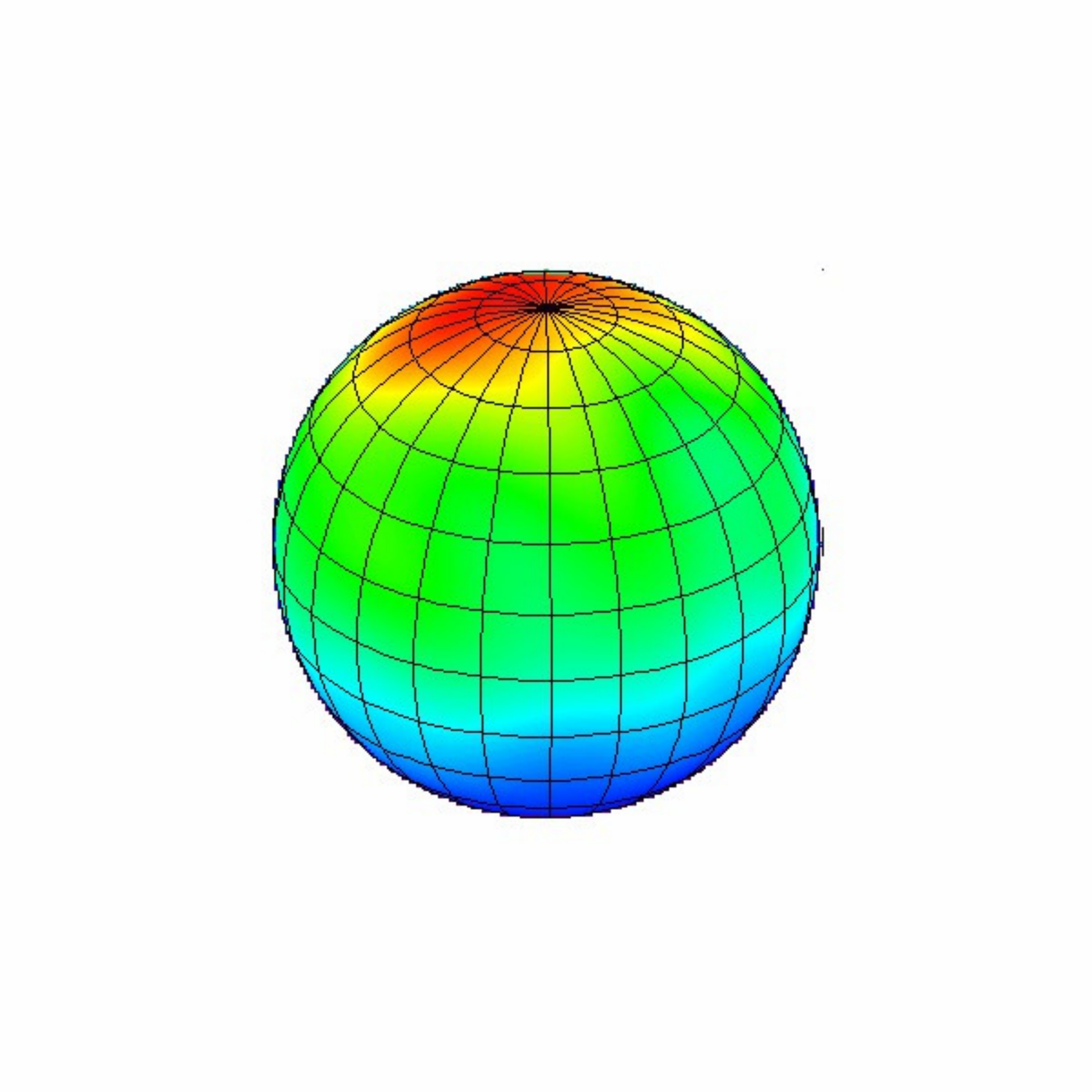}}
	\subfloat(b){\label{fig.edge-b}\includegraphics[width=44mm]{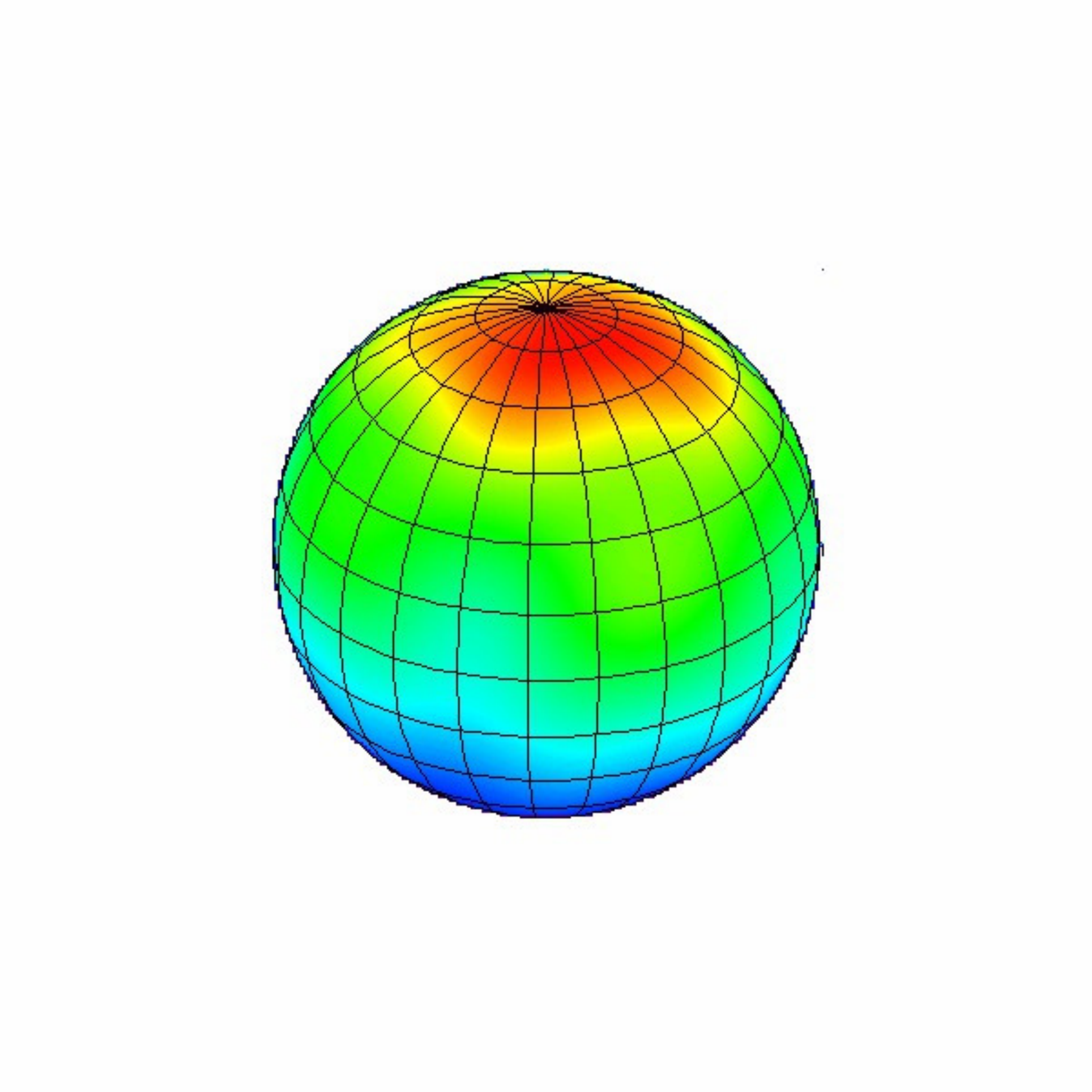}}
	\subfloat(c){\label{fig.edge-c}\includegraphics[width=44mm]{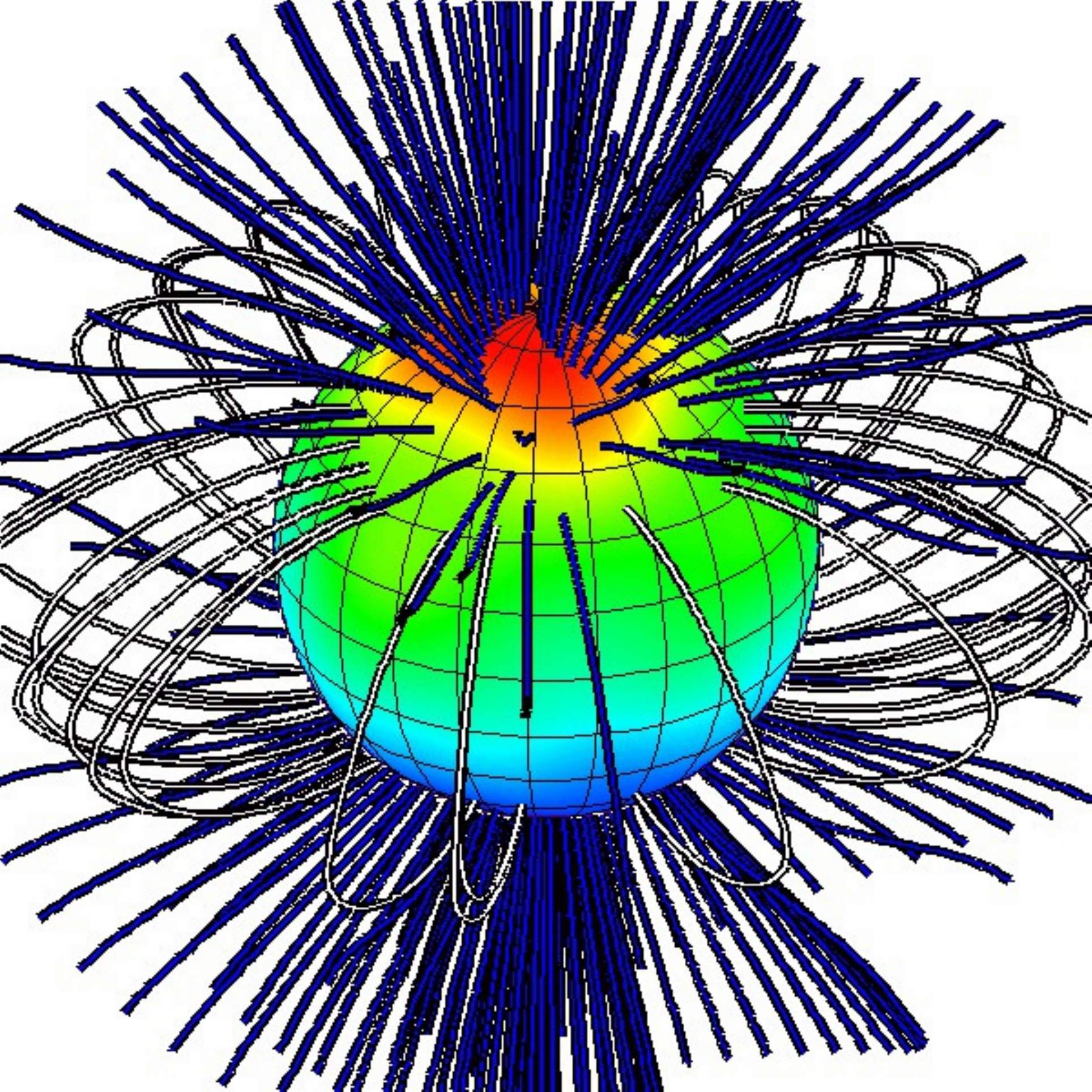}}\\
	\caption{(a) and (b): The reconstructed surface radial maps shown at longitudes $180^{\circ}$ apart to give the best viewing angle of both magnetic poles.  (c): The extrapolated field, using the PFSS method with $R_{ss}=2.5R_*$.  For each star, colours are scaled to the maximum and minimum values of the radial magnetic field component: blue represents negative flux and red positive flux.  (cont.)}
	\label{fig.surface_get_3}
	\end{center}
\end{figure*}

\section{Field Geometry}

\subsection{Potential-Field Extrapolation}\label{sec.FieldExtrapolation}

To calculate the magnetic field above the stellar surface, a potential-field source surface (PFSS) approach is applied \citep{Altschuler_Newkirk_1969}.  By assuming that the coronal magnetic field is current-free ($\underline \nabla \times \underline B = 0$), the magnetic field can be defined as the gradient of the scalar potential, $\underline B = - \underline \nabla \psi $.  Since the magnetic field is divergence free ($\underline \nabla \cdot \underline B = 0$), this results in Laplace's Equation: $\nabla ^2 \psi = 0$.  The solution to Laplace's equation in spherical harmonics allows for the determination of the coefficients of the associated Legendre polynomials $P_{lm}$ and, therefore, the coronal magnetic field components. 
\begin{equation}
B_{r}=-\displaystyle\sum_{l=1}^{N}\displaystyle\sum_{m=1}^{l}[la_{lm}r^{l-1} - (l+1)b_{lm}r^{-(l+2)}]P_{lm}(\cos\theta)e^{im\phi}
\label{eq.field_1}
\end{equation}
\begin{equation}
B_{\theta}=-\displaystyle\sum_{l=1}^{N}\displaystyle\sum_{m=1}^{l}[a_{lm}r^{l-1} + b_{lm}r^{-(l+2)}]\frac{d}{d\theta}P_{lm}(\cos\theta)e^{im\phi}
\label{eq.field_2}
\end{equation}
\begin{equation}
B_{\phi}=-\displaystyle\sum_{l=1}^{N}\displaystyle\sum_{m=1}^{l}[a_{lm}r^{l-1} + b_{lm}r^{-(l+2)}]P_{lm}(\cos\theta)\frac{im}{\sin\theta}e^{im\phi}
\label{eq.field_3}
\end{equation}
where $a_{lm}$ and $b_{lm}$ are the amplitudes of the spherical harmonics, \textit{l} is the spherical harmonic degree  and \textit{m} is the order or 'azimuthal number'.

To extrapolate the 3D coronal field and determine the amplitude of the spherical harmonics, $a_{lm}$ and $b_{lm}$, we apply two boundary conditions.  The upper condition is that at some distance from the star \textit{i.e. the Source Surface, $R_{ss}$}\citep{Schatten_SourceSurface_1969}, the field opens and is purely radial ($B_\theta=B_\phi=0$).  The amplitude of the spherical harmonics $a_{lm}$ and $b_{lm}$ can then be written as
\begin{equation}
\frac{a_{lm}}{b_{lm}}=-\frac{1}{\left(\frac{R_{ss}}{R_*}\right)^{2l+1}}    ,
\label{eq.coeffs}
\end{equation}
where $R_{ss}$ is chosen to be $2.5R_*$ ($R_*$ is the stellar radius), based on the value for the solar source surface and for simulations of V374Peg \citep{Vidotto_V374Peg_2011}.

The lower boundary condition completes the extrapolation by imposing the observed radial field to determine the values of $b_{lm}$ for all values of \textit{l} (up to 31 - for model resolution purposes) and \textit{m} considered.  The code used to extrapolate the field is a modified version of the global diffusion model developed by \citet{VanBallegooijen_Diffusion_1998}.\\

\subsection{Stellar Surface Maps}\label{sec.SurfaceMaps}

Fig.(\ref{fig.surface_get_1},\ref{fig.surface_get_2} and \ref{fig.surface_get_3}) show the reconstructed maps of the surface radial magnetic field, viewed from our line of sight with respect to the stellar inclination, $i(^{\circ})$ (see Table~\ref{tab.stellardata}), for the early-to-mid M dwarf sample.  The ZDI maps show significant structure at the stellar surface (a and b). This degree of structure decreases with height until on the largest scales the extrapolated fields are well described by tilted dipoles (c). Since D08, M08 and M10 report strong components of the poloidal field at low Rossby numbers, we investigate the polar field strength of the dipole component.

\begin{figure}
	\includegraphics[width=84mm]{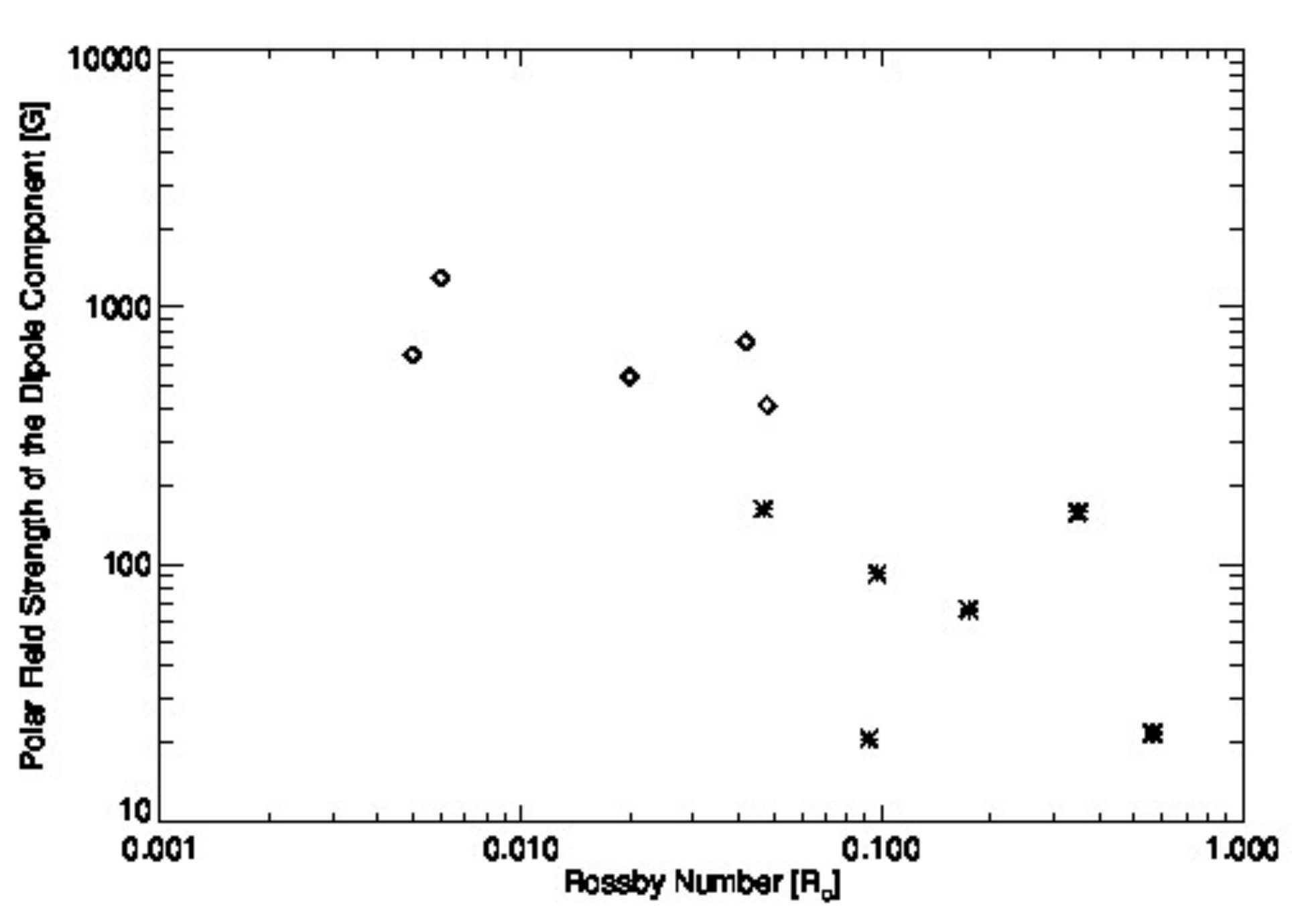}
\caption{Polar field strength of the aligned dipole component i.e. (m=0, l=1) as a function of Rossby number where asterisks represent $M > 0.4M_{\odot}$ diamond symbols represent $M \le 0.4M_{\odot}$.  Although no distinct transition is observed across the coronal saturation boundary ($Ro = 0.1$), it is clear that stars approaching full-convection (diamonds) have strong polar dipole components.  Note that the outlier in this plot is M0.5 dwarf DT Vir (GJ 494A) at Ro =0.0092. \label{fig.Bpole_Rossby} }
\end{figure}

Fig. \ref {fig.Bpole_Rossby} shows the polar field strength of the dipole component to increase with decreasing Rossby number.  Using this combined with Table~\ref{tab.stellardata}, we find that the most noticeable transition occurs in the stars with mass $\le 0.40M_{\odot}$, where the field strength increases from $\le200G$ to $\ge400G$.  This behaviour follows the results obtained by D08, M08 $\&$ M10 and indicates that the large-scale field has a stronger dipole component for fully convective stars.\\ 

\subsection{Surface Flux and Open Flux}\label{sec.Flux}

In order to investigate the coronal structure of our sample of stars, we determine the variations with Rossby number of both the total magnetic flux through the stellar surface and also the amount of that flux that is open.  The open flux is of particular interest because it channels the stellar wind.  Open field lines are therefore not only dark in X-rays (and hence contribute to any rotational modulation of the X-ray luminosity) but they are also responsible for the angular momentum loss by the stellar wind. 

The ratio of open flux to surface flux is given by
\begin{equation}
\frac{\Phi_{Open}}{\Phi_{Surface}}=\frac{R_{ss}^{2}\int|B_{r}(R_{ss},\theta,\phi)|d\Omega}{R_{*}^{2}\int|B_{r}(R_{*},\theta,\phi)|d\Omega}   ,
\label{eq.field_8}
\end{equation}
where $d\Omega$ is the solid angle.

To analyse the geometry of the field, which is described by the overall combination of \textit{l} and \textit{ m } modes, we compare the predicted total open flux of each star in the sample to that obtained by only single value modes.  By choosing only single modes, we can simplify equation (\ref{eq.field_8}) to

\begin{equation}
\frac{\Phi_{open}}{\Phi_{surface}} = \frac{(2l + 1)(\frac{R_{ss}}{R_*})^{l+1} }{l + (l + 1)(\frac{R_{ss}}{R_*})^{2l + 1}  }       ,
\end{equation}
and compare the predicted open-to-observed surface flux, calculated through the combination of \textit{l} and \textit{ m} modes for each of our stellar sample, to 3 single modes: a dipole (\textit{l=1}), a quadrupole (\textit{l=2}) and an octupole (\textit{l=3}).

\begin{figure}
	\includegraphics[width=84mm]{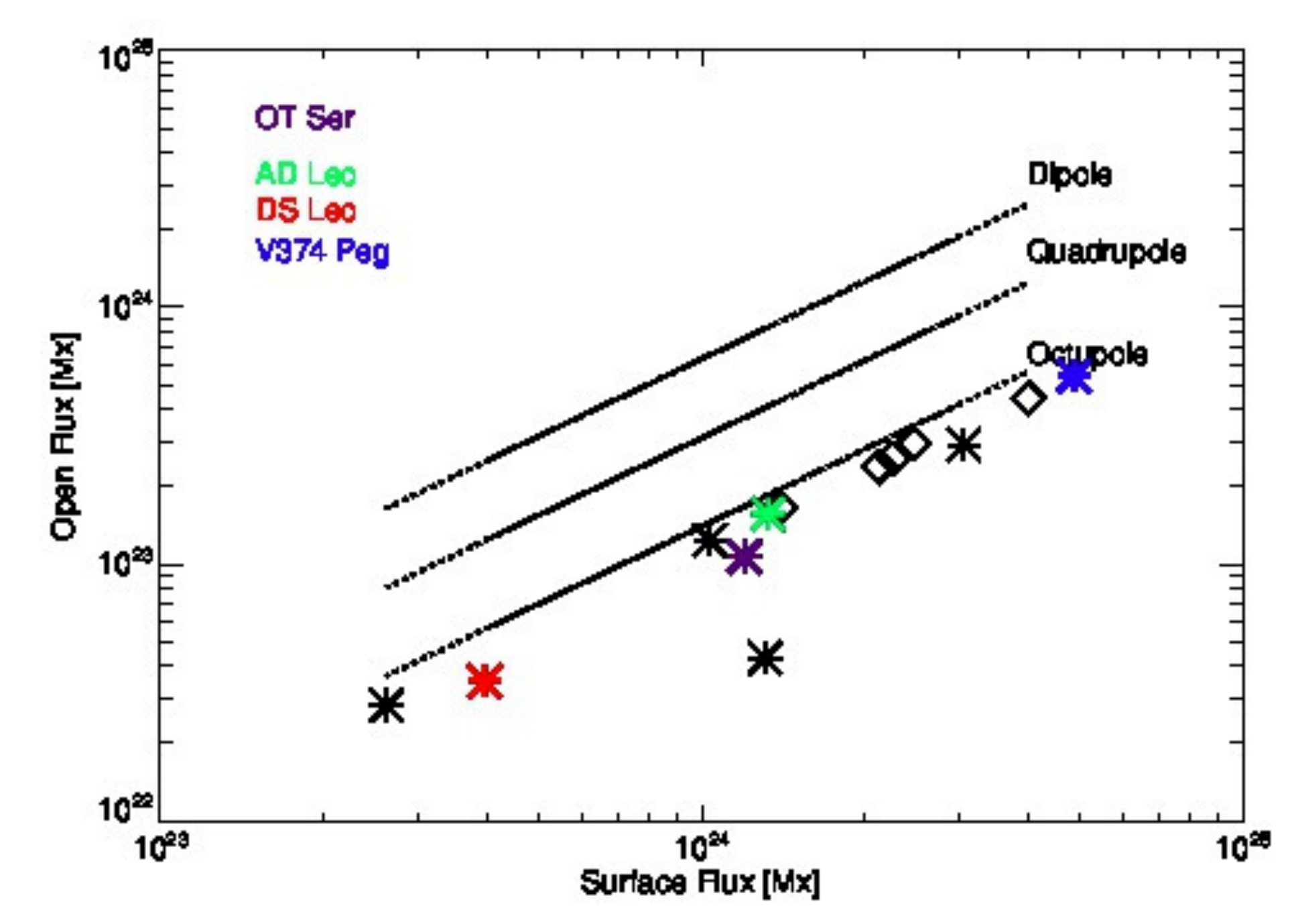}\\
	\includegraphics[width=84mm]{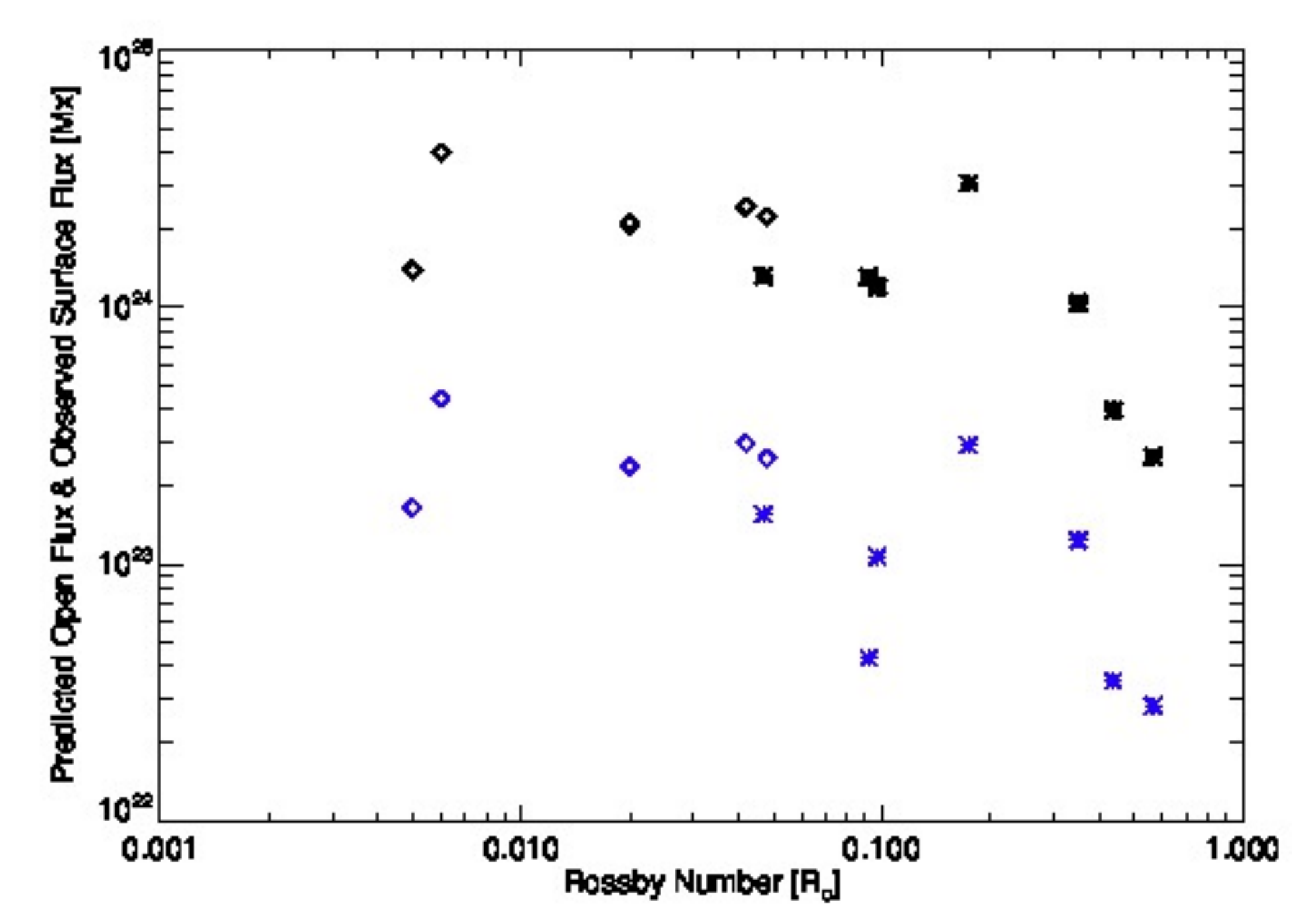}
\caption{ Top: The magnitude of the predicted open flux (where asterisks represent $M > 0.4M_{\odot}$ and diamond symbols represent $M \le 0.4M_{\odot}$) as a function of the observed surface flux is influenced by the higher multipole field components.  Although the open flux tracks the surface flux, which has a large dipole component (\textit{l=1}), the overall combination of \textit{l} and \textit{m} modes is closer to that of an octupole (\textit{l=3}).  Bottom: The open flux (shown in blue) and the surface flux (shown in black) as a function of Rossby number shows both to increase with lower Rossby number.  This follows the same trend as the polar field strength of the dipole component shown in Fig. \ref{fig.Bpole_Rossby} but there is a more pronounced flattening at the onset of coronal saturation i.e. $Ro < 0.1$. Note that again the outlier in this plot is M0.5 dwarf DT Vir (GJ 494A).  The stars shown coloured are discussed in detail in section \ref{sec.symmetries}.
\label{fig.Open_to_total_flux} }
\end{figure}

From Fig. \ref{fig.Open_to_total_flux} (top) we can deduce that although (near) fully-convective objects have increasingly large-scale, low multipole field topologies (D08, M08 $\&$ M10), demonstrated by the extrapolated 3D coronal fields (Fig.\ref{fig.surface_get_1}, \ref{fig.surface_get_2} and \ref{fig.surface_get_3}) and the strong polar field strengths of the dipole component (Fig. \ref{fig.Bpole_Rossby}), we cannot ignore the higher multipole components when modelling magnetic activity, or even the magnetic field itself.  For example, if the dipole component, which has the most open flux associated with it, was assumed to be the only significant magnetic field component, the amount of open flux could be overestimated by at least an order of magnitude.  This would have consequences for, say a Weber-Davis model \citep{Weber_Davis_Model_1967} of angular momentum loss, $\dot{J}$, due to the stellar wind,
\begin{equation}
\dot{J} \propto \Phi_{Open}^{2}    ,
\label{eq.WeberDavis}
\end{equation}
which could overestimate the angular momentum loss by 2 orders of magnitude.  We note here that the magnitude of the open flux is heavily dependent on the chosen value for the source surface as $R_{ss} =\rightarrow \infty$, $\Phi_{Open}\rightarrow0$. 

Fig. \ref{fig.Open_to_total_flux} (bottom) demonstrates that, as with the polar field strength (Fig. \ref{fig.Bpole_Rossby}), the amount of surface and therefore open flux increases with inverse Rossby number, albeit with a more notable flattening at the fully convective boundary.  This indicates that the lower mass stars (diamonds) exhibit a significant fraction of open flux [$10^{24}$Mx] and supports the idea that the large-scale dipole component is having an effect on the magnetic activity.  Although it is tempting to extrapolate from this result that, according to a Weber-Davis model \citep{Weber_Davis_Model_1967}, the increased flux on these stars should cause a high rate of angular momentum loss, due to the stellar wind (Eq. \ref{eq.WeberDavis}), we lack an appropriate stellar wind model that would enable us to calculate this loss and make any definite conclusions.  Furthermore, without stellar ages for our sample, it is difficult to determine whether these stars follow an age-rotation law e.g. (\ref{eq.age-rotation}) or a mass dependent rotational braking law (Eqn.\ref{eq.rotational-braking}), i.e. have high rotation rates because they are young.  

\citet{Vidotto_V374Peg_2011} modelled the angular momentum loss of the M4 dwarf V374 Peg through its stellar wind, and in conjunction with its rapid rotation rate, V374 Peg exhibits a high angular momentum loss, suggesting it is still young.  However, for comparison purposes, an approximate stellar age for V374 Peg has not yet been determined.

\section{Modelling the Emission}\label{sec.EmissionModel}

\subsection{The X-ray Emission Model}\label{sec.XrayModel}

Once the global magnetic field is determined, the gas pressure and density structure of the stellar corona can be estimated.   We assume the plasma is hydrostatic and isothermal ($T = 2\times10^{6}K$; \citet{Giampapa_temps_1996}) and that the gas pressure at the stellar surface is proportional to the magnetic pressure ($p_{o}\propto B_{o}^{2}$).  The pressure therefore varies along each field line according to

\begin{equation}
p=p_{o}e^{\int{\frac{\underline g \cdot \underline B    ds}{|B|}}}           , 
\label{eq.pressure_1}
\end{equation}
as described by \citet{Jardine_XrayCorona_2002} and \citet{Gregory_XrayRotMod_2006}.\\
Expanding the component of gravity along the field line ($\underline g \cdot \underline B$), we have 

\begin{equation}
p=\frac{\kappa B_{o}^{2}}{2\mu} exp\left[\int {\frac{\left(\frac{-\phi_{g}}{r^{2}} + \phi_{c} r \sin^{2}\theta \right)B_r + \left(\phi_{c} r \sin\theta \cos\theta \right)B_{\theta}} {\sqrt{B_{r}^{2}+B_{\theta}^{2}+B_{\phi}^{2}}}}\right]     ,
\label{eq.pressure}
\end{equation}
where $\kappa$ is a constant of proportionality and $B_{r}, B_{\theta}, B_{\phi}$ are the radial, meridional and azimuthal components of the magnetic field, respectively.  Finally, the ratios of centrifugal-thermal energy ($\phi_{c}$) and gravitational-thermal energy ($\phi_{g}$) are given by

\begin{equation}
\phi_{c}=m_{e} \left( \frac{(\Omega R_{*})^{2}}{k_{B} T} \right)
\label{eq.cent-thermal}
\end{equation}
\begin{equation}
\phi_{g}=m_{e} \left( \frac{G M_{*}}{R_{*}k_{B} T} \right)
\label{eq.grav-thermal}
\end{equation} 
where $\Omega$ is the stellar rotation rate, $k_{B}$ is the Boltzmann constant, G is the gravitational constant and $m_{e}$ is the electron mass.

To ensure that only regions of the closed stellar corona contribute towards the emission measure, the gas pressure along open field lines is taken to be zero.  In addition to this, if there is any over-pressure along the designated closed loops ($\beta\ge 1.0$), then the pressure at that grid point is also set to zero.  Here $\beta$ is the ratio of gas pressure ($p=2n_{e}kT$) to magnetic pressure ($p_{B}=B^{2}/2\mu$). 

\begin{figure}
	\includegraphics[width=84mm]{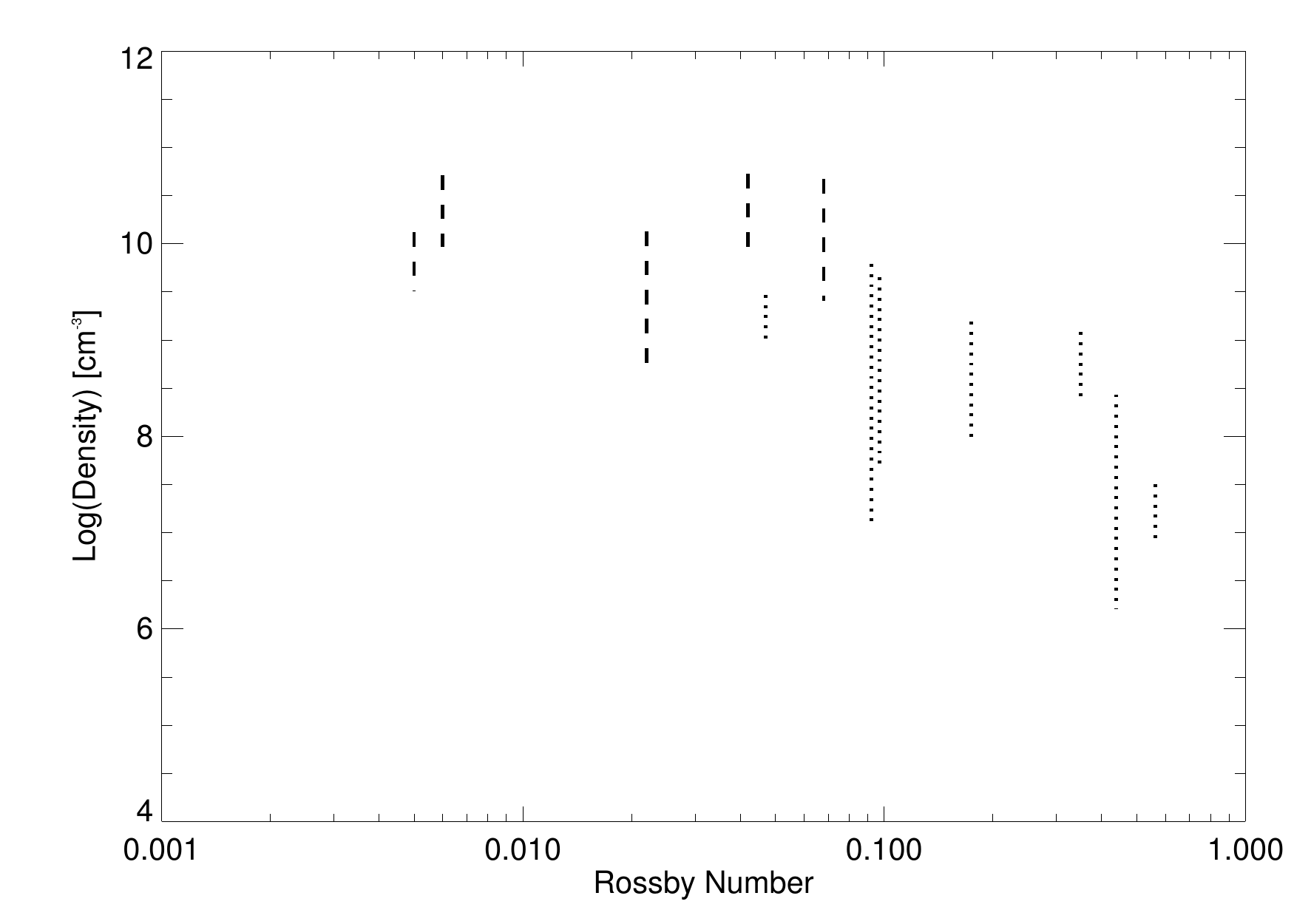}\\
	\includegraphics[width=84mm]{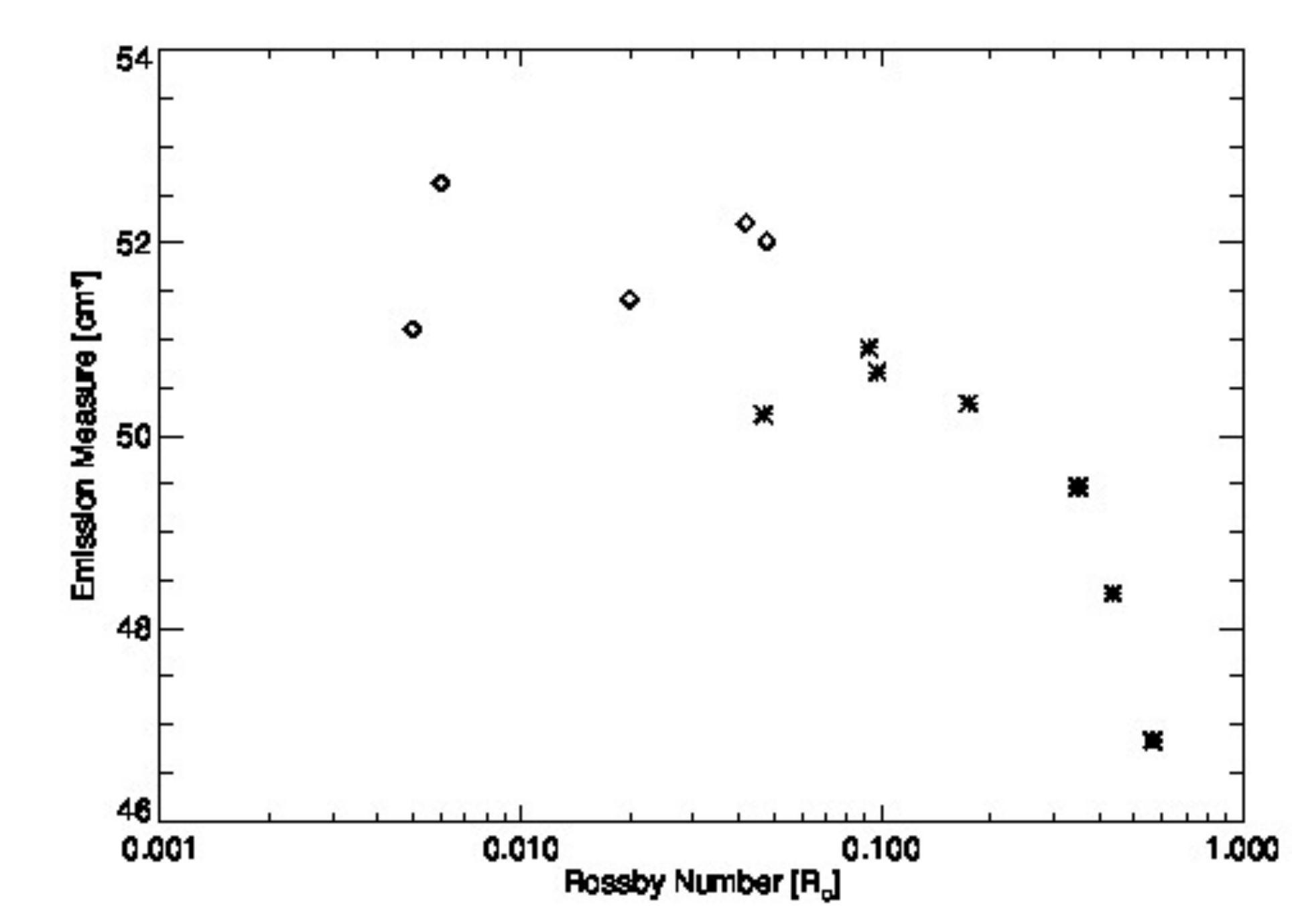}
	\caption{ Top: The maximum and minimum coronal density occuring on each the stars in the sample (where the long- dashed line represents $M \le 0.4M_{\odot}$ and the short-dashed line represents $M > 0.4M_{\odot}$.  Note the spread in the density values calculated for each star can extend over many orders of magnitude but represent different values for $(r,\theta,\phi)$.  Bottom: Emission Measure as a function of Rossby number (where asterisks represent $M > 0.4M_{\odot}$ and diamond symbols represent $M \le 0.4M_{\odot}$) .  This demonstrates the trend observed as the boundary of coronal saturation is approached i.e. $Ro < 0.1$.   Note the short dashes are the partly convective stars and the long dashes are the fully convective stars.\label{fig.Density_Rossby} }
\end{figure}

\begin{figure}
	\includegraphics[width=84mm]{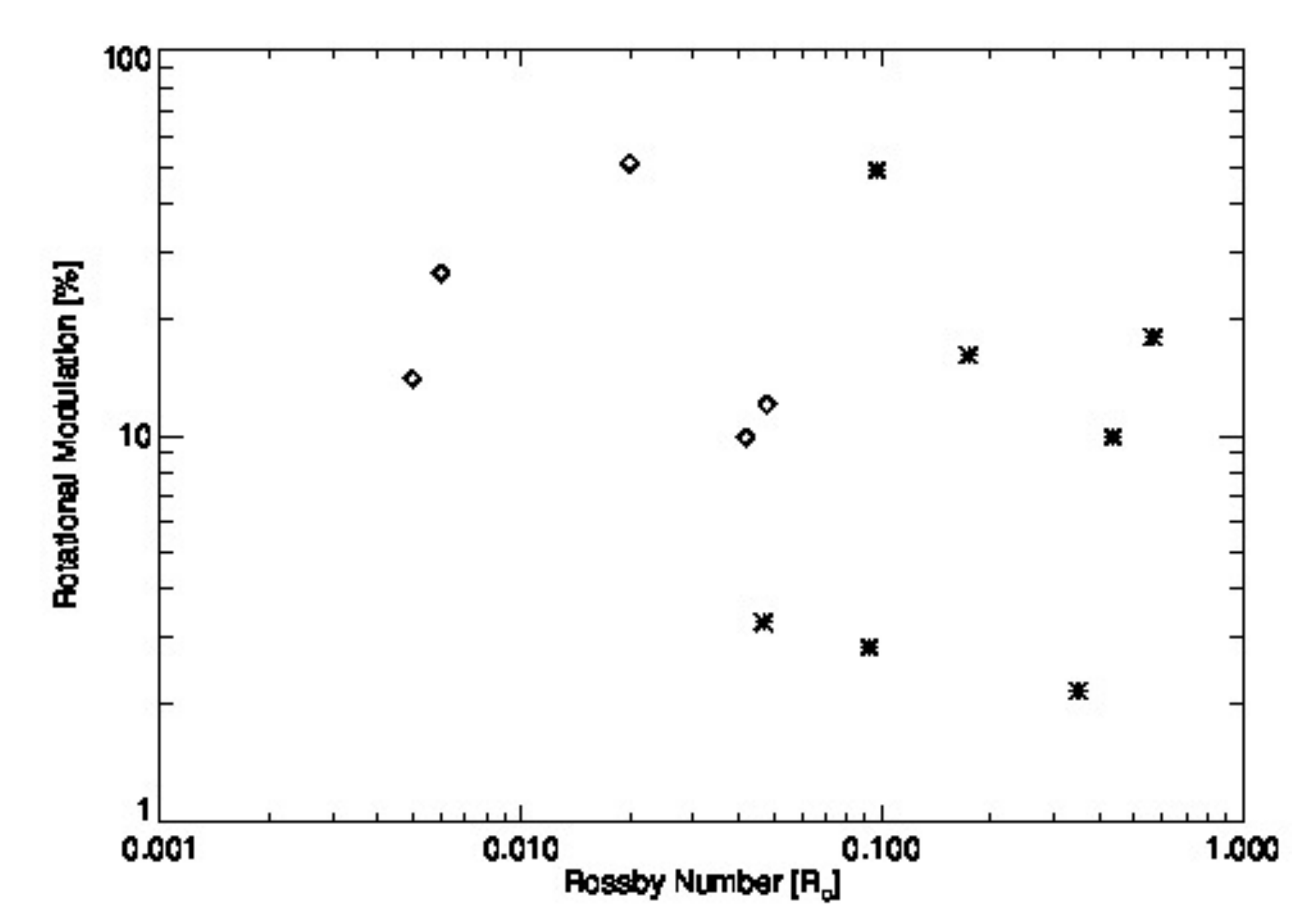}
	\caption{ The rotational modulation of the X-ray emission as a function of Rossby number (where asterisks represent $M > 0.4M_{\odot}$ diamond symbols represent $M \le 0.4M_{\odot}$) depicts the variation in emission measure during a full rotation of the star.  Even at low Rossby number, there is still enough structure present in the magnetic field to give significant rotational modulation. \label{fig.Rotmod_rossby} }
\end{figure}

The X-ray emission measure is then determined assuming the gas is optically thin, i.e.
\begin{equation}
EM(r) = \int n_e^{2}   dV       ,
\label{eq.EmissionMeasure2}
\end{equation}
and we consider both its magnitude (Fig. \ref{fig.Density_Rossby}) and rotational modulation i.e. variation in emission measure (Fig. \ref {fig.Rotmod_rossby}).

The high levels of rotational modulation indicate that even on stars where the magnetic field is mainly dipolar, i.e. low Rossby number, there is still enough structure present in the magnetic field to give significant rotational modulation.  Of course, we must also consider the large variation in stellar inclination angle and angle between the magnetic axis and rotation axis, i.e. $\beta_{M}$ (see Table~\ref{tab.stellardata}), when analysing the modulation to determine whether it is a good indicator of magnetic structure.  From Fig. (\ref{fig.surface_get_1},\ref{fig.surface_get_2} and \ref{fig.surface_get_3}) and Table 1, we can see that the magnetic axis and rotation axis are mis-aligned by $ \beta_{M} > 10^{\circ}$ for the vast majority of the sample and this, coupled with the stellar inclination angles, could account for the large variation between rotational modulation values within the sample.  

The free parameters in this model are the temperature (fixed at $T = 2\times10^{6}K$), the source surface (fixed at $R_{ss}=2.5R_{*}$) and the constant of proportionality $\kappa$ relating the base gas to magnetic pressure.  Changing $\kappa$ essentially scales the global plasma pressure up or down and we use observations of coronal density to fix $\kappa$. \citet{Ness_Density_ADLeo_2002,Ness_Density_2004} estimates coronal densities on the dM stars to be within the range of $10^{9}-10^{12}cm^{-3}$.  Fig. ~\ref{fig.Density_Rossby} (upper) shows the maximum and minimum density occurring in the closed corona of each sample star.  There is a great spread in the range of maximum-to-minimum densities calculated for each star, with some extending over as much as 4 orders of magnitude within the closed corona.  However, we note that these density ranges represent different $(r,\theta,\phi)$ values for each star.  The average coronal density calculated in our model for each star, which can be found in Table~\ref{tab.stellardata}, are estimated to lie within the range $10^{7}-10^{11}cm^{-3}$, representing early-dM to mid-dM respectively.  We note the former would appear to encompass densities comparable to solar values.

\section{Extrapolation Assumptions}\label{sec.symmetries}

\begin{figure}
	\begin{center}
	\subfloat(a) Antisymmetric{\label{fig.edge-a}\includegraphics[width=54mm]{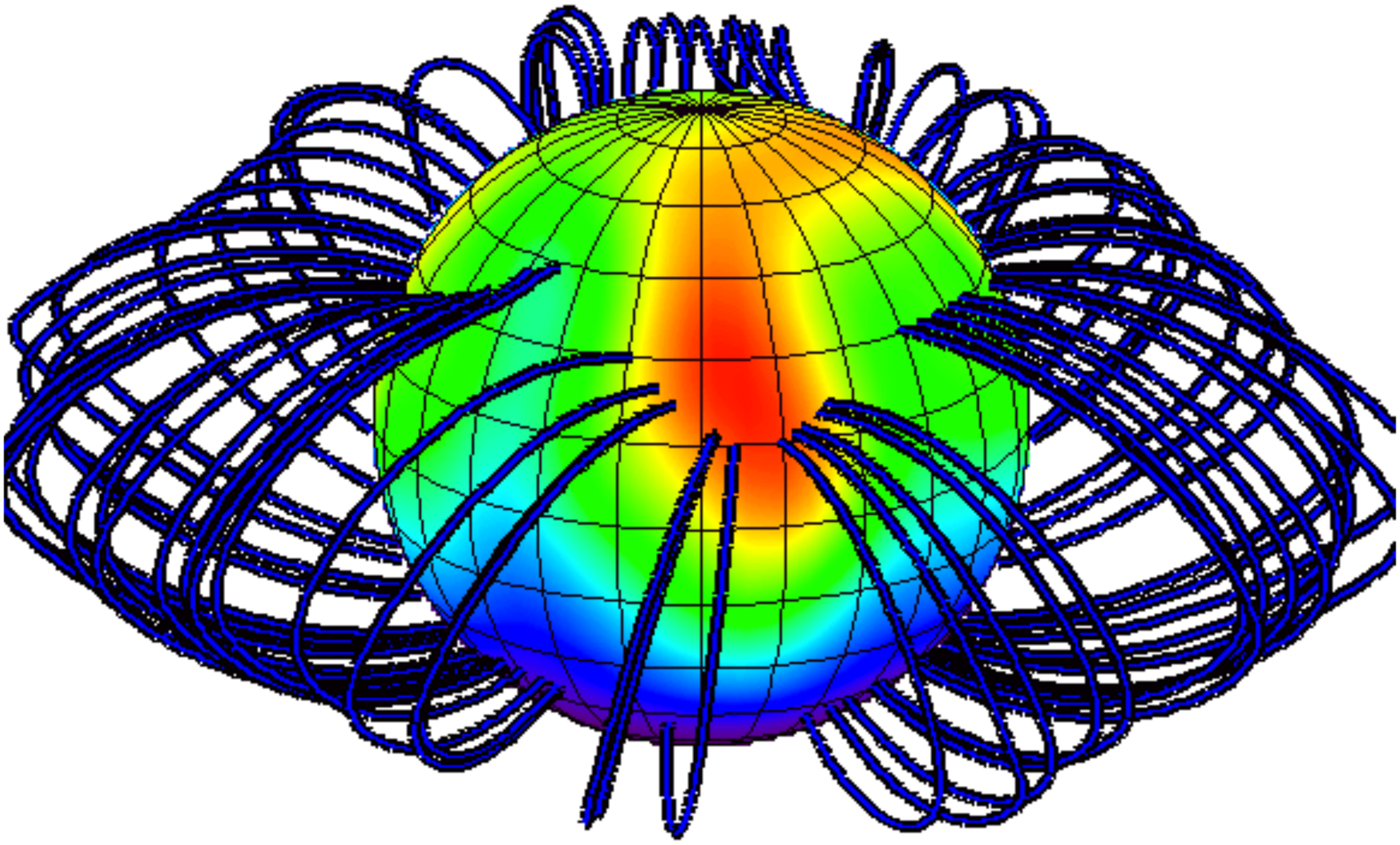}}\\
	\subfloat(b) Symmetric{\label{fig.edge-b}\includegraphics[width=54mm]{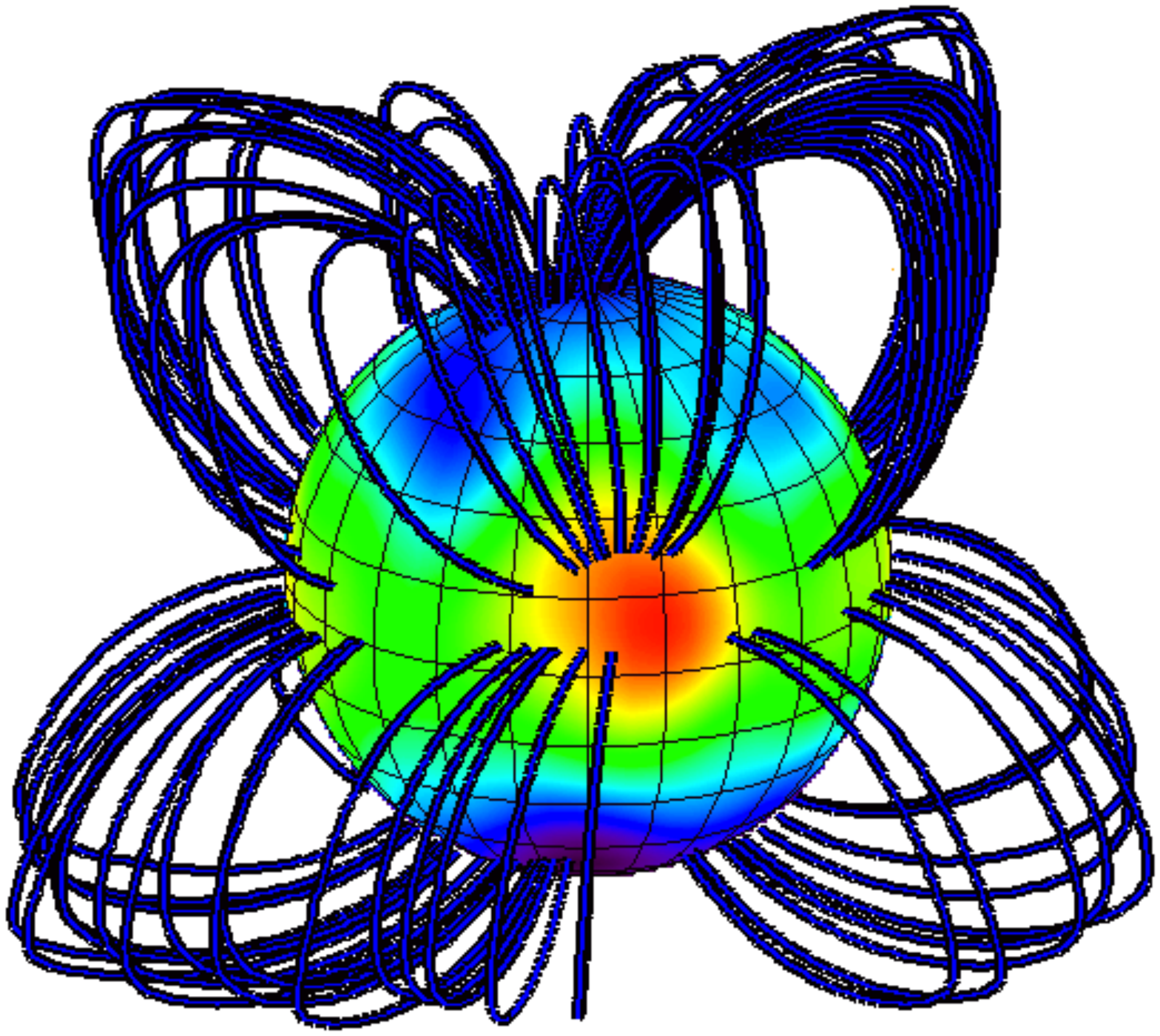}}\\
	\subfloat(c) Unconstrained{\label{fig.edge-c}	\includegraphics[width=54mm]{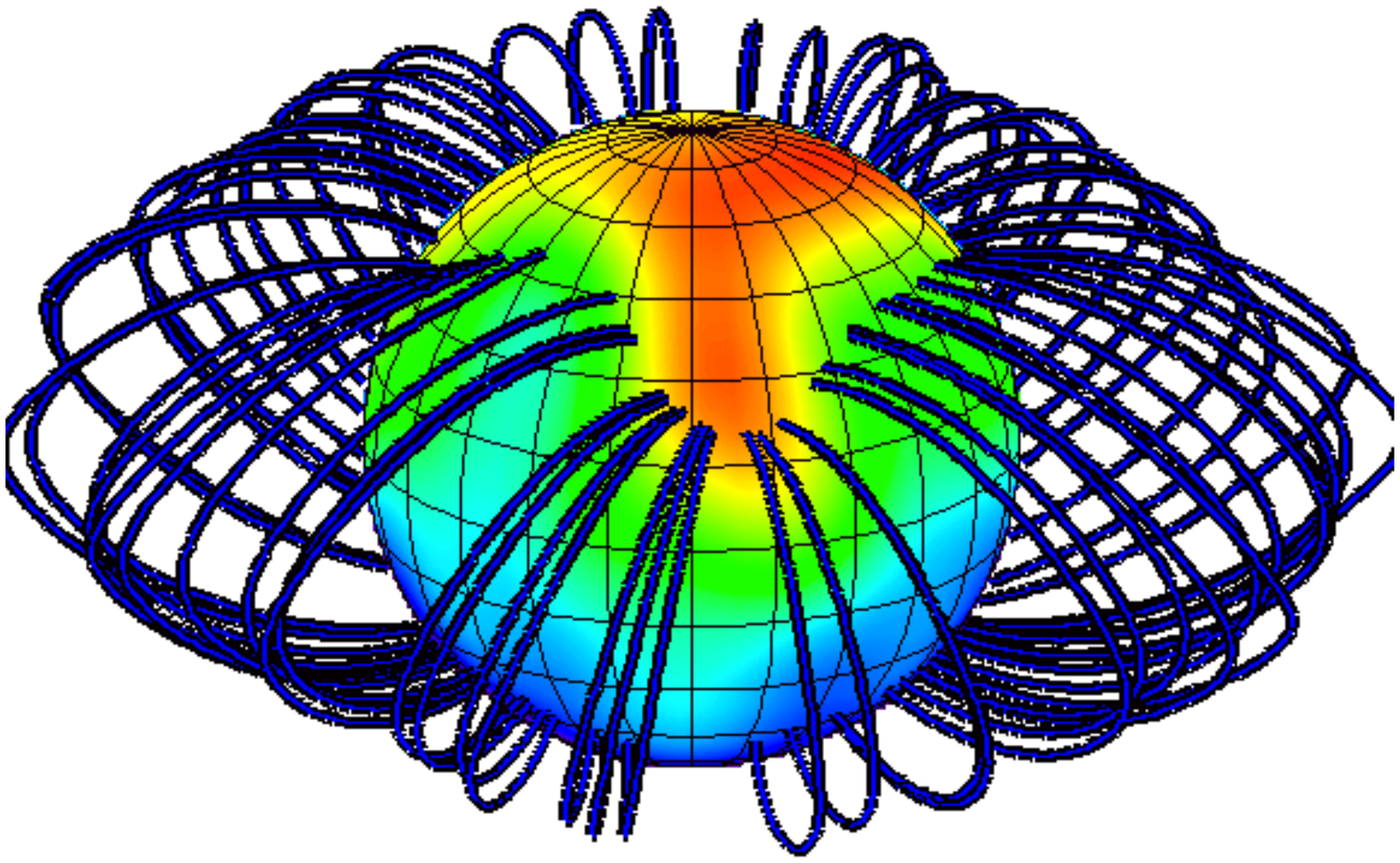}}\\
	\caption{Coronal extrapolation of the (a) antisymmetric, (b) symmetric, and (c) unconstrained ZDI solutions of the coronal extrapolation for V374 Peg.  Note that only the closed field lines are shown. \label{fig.Symmetries_Extrapolation} }
	\end{center}
\end{figure}

In the reconstruction of the surface magnetic field it is possible to push the solution towards field geometries that are either symmetric (e.g. a quadrupole) or antisymmetric (e.g. a dipole) about the centre of the star (see for example Fig.\ref{fig.Symmetries_Extrapolation}).  The magnetic modelling is described in detail in D08.  The principle of ZDI requires a way to choose between the many field configurations that fit the data equally well.  The imaging process is based on the principles of maximum entropy image reconstruction and in the absence of any prior information on the field topology, an entropy form that selects the solution with the lowest energy content and the lowest degree modes, is generally used.  Should we have a good physical reason to think that the field is predominantly symmetric or antisymmetric, we could impose such a constraint with a modified entropy.  In comparison to the unconstrained solution, magnetic energy is higher (and entropy lower) for both the symmetric and anti-symmetric cases because the magnetic field is on smaller spatial scales.  We remind the reader that the antisymmetric solution is not only dipolar. The ZDI solution with the lowest energy (weighted by the order of the spherical harmonics) that fits the data is selected. When one forces antisymmetry, the number of degree of freedoms in the model is decreased, and hence, if the unconstrained solution was not purely antisymmetric, the antisymmetric solution has a lower entropy/higher energy.  We investigate the effect of this choice on coronal extrapolations (and quantities derived hereafter).  In particular, we determine if the X-ray emission measure is sensitive to the nature of the constraint since comparison with the observed emission measure may then help to discriminate between choices in the way that for T Tauri stars, the location of accretion spots does \citep{Donati_AccretionV212oph_2011}.  

We explore the assumptions made for surface field images of 4 stars: M0, DS Leo; M1.5, OT Ser; M3 AD Leo; and M4 V374 Peg, which encompass high, mid and low inclination angles, as well as a range of stellar masses.  Figure \ref{fig.Symmetries_Extrapolation} shows the coronal extrapolation for the 3 cases: anti-symmetric, symmetric and unconstrained, for V374 Peg.  
We note here a few remarks on the reconstruction of the individual stars.  For AD Leo, which has a low inclination angle, $70^{\circ}$, a predominantly dipolar field is reconstructed with ZDI.  However, for the field in the invisible hemisphere instead of a clear pole of strong magnetic field, it is covered with field of average intensity with polarity opposite to that of the visible pole.  As a result the field lines from the visible pole connect on the whole hidden hemisphere whereas for a pole-like model they would connect on the well-defined opposite pole.  By imposing an antisymmetric field this effect is corrected for: the resulting ZDI map features a strong polar spot in the invisible hemisphere with opposite polarity with respect to the visible pole.  For DS Leo, the inclination angle is intermediate and the map does not feature any strong spot on the partly hidden hemisphere.  It is likely in this case, as for AD Leo, that we lack information and do not know what is on the less-visible hemisphere.  Imposing antisymmetry indeed results in spots of high magnetic field on this hemisphere, and also modifies more importantly the map in the visible hemisphere.  For a star with a well constrained dataset and high inclination angle like V374 Peg, the map is already well constrained and features an overall dipolar structure with a strong polar spot on the hemisphere orientated away from the observer.

\begin{figure}
	\includegraphics[width=84mm]{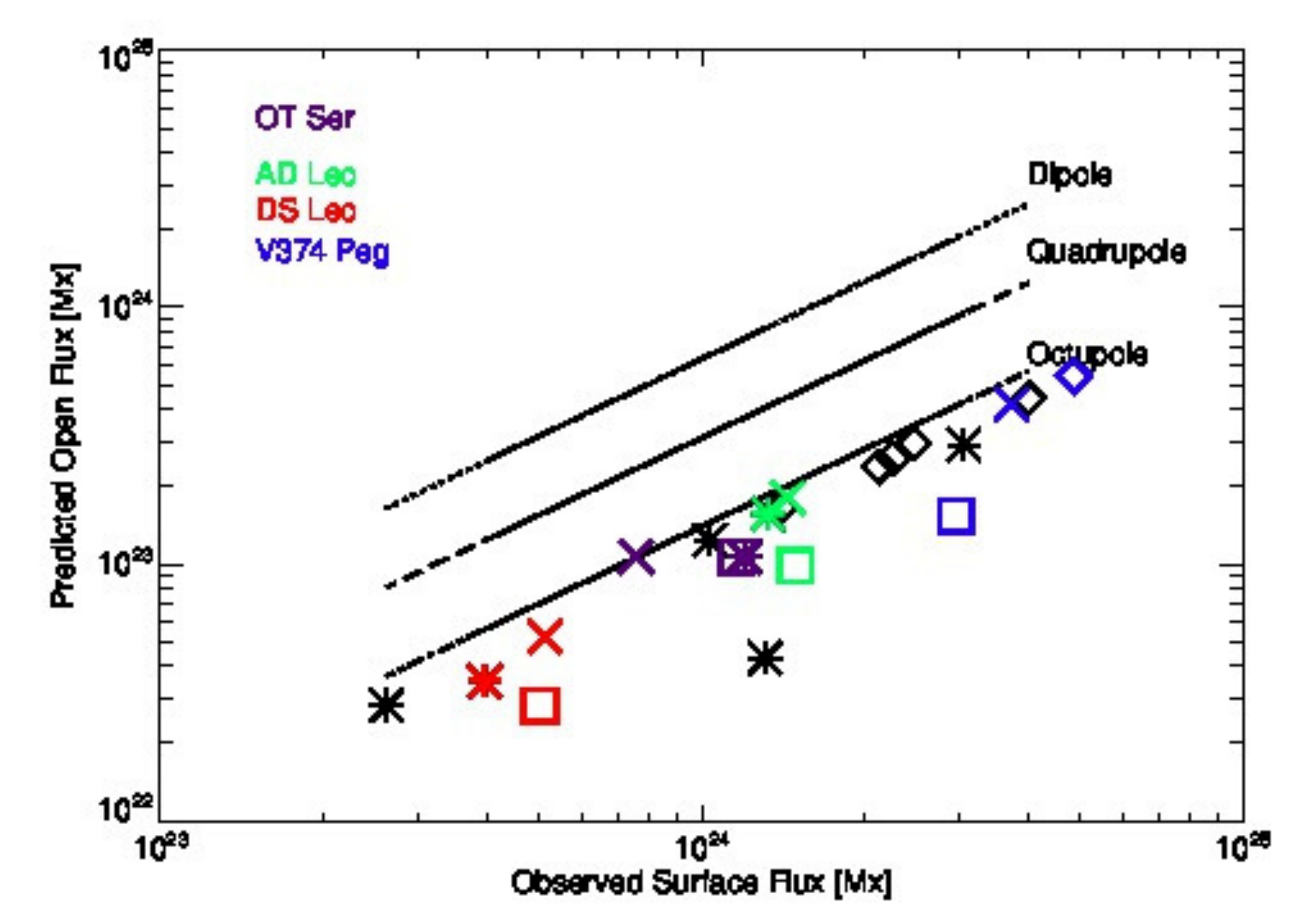}
	\caption{The variation in the open flux as a function of the surface flux when different configurations of the magnetic field are assumed.  The solutions are represented as follows: antisymmetric, crosses; symmetric, square.  The unconstrained solutions are represented by asterisks if $M > 0.4M_{\odot}$and  diamond symbols if $M \le 0.4M_{\odot}$).  Artificially imposing a symmetric or antisymmetric field increases the reconstructed magnetic energy and hence both the observed surface flux and predicted open flux.  The symmetric solution (squares) has the lower open flux, which is to be expected since imposing a symmetric field shifts magnetic energy from order \textit{l} to (at least) \textit{l+1}. \label{fig.Symmetries_Open_Flux_Surface_Flux} }
\end{figure}

Whether the unconstrained or antisymmetric solution is forced we find a similar result in terms of the predicted open flux (Fig. \ref{fig.Symmetries_Open_Flux_Surface_Flux}) and the magnitude of X-ray emission measure.  Imposing a symmetric field, one artificially imposes smaller spatial scales e.g. \textit{$l_{min}=2$}, i.e. the lowest degree mode available becomes the quadrupole, which shifts magnetic energy from order \textit{l} to (at least) \textit{l+1} and automatically decreases the amount of open flux.  This can be considered as a test on a star for which a good dataset is available, e.g. V374 Peg.  We note that even in the cases of forcing a solution (either symmetric or antisymmetric), the relation between the open flux and surface flux is still followed.

\begin{figure}
	\includegraphics[width=84mm]{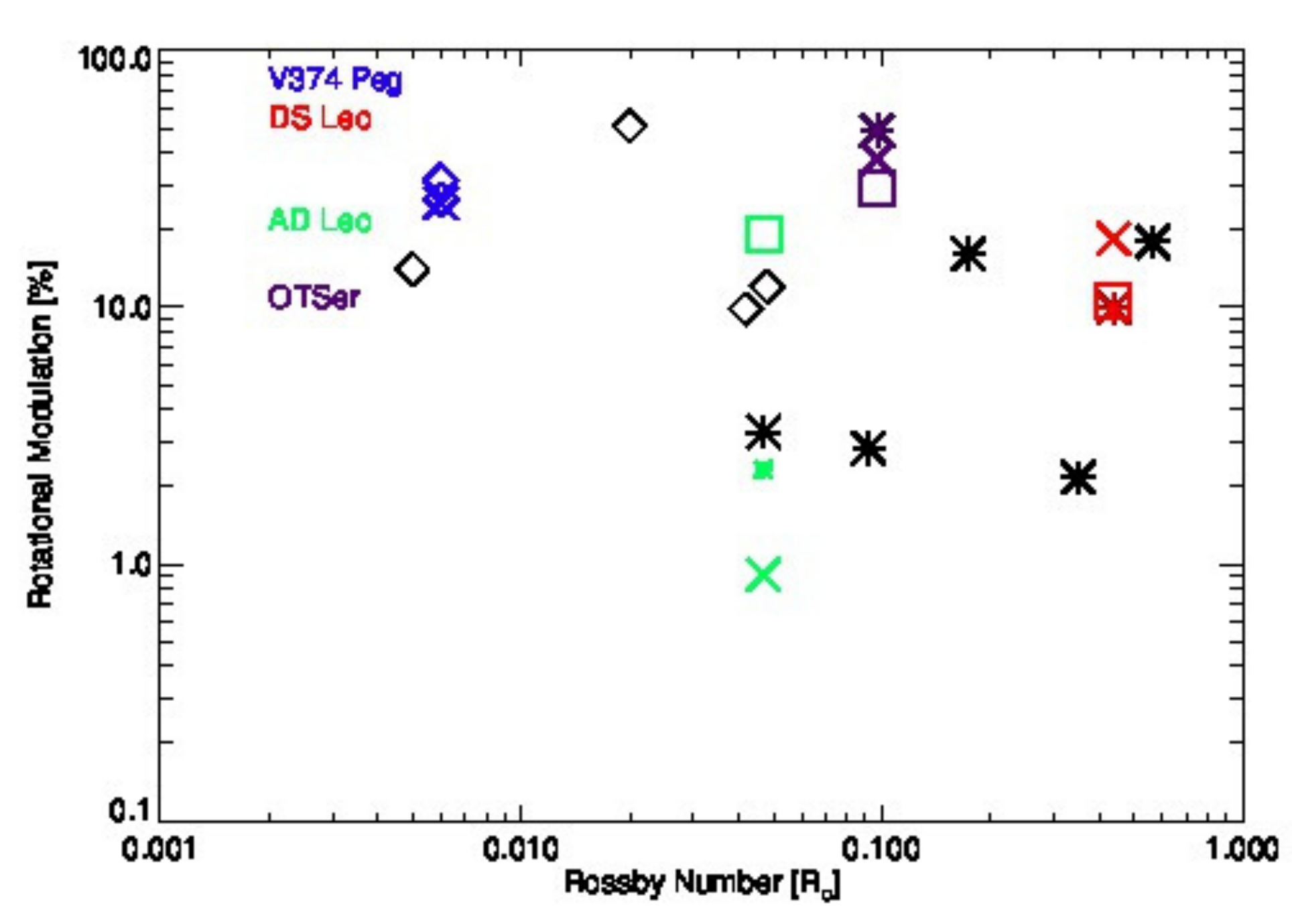}
		\caption{ The effect of imposing different magnetic field symmetries on the X-ray rotational modulation.  The solutions are represented as follows; antisymmetric, crosses; symmetric, square.  The unconstrained solutions are represented by asterisks if $M > 0.4M_{\odot}$and  diamond symbols if $M \le 0.4M_{\odot}$).  Although we have shown that rotational modulation is not a good indicator of field structure when considering a large sample of stars with different inclination, etc, it could be useful when considering a single object, e.g. checking the symmetry and providing independent confirmation of the  predominant mode. \label{fig.Symmetries_Rot_Mod} }
\end{figure}

The effect of imposing different solutions has very little or no effect on the X-ray Emission Measures or coronal densities calculated.  We do however note a change in the X-ray rotational modulation between solutions (Fig. \ref{fig.Symmetries_Rot_Mod}).  An observation of the X-ray rotational modulation (such as that described in \citet{Hussain_RotMod1_2004,Hussain_RotMod2_2005}, may discriminate between fields that are predominantly symmetric or antisymmetric.  As we discussed in Section \ref{sec.EmissionModel}, this depends on the inclination of the magnetic pole and the inclination of the star itself.  For some stars e.g. V374 Peg, this is not so useful, whereas for other like AD Leo there is a large variation in the rotational modulation making it a good candidate.

\section{Summary and conclusions}\label{sec.Summary}

We have used reconstructed maps of the radial magnetic field at the stellar surface for a sample of early-to-mid M dwarfs to extrapolate their 3D coronal magnetic field (using the PFSS method).  We have investigated the topology of the large-scale magnetic field at the stellar surface and the structure of the extrapolated 3D corona. By assuming a hydrostatic, isothermal corona, we have modelled the density structure within the corona and hence determined the X-ray emission measure. 

We have focussed in particular on variations with Rossby number. We find the following:

1. As the Rossby number decreases, the polar field strength of the dipole component of the field increases and then appears to saturate. Stars with low Rossby numbers have strong, mainly dipolar fields.

2. A similar variation with Rossby number is seen in both the total (unsigned) surface magnetic flux and the flux of open field (which can carry the stellar wind). The role of the topology of the large-scale field is apparent when we calculate the magnitude of the open flux. This is significantly less than would be predicted if all of the surface magnetic flux were contained in a purely dipole field. The contribution of the higher multipoles therefore reduces the open flux and may also significantly influence the angular momentum loss rate, which for a Weber-Davies model scales as the square of the open flux. Both the strength and also the topology of the large-scale field are therefore important in angular momentum loss.

3. As is observed, a rise and then saturation of the X-ray emission measure with decreasing Rossby number is also found. The stellar coronae are compact, with most of the emission originating from regions below approximately 1.5 stellar radii. Our sample contains a large range of both the inclinations of the stellar rotation axis and also the tilt of the magnetic axis. As a result, there is a large spread in the values of the rotational modulation of the X-ray emission and no clear trend with Rossby number can be detected.

For low-mass stars, the observed variation in X-ray emission with Rossby number results naturally from the observed variation in the surface magnetic field. We note that we choose the parameter $\kappa$ that scales the surface pressure in such a way that we reproduce typical X-ray fluxes and we do not model the ionisation fraction sometimes invoked in the atmospheres of later spectral types. While there is a range of magnetic topologies within our sample, the spread of values for the rotational modulation of the X-ray emission is too great for it to be a useful indicator of the field structure.  Although not a good indicator of field structure when considering a large sample of stars with different inclination, it could be useful when considering a single object.  For example, checking the symmetry and providing independent confirmation of the predominant mode (e.g. dipole versus quadrupole), or observing a magnetic cycle with a change from predominantly quadrupolar to predominantly dipolar as on the Sun \citep{Sanderson_Sun_2003}.

We find that both the strength of the field and its geometry, however, affect the magnetic flux that is open (wind-bearing) and which therefore allows the star to lose mass and angular momentum. The magnitude of this open flux is significantly reduced by departures from a purely dipolar field. This suggests that simple scalings for angular momentum losses based on dipolar field geometries may not be sufficient to explain the angular momentum evolution of low mass stars. The high values of open flux for stars with the lowest Rossby numbers may indicate that they have significant angular momentum loss rates.

\begin{table*}
\caption{Data for stellar sample of early-to-mid M dwarfs.  Mass, radius, rotation period, inclination and, where available, $B_{V}$, the average large-scale magnetic field derived from spectropolarmetric measurements are provided by \citet{Donati_EarlyM_2008,Morin_MidM_2008}.  $B_{I}$, is the average magnteic field (i.e. small + large-scale field) derived from unpolarised spectroscopy, supplied by \textit{a)} \citet{Reiners_Basri_MagneticTopology_2009}, \textit{b)} \citet{Saar_Recent_1996}, \textit{c)} \citet{ReinersBasri_FirstDirect_2007}, \textit{d)} \citet{JohnsKrull_Valenti_2000}. Rossby number are from \citet{Donati_EarlyM_2008,Morin_MidM_2008} and were computed from empirical $\tau_{c}$ suited to the stellar mass from \citet{Kiraga_Stepien_MDwarfs_2007}.  $\beta$, the estimated angle between the rotation and magnetic axis, are from this paper, along with the predicted values for emission measure (both magnitude and rotational modulation) and coronal density.   \label{tab.stellardata}}
\centering
\begin{tabular}{cccccccccccccc}
\hline
Star & Sp Type & Mass ($M_\odot$) & Radius ($R_\odot$) & P (days) &Ro($10^{-2}$)  &i($^{\circ}$) &$\beta_{M}$ ($^{\circ}$) & $B_{V}$ (kG) &$B_{I}$ (kG)&LogEM ($cm^{-3}$)&Rot Mod &Log$\overline n_{e}$ ($cm^{-3}$)
\\
\hline
\hline
GJ 182 & M0.5 & 0.75 & 0.82 & 4.35 & 7.44  & 60 & 41.1 & 0.172 & $2.5^{a}$ & 50.33&12.18&8.62\\
DT Vir & M0.5 & 0.59 & 0.53 & 2.85 & 9.2  & 60 & 83.6 & 0.149 & $3.0^{b}$ &50.91&2.82& 9.26\\
DS Leo & M0 & 0.58 & 0.52 & 14.0 & 43.8  & 60 & 41.1 & 0.087 & - & 48.36&9.93&7.98\\
GJ 49 & M1.5 & 0.57 & 0.51 & 18.6 & 56.4  & 45 & 10.9 & 0.027& - & 46.83&18.02&7.05\\
OT Ser & M1.5 & 0.55 & 0.49 & 3.40 & 9.7  & 45 & 12.1 & 0.123 & - &50.66& 48.83&9.22\\
CE Boo & M2.5 & 0.48 & 0.43 & 14.7 & 35.0  & 45 & 7.4 & 0.10 & $1.8^{a}$ & 49.47&2.17&8.52\\
AD Leo & M3 &0.42 & 0.38 &2.3399 & 4.7 & 20 & 4.5& 0.19& $2.9^{c}$&50.22&3.27&8.95\\
EQ Peg A & M3.5 & 0.39 & 0.35 & 1.061 & 2.0  & 60 & 25.9 & - & - & 51.42&51.02&9.66\\
EV Lac & M3.5 & 0.32 & 0.30 & 4.3715 & 6.8  & 60 & 45.8 & 0.53 & $3.9^{d}$ &52.02&12.05&10.14\\
YZ CMi & M4.5 & 0.31 & 0.29 & 2.7758 & 4.2  & 60 & 24.8 & 0.50 & $\ge 3.9^{c}$ &50.21&9.89& 10.27\\
V374 Peg & M4 & 0.28 & 0.32 & 0.44565 & 0.6  & 70 & 9.3 & - & - &52.62&26.49&10.26\\
EQ Peg B & M4.5 & 0.25 & 0.25 & 0.404 & 0.5  & 60 & 6.5 & -& - & 51.11&14.05&9.64\\
\\
\hline
\end{tabular}
\end{table*}

\section*{Acknowledgements}
The authors would like to thank Aad Van Ballegooijen for the use of this global diffusion model for extrapolating the coronal 3D magnetic fields.  PL acknowledges support from an STFC studentship.  JM acknowledges support from a fellowship of the Alexander von Humboldt foundation.  AV acknowledges support from a fellowship of the Royal Astronomical Society.

\bibliography{MDwarfs.bib}
\bsp
\label{lastpage}
\end{document}